\pdfoutput=1
\documentclass[twocolumn]{aastex6}
\usepackage{amsmath}
\usepackage{natbib}
\usepackage{upgreek}
\usepackage{cases}
\usepackage{url}
\usepackage{graphicx}
\usepackage[toc,page]{appendix}
\usepackage{subcaption}
\hypersetup{linkcolor=red,citecolor=blue,filecolor=cyan,urlcolor=magenta}

\begin{document}

\title{Dust Growth and Dynamics in Protoplanetary Nebulae: Implications for Opacity, Thermal Profile and Gravitational Instability}

\author{Debanjan Sengupta\altaffilmark{1,2}}
\author{Sarah E. Dodson-Robinson\altaffilmark{1,3}}
\author{Yasuhiro Hasegawa\altaffilmark{2}}
\author{Neal J. Turner\altaffilmark{2}}

\altaffiltext{1}{Department of Physics \& Astronomy, University of Delaware, Newark, DE 19716 USA}
\altaffiltext{2}{Jet Propulsion Laboratory, California Institute of Technology, 4800 Oak Grove Drive, Pasadena, CA 91109}
\altaffiltext{3}{Bartol Research Institute}

\begin{abstract}

In spite of making a small contribution to total protoplanetary disk mass, dust affects the disk temperature by controlling absorption of starlight. As grains grow from their initial ISM-like size distribution, settling depletes the disk's  upper layers of dust and decreases the optical depth, cooling the interior. Here we investigate the effect of collisional growth of dust grains and their dynamics on the thermal and optical profile of the disk, and explore the possibility that cooling induced by grain growth and settling could lead to gravitational instability. We develop a Monte Carlo dust collision model with a weighting technique and allow particles to collisionally evolve through sticking and fragmentation, along with vertical settling and turbulent mixing. We explore two disk models, the MMEN (minimum-mass extrasolar nebula), and a ``heavy'' disk with higher surface density than the MMEN, and perform simulations for both constant and spatially variable turbulence efficiency profile $\alpha(R,z)$.  We then calculate mean wavelength-dependent opacities for the evolving disks and perform radiative transfer to calculate the temperature profile $T(R,z)$. Finally, we calculate the Toomre Q parameter, a measure of the disk's stability against self-gravity, for each disk model after it reaches a steady state dust-size distribution. We find that even weak turbulence can keep sub-micron sized particles stirred in the disk's upper layer, affecting its optical and thermal profiles, and the growth of large particles in the midplane can make a massive disk optically thick at millimeter wavelengths, making it difficult to calculate the surface density of dust available for planet formation in the inner disk. Also, for an initially massive disk, grain settling and growth can produce a drop in the Toomre Q parameter, driving the disk to $Q < 1.4$ and possibly triggering spiral instabilities.  
\end{abstract}

\section{Introduction}\label{sec:intro}

While most planets form ``bottom-up'' from dust particles accumulating into pebbles, planetesimals, and then solid cores \citep{lissauer93, pollack96, morbidelli12}, some massive giant planets and brown dwarfs may form by top-down collapse in fragmenting protostellar disks \citep{kratter16, boss97}. Despite inferred low disk masses \citep{andrews13, ansdell16, pascucci16}  and stringent cooling requirements for fragmentation \citep{gammie01, boley06, stamatellos08, stamatellos09} observational evidence has been emerging that suggests some disks are gravitationally unstable \citep{kwon11, jin16, perez16, tobin16}. Furthermore, disk masses may be substantially underestimated due to the assumed value of the gas-to-dust ratio \citep{bergin13, mcclure16, miotello17, tsukamoto17, yu17}, and the companion mass-ratio distribution for B- and A-type primaries is separation-dependent, suggesting that close companions may originate in circumprimary disks rather than cloud core fragments \citep{gullikson16}. Evidence that instability and fragmentation are taking place in at least a few astrophysical systems gives theorists a mandate to identify plausible ways to trigger them, at least in disks with high gas masses.

Disk cooling, which must occur on dynamical timescales for fragments to form \citep{gammie01}, is regulated by opacity \citep{cai06, boley10, cossins10, podolak11, lin16}. The odds of fragmentation increase when the disk becomes optically thin to its own thermal radiation, allowing it to cool quickly \citep{meru10}. Grain growth, which significantly lowers disk opacity, proceeds rapidly: even some Class 0 YSOs, which have ages under 200,000 years \citep{enoch09},  show some degree of dust growth via the core-shine effect \citep{steinacker10, steinacker15}, or have non-ISM spectral indices \citep{jorgensen07, ricci10, chiang12}. As disks evolve, the largest observed (or inferred) grain sizes increase from millimeter in the Class-I phase \citep{miotello14} to centimeter in the T-Tauri phase \citep{perez12, perez15, tazzari16}. Here we examine the extent to which grain growth alone---with no other triggers such as infall---can alter a disk's gravitational stability to axisymmetric perturbations.

The effect of self-gravity in a protoplanetary disk is multifaceted. Apart from implications for planet formation, gravitational instability (GI) can contribute to angular momentum transport by producing turbulent stresses \citep{gammie01, baehr17}. Our work thus also helps address the broader question of how dust can affect gas dynamics in disks.

 This paper is organized as follows: In \S \ref{sec:diskmodel} we discuss our models of the gas disk and dust sub-disk. In \S \ref{sec:dustmodel} we describe our prescription for collision speeds and outcomes. \S \ref{sec:numericalalgorithm} explains our Monte Carlo method for simulating dust growth and settling, while in \S \ref{sc:results} we present results from each disk model. In \S \ref{sec:opacity} we describe our opacity-calculation method and radiative transfer simulations, and in \S \ref{sc:discussion} we discuss the implications of our results for opacities, thermal profiles and disk instability, and examine the limitations of our model. We present our conclusions in \S \ref{sc:conclusion}.

\section{Disk Model: Gas and Dust} \label{sec:diskmodel}

We describe our disk models that are used for dust growth and settling calculations. All the key quantities are summarized in Table 1. The central star is assumed to be a pre-main-sequence classical T-Tauri star with a mass $M_{\star} = 0.95 M_{\odot}$.  In all our simulations, the disk is represented in a cylindrical coordinate system $\left(R,\phi,z\right)$ with $R$ being the distance from the central  star and $z$ the height above the midplane.  We assume that the disk is axisymmetric and vertically symmetric with respect to its  midplane. We make a $1+1D$ disk model in $(R,z)$ by decoupling the  radial and vertical dimensions and simulating every vertical column  independently. Here we do not perform gas evolution; the dust evolves against the background of a fixed gas disk with turbulent speeds  specified analytically (\S \ref{ssc:alpha}). We assume the gas and dust  temperatures are equal, with the dust opacity regulating the radiative transfer. 

 We are primarily  focused on an accurate temperature structure, which plays a significant role in determining the Toomre-Q, a measure of stability against self-gravity \citep{toomre64}:
\begin{equation}
Q=\frac{c_s\Omega}{\pi G \Sigma_g}.
\label{eqn:tq}
\end{equation}
In Equation \ref{eqn:tq}, $c_s$ is the local sound speed, $\Omega$  is the local angular frequency, and $\Sigma_g$ is the gas surface density. The parameter $Q$ is the measure of stability of the disk under self-gravity against thermal and shear effects. Theoretically $Q=1$ is the exact threshold in the linear stability analysis for axisymmetric perturbations. However, for non-axisymmetric perturbations the critical value for $Q$ is slightly higher than $1$ and the instability gives rise to spiral modes instead of ring-like structure \citep[e.g.][]{papa91,nelson98,mayer02,johnson03,pickett03}. \citet{nelson98} reported the value of $Q=1.5$ for the onset of spiral instabilities, while isothermal simulations  by \citet{johnson03} achieved  fragmentation at $Q=1.4$. Similarly the SPH simulations by \citet{mayer02} find the growth of a two-armed mode until fragmentation takes place at $Q=1.4$. In this paper we shall use the value $1.4$ as the critical value of $Q$ for which instability sets in. However, we caution that the disk's vertical thickness, which mimics a pressure term, may also provide support against self-gravity, lowering the threshold value to Q $\sim 0.7$ \citep{kratter16, baehr17}. 

Two important assumptions of our model are:
 \begin{enumerate}
\item Although the disk is turbulent and the turbulent speeds help determine the particle collision speeds, we do not include viscous heating: we assume stellar illumination is the dominant heat source \citep[e.g.][]{yu16}. 
\item We assume no radial drift for the dust particles. For the parameters we consider here, the radial drift timescale is long compared to the growth and settling timescales of dust grains.

\end{enumerate}

\subsection{The Gas Disk}\label{sc:gasdisk}

To construct our disk models at $t=0$, we assume a power law temperature profile in the radial direction as
\begin{equation}
T(R)=280 \times \left(\frac{R}{1{\text{au}}}\right)^{-1/2}.
\label{eqn:tempt0}
\end{equation}
  We also assume that each vertical column is isothermal at $t=0$. The isothermal assumption is used only to generate the initial setup; after the simulation is initiated,  the temperature profile of the disk is governed by the evolving dust opacity. Assuming vertical hydrostatic equilibrium, we write the gas density profile as 
\begin{equation}\label{eqn:gaussiangas}
\rho_g\left(R,z\right)=\rho_0\left(R\right) e^{-\left(z^2/2h_g^2\right)}
\end{equation} 
where $\rho_0\left(R\right) = \Sigma_g/\sqrt{2 \pi} h$ is the midplane density and $h_g$ is the local gas scale height, given by
\begin{equation}
h_g=c_s/\Omega ,
\label{eqn:scaleheight}
\end{equation}
where $\Omega$ is the Keplerian angular speed and $c_s=(k_bT/\mu m_p)^{1/2}$ is the local isothermal sound speed with $k_b$ the Boltzmann constant, $\mu$ the mean molecular weight, taken as $2.33$, and $m_p$ the proton mass. 

\begin{figure*}
	\begin{subfigure}{0.49\linewidth}
	\centering
	\includegraphics[scale=0.45]{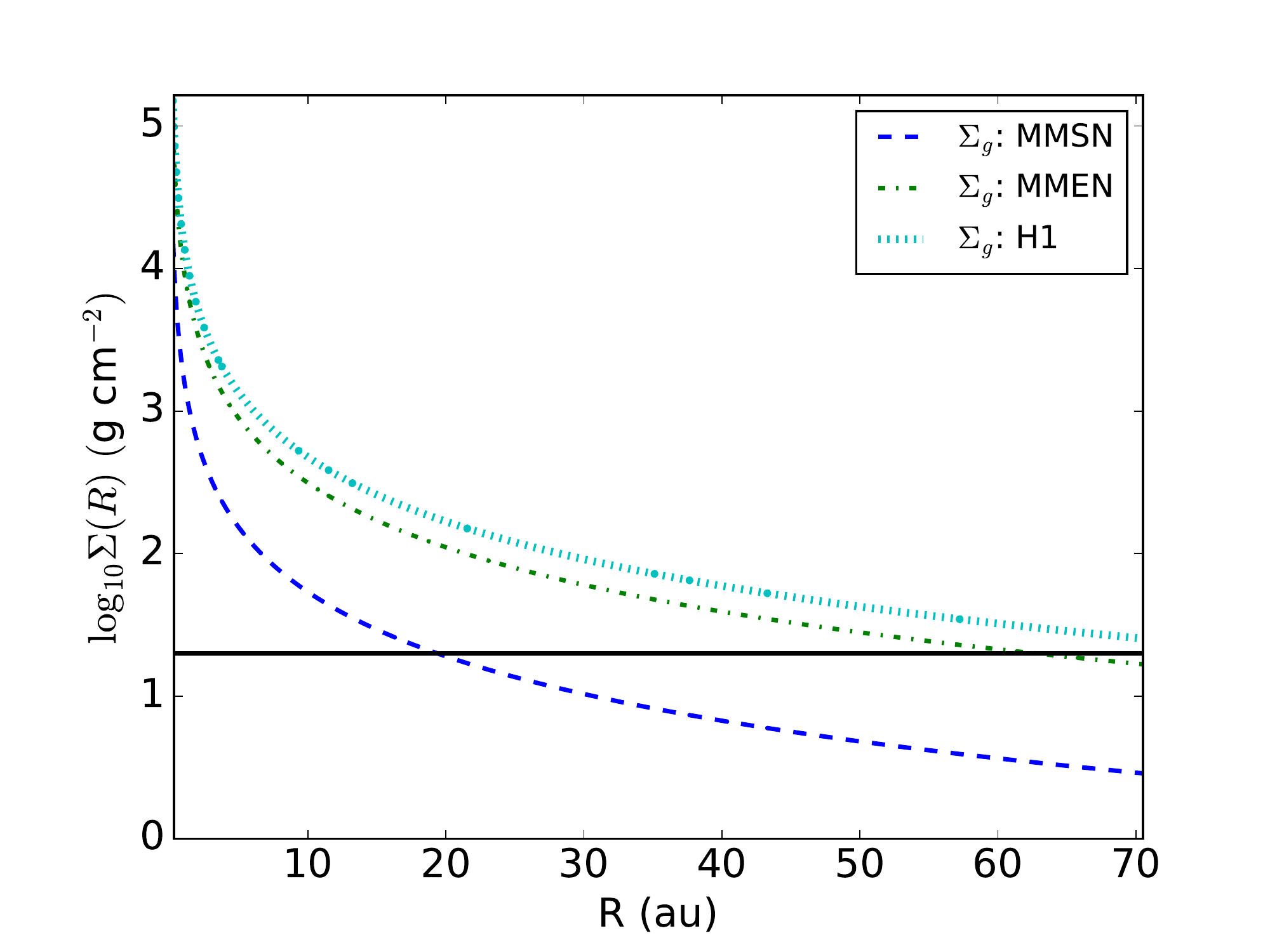}
	\end{subfigure}	
	\begin{subfigure}{0.49\linewidth}
	\centering
	\includegraphics[scale=0.45]{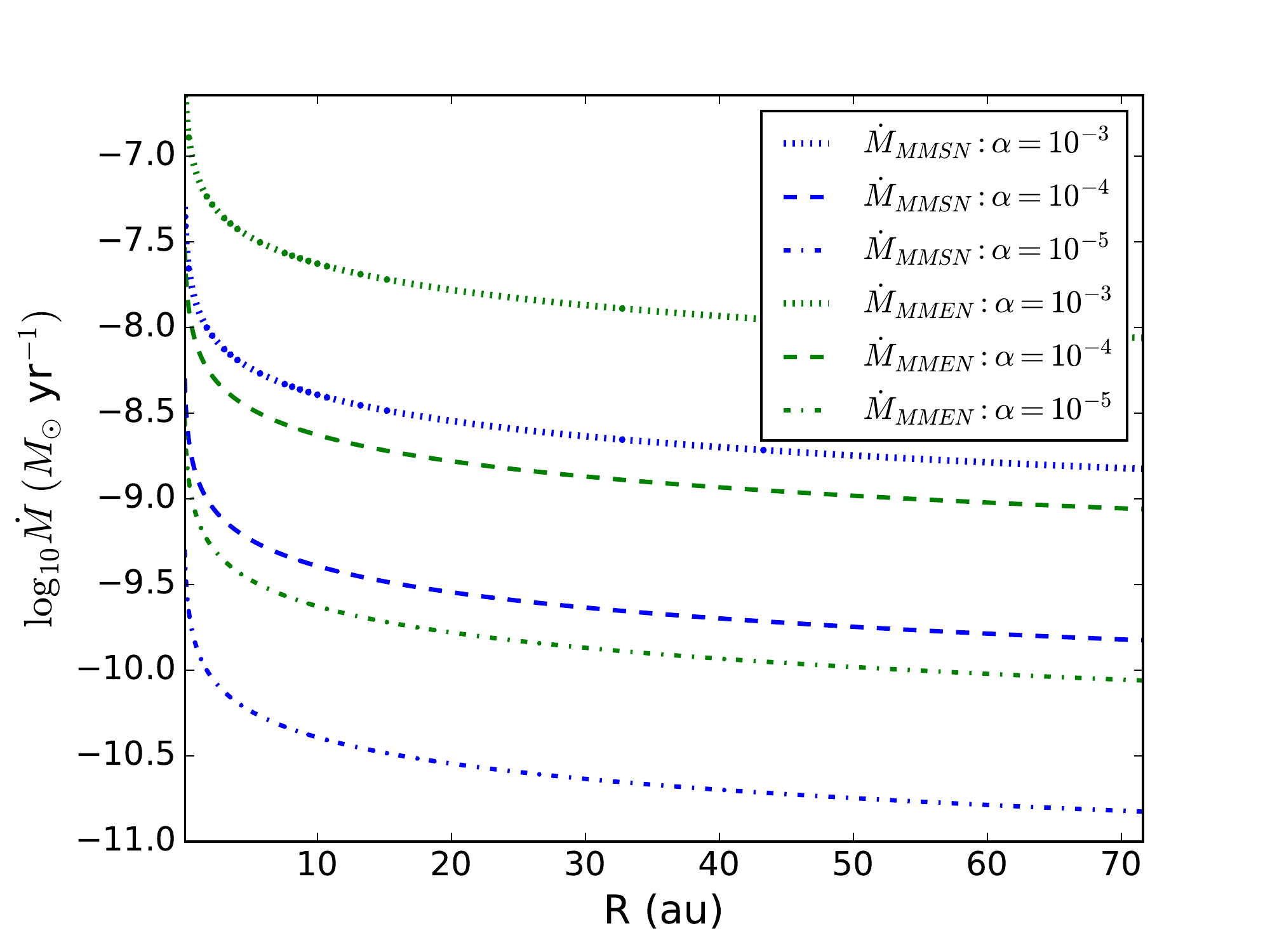}
	\end{subfigure}	
\caption{{\bf Left:} Surface density profile for the disk models: MMSN, MMEN and H1. The black horizontal line corresponds to $\Sigma_g=20$ g~cm$^{-2}$ which is the surface density threshold at the outer edge of the dead-zone. As can be seen from the plot, the surface density is more than $20$ g~cm$^{-2}$ out to $\sim 65$ au for the MMEN model. For H1 model the surface density is more than the threshold for the full radial extent of our simulations. {\bf Right:} The mass accretion rates with constant $\alpha = 10^{-3}$, $10^{-4}$, and $10^{-5}$ respectively for MMSN (red) and MMEN (green) disk models. }
	\label{fig:sigmaprof}
\end{figure*}

We investigate the gravitational stability of two different disk models and use an additional model for code tests.  The minimum-mass solar nebula \citep[MMSN;][]{hayashi81} is our test laboratory; we conducted simulations to compare with literature results, but mention {\em a priori} that grain growth and settling cannot trigger instability in the low-mass MMSN. For science simulations we adopt the minimum-mass extrasolar nebula \citep[MMEN;][]{ chiang13}, which is substantially heavier than the MMSN but has the same surface density power law index. The model surface densities are as follow:
\begin{eqnarray}
\Sigma_g(R) &\sim 1.7 \times 10^3\left(\frac{R}{1{\text{au}}}\right)^{-3/2} {\text{g~cm}}^{-2}    & {\text{(MMSN)}}\\\label{eqn:mmsn}
\Sigma_g(R) &\sim 10^4\left(\frac{R}{1{\text{au}}}\right)^{-3/2}  {\text{g~cm}}^{-2}   & {\text{(MMEN)}}\label{eqn:mmen},
\end{eqnarray}
where $\Sigma_g(R)$ is the surface density at radius $R$ (see Table \ref{tbl:simulations} for variable definitions). Finally, we simulate a heavy disk model which is only marginally stable at $t=0$ with the surface density profile 
\begin{equation}\label{eqn:heavydisk}
\Sigma_g(R) \sim 1.5\times 10^4 \left(\frac{R}{1{\text{au}}}\right)^{-3/2} {\text{g~cm}}^{-2}.
\end{equation}
In the text, test simulations of the MMSN are identified by `T' and those of the MMEN by `F' (see Table \ref{tbl:simulations}, which lists the simulations performed for this paper). The heavy disk model is named H1. The surface density profiles $(\Sigma_g(R))$ for all disk models are shown in the figure \ref{fig:sigmaprof}. We simulate a radial range of $R_{min} = 0.1$~au to $R_{max} = 75$~au. With these radial extent, the disk masses are approximately $0.018$, $0.12$ and $0.18$ $M_{\odot}$ for MMSN, MMEN and H1 respectively. In the vertical direction, we extend the grid to $4 h_g$ above the midplane in each radial grid zone; we ignore regions with $z > 4 h_g$ as dust density above that height is less than $0.1$\% of that in the midplane even at $t=0$. The 40 radial zones are equispaced in $\log(R)$, and function independently: particles do not move between vertical columns due to the omission of radial drift. We divide each column into $32$ cells equispaced in $z$, $8$ cells per scale height. The typical mass accretion rates $(\dot{M})$ for MMSN and MMEN models, calculated according to the classical accretion theory \citep{hartmann98}, are also shown in figure \ref{fig:sigmaprof}.

\subsection{Turbulence Efficiency $(\alpha)$}\label{ssc:alpha}

It is believed that a protoplanetary nebula is turbulent due to several proposed hydrodynamic \citep{lovelace99, lyra14, nelson13, marcus15} and magneto-hydrodynamic \citep{balbus91, neal14} instabilities. However,  we assume that the magnetorotational instability (MRI) is the source of turbulence in all our disk models. For our first set of simulations (T1, T2, T3, and F1 - F6) we follow the spatially uniform $\alpha$ viscosity prescription \citep{ss73}
\begin{equation}\label{eqn:nu}
\nu=\alpha c_s h_g,
\end{equation} 
 where $\nu$ is the turbulent viscosity. For simulations with variable  $\alpha(R,z)$ (T4, F7, F8 and H1), we simulate a disk with layered accretion \citep[e.g.][]{gammie96}. MHD  turbulence depends on how the gas is coupled to the magnetic field,  which strongly depends on the degree of ionization. For simulations  with variable $\alpha$, we adopt the ionization prescription of  \citet{landry13}, who consider cosmic rays, stellar X-rays and  radionuclides as the ionization sources. The model first  calculates the equilibrium abundances of charged species by solving a  simplified set of chemical reactions, including grain surface reactions  and the metal atoms' adsorption and desorption on the grains, adopted  from \citet{ilgner06}. In the regime where recombination mostly occurs  on the grain surface, the simplified model gives similar results to a  detailed chemical model. Subsequently, the Ohmic $(\eta_{O})$ and   ambipolar $(\eta_A)$ diffusivities are calculated and  $\alpha(R,z)$ is computed. The minimum turbulent efficiency,  $\alpha_{min}$, due to large scale fields in the dead zone, is taken as  $10^{-5}$ \citep{turner07}. For details of how we compute the spatially non-uniform  $\alpha$ profile see \citet{landry13}.  We note that hydrodynamic instabilities can provide viscosity even in magnetically inactive regions  \citep{lyra14,nelson13}. These instabilities can maintain a higher value of $\alpha$ which will affect the global dust evolution in the disk. To test how higher $\alpha_{\rm min}$ affects the size distribution, we have chosen one model (F8) with a minimum value for $\alpha=10^{-4}$ at the midplane.
 
 After the initial calculation of turbulence efficiencies, we do not evolve the $\alpha(R,z)$ profile with time in the course of our simulations. The initial prescription from \citet{landry13} assumes a nominal $1\upmu$m grain size. Due to grain growth and settling, the gas-to-solid ratio decreases at the midplane by almost an order of magnitude from its initial value. This evolving gas-to-solid ratio would alter the height of the dead-zone above the midplane as the disk evolves. \citet{okuzumi12} have found a similar trend with their grain evolution model in which the dead-zone initially shrinks, with its upper boundary contracting towards the midplane, and then extends vertically again. We note that for a self-consistent treatment, varying $\alpha(R,z)$ and hence the thickness of the dead-zone would be necessary. We leave the improved $\alpha(R,z)$ prescription for future work.

\begin{deluxetable}{cl}
\tabletypesize{\footnotesize}
\tablecolumns{2}
\tablewidth{0pt}
\tablecaption{Variables used in theoretical modeling}\label{tbl:vartheory}
\tablehead{
\colhead{Variable} \vspace{-0.2cm} & \colhead{Meaning}\\
&  }
\vspace{-0.2cm}
\startdata
\vspace{-0.2cm}\\
$c_s$ & local isothermal sound speed \\
$\Sigma_g$ & gas surface density\\
$\kappa$ & epicyclic frequency \\
$\Omega$ & Keplerian frequency \\
$R$ & orbital distance from central star \\
$\alpha$ & turbulence strength \\
$M_{\star}$ & stellar mass (mass of central star) \\
$\rho_g$ & gas volume density \\
$\rho_d$ & dust volume density \\
$\rho_m$ & material density of dust \\
$h_g$ & local gas pressure scale height \\
$h_d$ & dust scale height \\
$a$ & radius of dust particles \\
$\eta$ & dust to gas mass ratio \\
$t_{fric}$ & friction/stopping time \\
$v_{rel}$ & relative velocity of collision \\
$v_{frag}$ & fragmentation threshold velocity \\
$v_{dm}$ & relative speed of dust grains and gas molecules \\
$V_g$ & gas dispersion velocity \\
$\lambda_{mfp}$ & mean free path\\
$St$ & Stokes number \\
$\xi$ & fragmentation power law index \\ 
$D_g$ & gas diffusion coefficient \\
$D_d$ & dust diffusion coefficient \\
$\kappa_{\lambda}$ & monochromatic opacity \\
$\langle\kappa\rangle_{\rho_d}$ & density weighted opacity \\
$t_\eta$ & smallest eddy turnover time \\
$t_L$ & largest eddy turnover time \\
$Re$ & Reynolds number \\
$\lambda$ & wavelength of photon radiation \\
\vspace{-0.2cm}\\
\enddata
\end{deluxetable}

\begin{deluxetable}{cccccc} 
\tabletypesize{\footnotesize} 
\tablecolumns{4} 
\tablewidth{0pt} 
\tablecaption{ Simulations Performed \label{tbl:simulations}} 
\tablehead{ 
\colhead{Simulation} \vspace{-0.2cm} & \colhead{$\Sigma(R)$} & \colhead{} & \colhead{$v_{frag}$} & \colhead{} & \colhead{}  \\
 \vspace{-0.2cm} 
 & &  \colhead{$\alpha$} & & \colhead{$M_{disk}/M_{\star}$} & \colhead{$\alpha_{min}$}\\ 
\colhead{Name} & \colhead{profile} & \colhead{} & \colhead{cm~s$^{-2}$} }
\startdata 
\vspace{-0.2cm}\\
 T1\tablenotemark{a} & MMSN & $10^{-3}$ & $100$ & $0.018$ & \nodata \\
 T2\tablenotemark{a} & MMSN & $10^{-4}$ & $100$ & $0.018$ & \nodata \\
 T3\tablenotemark{a} & MMSN & $10^{-5}$ & $100$ & $0.018$ & \nodata \\
 T4\tablenotemark{a} & MMSN & variable & $100$ & $0.018$ & $10^{-5}$ \\
 \vspace{-0.2cm}\\
 \hline
 \vspace{-0.2cm}\\
 F1 & MMEN & $10^{-3}$ & $50$ & $0.12$ & \nodata \\
 F2 & MMEN & $10^{-4}$ & $50$ & $0.12$ & \nodata\\
 F3 & MMEN & $10^{-5}$ & $50$ & $0.12$ & \nodata\\
 F4 & MMEN & $10^{-3}$ & $100$ & $0.12$ & \nodata \\
 F5 & MMEN & $10^{-4}$ & $100$ & $0.12$ & \nodata\\
 F6 & MMEN & $10^{-5}$ & $100$ & $0.12$ & \nodata\\
 F7 & MMEN & variable & $100$ & $0.12$ & $10^{-5}$\\
 F8 & MMEN & variable & $100$ & $0.12$ & $10^{-4}$\\
 \vspace{-0.2cm}\\
 \hline
 \vspace{-0.2cm}\\
 H1 & equation \ref{eqn:heavydisk} & variable & $100$ & $0.18$ & $10^{-5}$\\
 \vspace{-0.2cm}\\
 \enddata 
\tablecomments{Science simulation set: Two different disk surface
density profiles with $\alpha=10^{-3}$, $10^{-4}$, $10^{-5}$ and
variable.}
\tablenotetext{a}{Code test}
\end{deluxetable}

\subsection{The Dust Distribution at $t=0$}\label{sc:dustdisk}

 Dust grains in the disk experience an aerodynamic drag which plays a significant role in setting their collision speeds. The coupling between gas and dust is defined by the friction time-scale, $t_{fric}$, which is the ratio of the particle momentum to the drag force and gives an estimate of the time required to change the relative velocity between gas and dust substantially. The friction time-scale is
	\begin{numcases}{t_{fric}=}
			\frac{\rho_m}{\rho_g}\frac{a}{c_s}, &  \text{if
			$ a \leqslant
			\frac{9}{4}\lambda_{mfp}$},\label{eqn:tfrice}\\
			\frac{8}{3}\frac{\rho_m}{\rho_g}\frac{a}{C_D
			v_{dm}}, & \text{otherwise}.\label{eqn:tfrics}
	\end{numcases}
Equation \ref{eqn:tfrice} applies in the Epstein regime for the small grain size limit and equation \ref{eqn:tfrics} applies in the  Stokes regime for the particle size of $a>(9/4)\lambda_{mfp}$, where $\lambda_{mfp}$ is the mean free path of gas molecules. Here, $\rho_m$ is the material density of dust grains, $v_{dm}$ is the relative velocity between dust and gas, and $C_D$ is the drag coefficient which depends on the Reynolds number. We select the dependence appropriate for spherical grains. 

For better comparison of the coupling for particles of different size and gas density, the dimensionless Stokes number is defined as
\begin{equation}
St=t_{fric} \Omega.
\label{stokesnumber}
\end{equation}
Particles with Stokes number unity come to match the gas velocity in one local orbital period. For a wide range of gas densities, sub-micron dust grains have $St \ll 1$ and hence come quickly to rest in the gas reference frame.

The dust scale height $(h_d)$ in a turbulent protoplanetary nebula can be calculated following \citet{dubrulle95} and \citet{youdin07}:
\begin{equation}
h_d=h_g\left(1+\frac{St}{\alpha}\right)^{-1/2}\label{eqn:dustsh}
\end{equation}
For tightly coupled particles with $St \ll 1$, $h_d\approx h_g$ and we approximate the dust scale height by that of the gas at $t=0$. To model the dust size distribution at $t=0$, we adopt a grain-size distribution with an  MRN \citep{mrn77} power-law index, $N(a) \propto a^{-3.5}$ where $N(a)\,da$ is the number of dust particles of radii between $\left[a,a+\,da\right]$. We consider that the dust grains already grow beyond the ISM size in the molecular cloud phase \citep{suttner01} and adopt the maximum and minimum size of the dust size distribution at $t=0$ as $a_{max}=1.0\upmu$m and $a_{min}=0.1\upmu$m. To begin our grain growth and settling  simulation, we make two assumptions:
\begin{enumerate}
\item Gas and dust of all sizes are dynamically coupled and well
mixed at $t=0$, with $St_{t=0} \ll 1$ (this will not be true at
later times);
\item The dust-to-gas mass ratio is $\eta = 0.01$, similar to the
interstellar medium (ISM).

\end{enumerate}

\section{Dust Evolution Model}\label{sec:dustmodel}

The main objective of this work is to obtain the disk temperature
profile, which is controlled by the dust opacity. 
To simulate dust evolution, we introduce a hybrid 
model in which the collisional dust growth is implemented through a weighted Monte Carlo method along with a Lagrangian Monte Carlo prescription for vertical settling and diffusion.

\subsection{Collision Model}\label{sc:collisionmodel}

The outcome of a model collision between two dust particles in a protoplanetary environment has many possibilities according to laboratory experiments. On the experimental side, \citet{guttler10} presented $19$ possible collisional outcomes for particles with various mass ratios, speeds, and porosities. However, it is prohibitively computationally expensive to include all possibilities in a global disk model. For simplicity, we adopt a collisional model that includes only sticking and fragmentation. We treat collisions as a binary process, identifying the smaller mass as the projectile $(m_p)$ and the bigger mass as the target $(m_t)$.  The collision outcome is determined by the relative velocity (see \S \ref{sc:relativevelocity} for a description of our velocity computation). If the particles collide with a velocity less than a threshold velocity $v_{frag}$, they stick and form a new particle with mass $m_{final} = m_p+m_t$. When $v_{rel} > v_{frag}$, the collision results in fragmentation.

As particles grow by sticking, their eddy-crossing times drop, leading to lower coupling with the gas and higher collision speeds \citep{ormel07b}.  When the collision speed reaches $v_{frag}$, instead of sticking, both particles fragment. For such an event the combined mass of the target and the projectile is made to follow a mass distribution $f(m)\,dm \propto m^{-\xi}\,dm$ with $\xi$ being the fragmentation distribution power law index. Here we adopt $\xi = 11/6$ \citep{windmark12b,drazkowska14,krijt16a}, though we note that some experiments predict a shallower fragment size distribution with $\xi = 9/8$ \citep{blum00,guttler10}. The smallest fragments are monomers of $0.1 \upmu$m. The largest body in the fragment mass distribution is set equal to the target mass as in \citet{drazkowska14} (personal communication).

  We note that setting the mass of the largest particle of the fragmenting distribution equal to that of the target is equivalent to assuming that the target is immune to fragmentation. Laboratory experiments, in fact, show that the mass of the largest fragment  is dependent on  the  collision velocity \citep{guttler10}. However, in our Monte Carlo collision model (see \S \ref{sec:numericalalgorithm} below) the mass contained in each bin is updated after each collision. As in the case of a fragmentation event, the total mass of the target and the projectile is distributed in a power-law distribution, after each fragmentation event the fractional contribution from the bin containing the target gets reduced. Moreover, mass-transfer, where a fraction of the projectile mass is transferred to the target, can extend the size distribution beyond the fragmentation barrier \citep{windmark12a,drazkowska13,estrada16}. Traditionally, mass transfer is thought to be a possible way to overcome growth barrier \citep{windmark12a,drazkowska13} and planetesimal formation through collisional growth. As the primary objective of this paper is not the formation of planetesimals and we do not run our simulations for extended timescales over which growth of solid bodies upto several meters would be relevant, we chose not to include mass transfer in our model.

We note that other collisional outcomes besides sticking and fragmentation are physically possible, most notably bouncing, erosion and mass-transfer. Inclusion of bouncing in the model slows down the growth process and the growth timescale may even become comparable to the timescale for radial drift \citep{estrada16}, which is not included in our model yet. Furthermore, bouncing effect restricts the growth of particle \citep{windmark12a} restricting the maximum {\em Stokes} number of the evolving size distribution \citep[See figure $2$ of][]{estrada16}. Inclusion of mass transfer helps dust particles to grow indefinitely and is considered a possible pathway to planetesimal formation \citep{drazkowska13} if drift is neglected.  \citet{estrada16}, on the other hand, have shown that under more realistic conditions where radial drift is included, the effect of mass-transfer is limited.  Given that the primary objective of this work is to examine how the disk's temperature profile responds to grain growth and settling, and grains of size similar to the peak wavelengths of star and disk emission control the temperature structure, we do not include the mass transfer/planetesimal formation pathway. Furthermore, although target erosion is the most likely outcome of high-speed collisions between particles of significantly different masses \citep[e.g.][]{windmark12a}, we neglect erosion due to computational constraints. Instead, we assume that all collisions with $v_{rel} > v_{frag}$ lead to fragmentation, as in the ``SF'' simulations of \citet{windmark12b}. We also assume that the dust particles remain compact spheres throughout their growth and fragmentation.

To ensure accuracy, we have tested our code by computing a steady-state particle size distribution in a single grid zone with parameters matching \citet{windmark12b}, and a vertically averaged steady-state size distribution at a single distance from the star with parameters matching \citet{drazkowska14}. Appendix \ref{apn:codetest} compares our results against the literature results and explains how our code conserves the dust mass.

\subsection{Collision Velocity}\label{sc:relativevelocity}

\begin{figure}
\centering
	\includegraphics[scale=0.33]{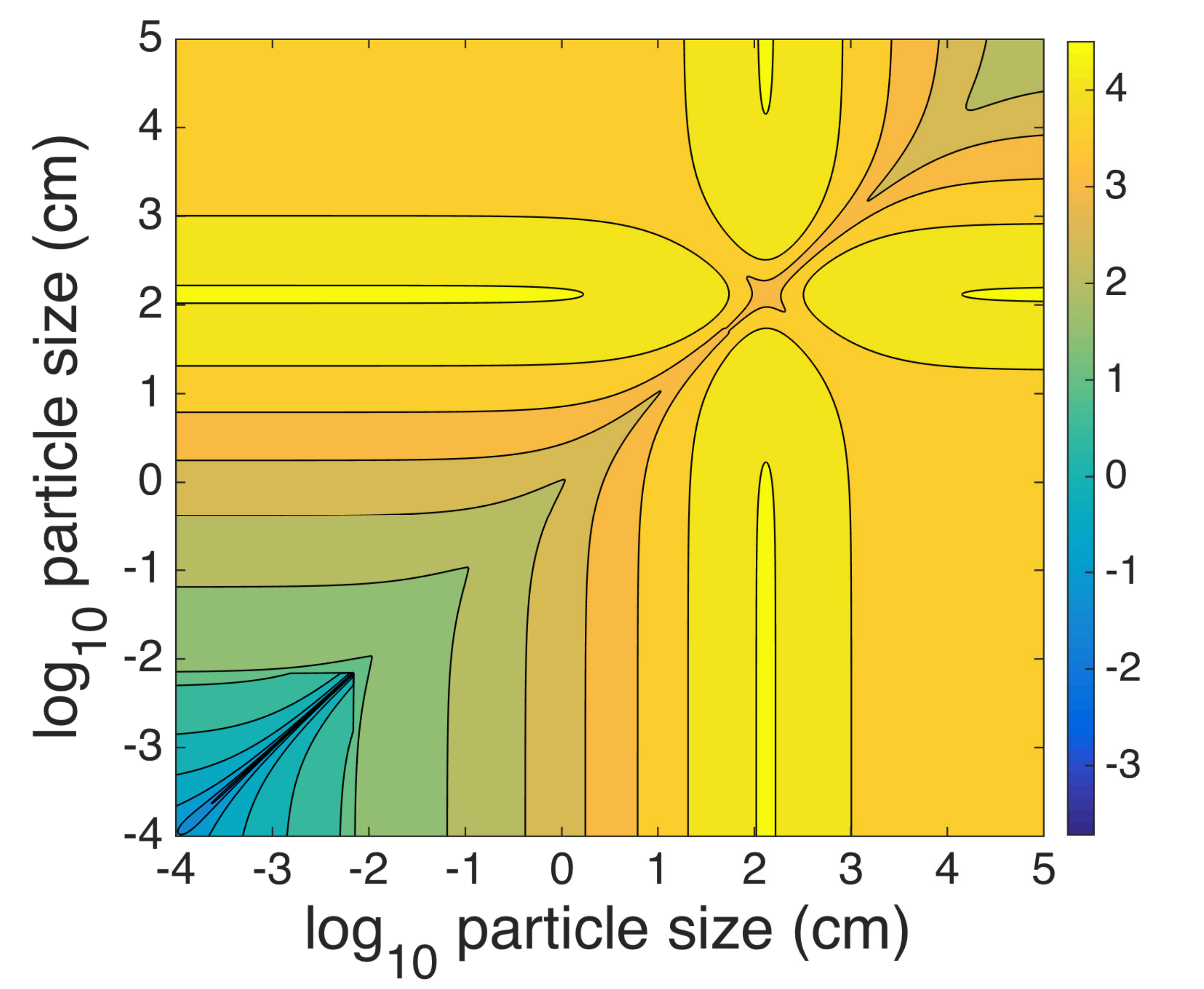}
	\caption{The relative velocity between different particle sizes (equation \ref{eqn:relativev}) with contributions from Brownian motion, turbulence, settling, radial and azimuthal drift as mentioned in section \S \ref{sc:relativevelocity}. The velocity profile is plotted for $\Sigma_g=330$g~cm$^{-2}$, $\eta=0.01$, $T=115$K, and $\alpha=10^{-3}$ at a distance $3$ au from the central star. Parameters listed above are directly taken from \citet{windmark12a}; see their Figure 6.}
	\label{fig:relativevel}
\end{figure}

We consider five different contributions to the particle relative velocity (see figure \ref{fig:relativevel}): Brownian motion $(\delta v_B)$, turbulent motion $(\delta v_t)$, vertical settling $(\delta v_z)$, radial drift $(\delta v_r)$ and azimuthal motion $(\delta v_{\phi})$. The relative velocity of collision is calculated as
\begin{equation}\label{eqn:relativev}
v_{rel}=\sqrt{\sum \delta v_i^2},
\end{equation}
where $i$ represents each of the five velocity contributions mentioned above.  Our simulations are azimuthally symmetric and we do not allow particles to move between radial grid zones. However, we include $\delta v_r$ and $\delta v_{\phi}$ contributions to $v_{rel}$ to improve the accuracy of our collision outcomes. Although it may seem physically inconsistent to include $\delta v_r$ in the velocity calculation while forbidding radial motion in our grid, consistent with the findings of \citet{estrada16}, $\delta v_r$ and $\delta v_{\phi}$ contribute significantly to collision velocities only for $a \geqslant 10$ cm which is rare in our simulations, and radial drift is important over a timescale much larger than where we found our steady states. (Also, see the discussion of our code's limitations in \S 7.)  



For the smallest particles, Brownian motion is the dominant contribution to $v_{rel}$, giving $\delta v_B = \surd{[8 k T (m_p + m_2) / (\pi m_p m_t)]}$ (where $k$ is the Boltzmann constant).  The collision velocities of dust particles with radii beyond a few $\upmu$m are dominated by the gas turbulence.
To calculate $\delta v_t$, we follow the closed-form velocity prescription of \citet{ormel07b} (their equations $26$, $28$ and $29$). \S \ref{sc:verticaldustmotion} describes our algorithm for computing $\delta v_z$. We compute $\delta v_r$ and $\delta v_{\phi}$ using equations 6, 7, and 18 of \citet{okuzumi12}.

\subsection{Vertical Dust Transport}\label{sc:verticaldustmotion}

\begin{figure}
\centering
\includegraphics[scale=0.2]{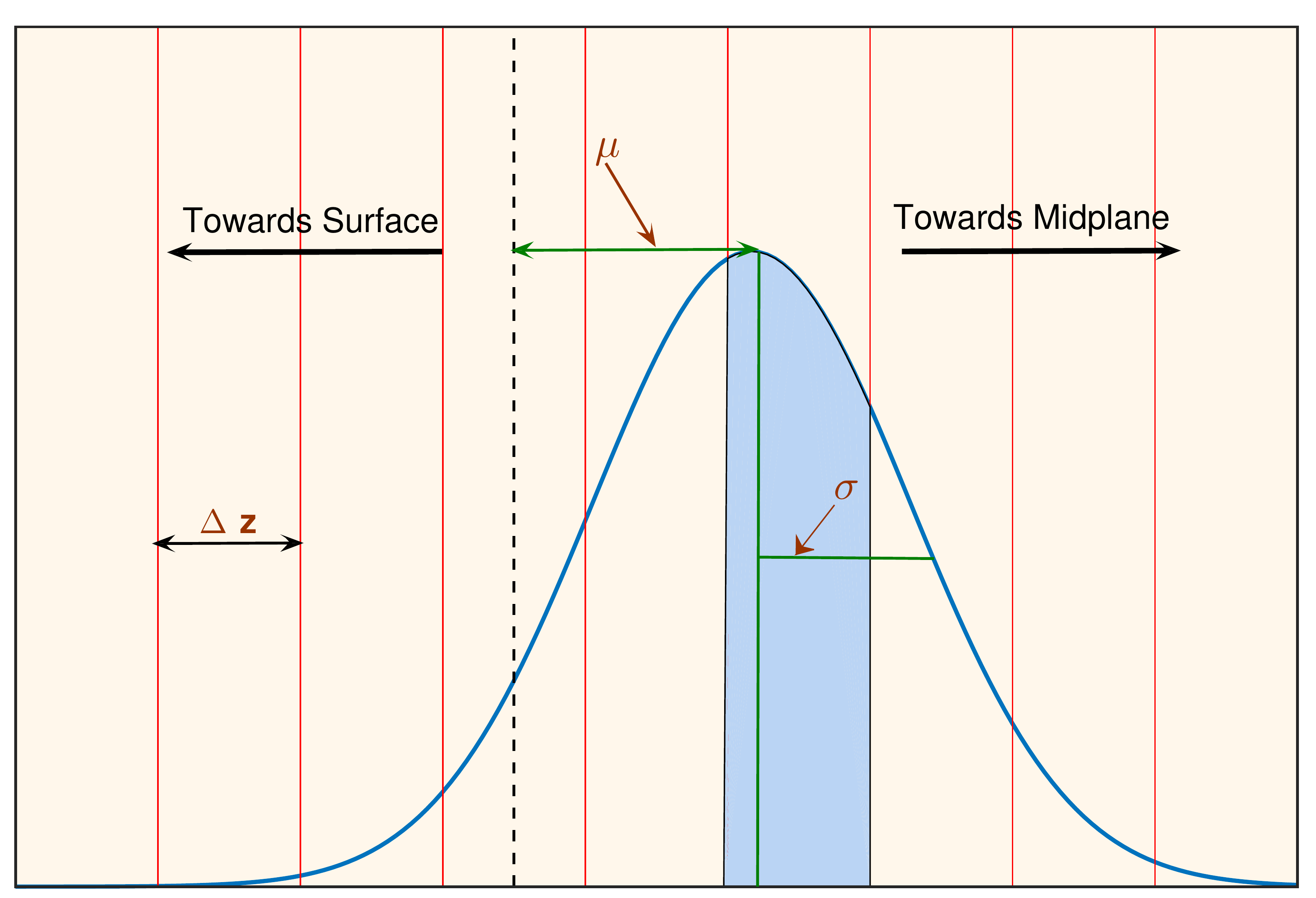}
\caption{A schematic of the settling algorithm implemented in our work. The vertical dashed black line is the height of {\it middle} of the cell the particle inhabits. Before each settling step, dust particles of radius $a$ are spread from the top to the bottom of the vertical column according to the background dust distribution.  The particle of size $a$ is then redistributed according to the prescription given by Equations \ref{eqn:mu} and \ref{eqn:sigma} (solid blue line). Red lines mark boundaries between cells, and the shaded region shows the probability that the particle will be  moved from the original cell to that particular cell. A similar Gaussian exists for each of the $N_s$ dummy particle used in the settling algorithm. }
\label{fig:settlingcartoon}
\end{figure}

Vertical dust settling plays a significant role in determining the dust abundance as a function of height, which in turn affects the collision
frequency, grain size distribution, and opacity. The continuity equation for vertical dust dynamics in 1D is
\begin{equation}
\partial_{t}\rho_d+\partial_{z} F=0\label{eqn:continuity}
\end{equation}
where $F$, the total diffusive flux of dust particles in g~cm$^{-2}$~s$^{-1}$, is composed of three components: vertical settling due to stellar gravity, diffusion of dust towards density maxima, and stirring of dust by gas turbulence. To model all three processes we follow the prescription given by \citet{charnoz11} with the addition of a settling term in their equation $22$. We compute the distance over which a particle settles by selecting randomly from a Gaussian distribution with
\begin{equation}
\mu=\Delta z_s+\left[\frac{D_d}{\rho_g}\partial_z\rho_g+\partial_zD_d(z)\right]\delta t \label{eqn:mu}
\end{equation}
\begin{equation}
\sigma^2=2D_d(z)\delta t+\left[(\partial_zD_d) \delta
t\right]^2.
\label{eqn:sigma}
\end{equation}
In Equation \ref{eqn:sigma}, the dust diffusion coefficient (given by \citet{youdin07}) is
\begin{equation}
D_d=\frac{\nu}{1+St^2},
\label{eqn:Dd}
\end{equation}
with $\nu$ defined by Equation \ref{eqn:nu}. $\Delta z_s$ is the mean distance dust particles move in time $\delta t$ due to the vertical component of stellar gravity as mean field:
\begin{equation}\label{eqn:settlingspeed}
\Delta z_s = v_z \delta t =
\frac{a}{c_s}\frac{\rho_m}{\rho_g}\Omega^2 z \delta t= t_{fric}
\Omega^2 z\delta t,
\end{equation}
where $v_z$ is the settling speed of the dust particle.  The second term in equation \ref{eqn:mu} is a correction term for the net diffusive flux $D_d/\rho_g\times \nabla \rho_g$ of dust particles, directed towards higher density region and induced by the non-zero gradient in gas density. This is a systematic velocity term that captures the effect of non-homogeneous diffusion in the presence of a non-uniform gas density distribution (See equation $18$ of \citet{drazkowska13}). The last terms in equations \ref{eqn:mu} and \ref{eqn:sigma} arise due to variations in the dust diffusion coefficient $D_d$. The first term in $\sigma^2$, $2D_d \delta t$, comes from turbulent diffusion and is responsible for particle stirring. See Figure \ref{fig:settlingcartoon} for a schematic of the contributions to particle vertical speeds. We discuss the calculation of settling timestep $\delta t$ in \S 5.  We note that equation \ref{eqn:dustsh} also gives an approximation to the steady state dust scale height (See appendix \ref{apn:codetest}) and has been used for our initial setup. \citet{estrada16} use the same prescription (equation \ref{eqn:dustsh}) for dust scale height to distribute solids in the vertical direction which extends their model to $1+1$D from a $1$D gas diffusion model. This method works perfectly fine as the vertical diffusion timescale is small compared to the inward drift timescale. However, \citet{mulders12} showed that the midplane approach of equation \ref{eqn:dustsh} from \citet{dubrulle95} estimates a higher dust abundance towards the disk surface compared to the abundance obtained using local gas parameters. Hence, to compute a more realistic vertical structure in our disk models with layered accretion, parameters such as Stokes number for individual particles are calculated locally. Thus, the prescription from \citet{charnoz11} gives a more accurate result.

\begin{figure*}
\centering
\includegraphics[scale=0.5]{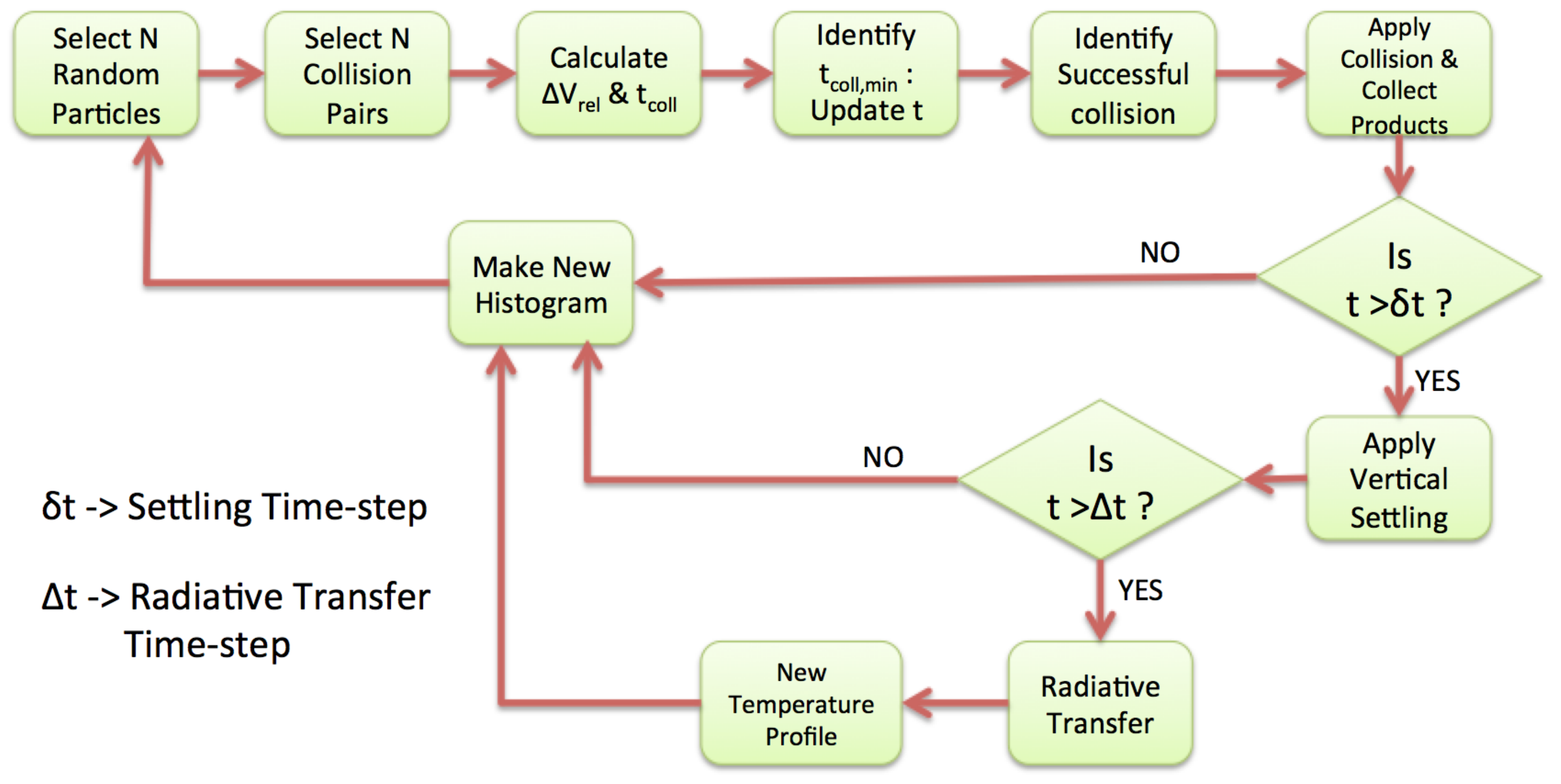} \caption{A pictorial depiction of the numerical algorithm implemented in this work. In this work, the gas density is held constant and we do not update the gas scale height of the disk through the course of our simulations.}
\label{fig:algorithm} 
\end{figure*}

\section{Monte Carlo Dust-Growth Model}\label{sec:numericalalgorithm}

Collisional dust growth and dynamics in a planet forming disk are generally modeled by either solving Smoluchowski's equation \citep{nakagawa81, birnstiel10} or with a Monte Carlo simulation \citep{ormel08, zsom08}, or using moment of the dust distribution \citep{estrada08}.  Although Smoluchowski's method is numerically less expensive, implementation with low resolution can lead to unphysical growth of dust particles \citep{ohtsuki90, drazkowska14}.  \citet{drazkowska14} showed in a comparative study that Monte Carlo techniques are not as sensitive to resolution in the particle size distribution. However, for simulating dust behavior throughout the disk over 10~kyr timescales, Monte Carlo methods can be computationally expensive. They also lack the dynamical range that can be easily achieved by Smoluchowski's method. Here we present a new Monte Carlo model which is fast and can achieve a larger dynamical range by using a weighting technique. The schematic plot of our algorithm is shown in Figure \ref{fig:algorithm}, and the key quantities are listed in Table \ref{tbl:varsim}.

\subsection{Selecting Collision Pairs}
\label{sec:montecarlo}

We start by dividing the total mass range of dust grains {\it in each grid zone} into $N_H$ equally spaced logarithmic histogram bins. At any given instant $t$, $N$ particles are drawn randomly from the particle mass distribution in that grid zone. We found converged results for $N = 60000$ and adopted that value for the simulations presented here. We denote the first array of Monte Carlo particles in any given grid zone by subscript `q'. (Below we will describe the selection of a second set of Monte Carlo particles in each grid cell to serve as potential collision partners.) If $N_i$ is the number of particles drawn from the $i^{th}$ bin in the particle mass distribution, we have
\begin{equation}
\sum_{i=1}^{N_H} N_i = N.
\label{eqn:massdist1}
\end{equation}
Given that $\rho_{d,i}$ is the dust mass per unit disk volume represented by bin $i$ in the mass distribution and
$\log m_i$ is the value of $\log m$ at the center of $i^{th}$ bin. The {\it number density of
particles per cm$^3$} contributed by bin $i$ is $n_{d,i} =
\rho_{d,i}/m_i$.  Finally,
\begin{equation}
n_{d,i}=f_i n_d,
\label{eqn:massdist2}
\end{equation}
where $n_d$ is the total number density of the dust grains {\it of all masses} in the grid cell and
$f_i$ is the fractional abundance of dust grains of mass $m_i$, such
that
\begin{equation}
\sum_{i=1}^{N_H} f_i=1.
\label{eqn:massdist3}
\end{equation}
In the same timestep and grid cell, another $N$ particles, denoted by subscript `k', are drawn randomly from the same particle mass distribution to be the possible collision partners. The dust mass distribution after a particular timestep is determined by the outcome of a collision chosen from these $N$ particle pairs (see \S \ref{sc:collisionmodel}).  Unlike \citet{ormel07a} or \citet{zsom08}, we only partially trace the evolution of a specific set of particles over time (See \S \ref{sec:randomselection}).

\begin{deluxetable}{cl}

\tabletypesize{\footnotesize}
\tablecolumns{2}
\tablewidth{0pt}
\tablecaption{Variables used in numerical algorithm \label{tbl:varsim}}
\tablehead{
\colhead{Variable} \vspace{-0.2cm} & \colhead{Meaning}\\
&  }
\vspace{-0.2cm}
\startdata
\vspace{-0.2cm}\\
$N_H$ & number of bins in mass histogram \\
$N$ & number of Monte Carlo particles used in each grid zone \\
$m_i$ & mass at the center of $i^{th}$ bin of mass
histogram \\
$N_i$ & number of particles from $i^{th}$ bin of mass
histogram\\
$n_{d,i}$ & number density of particles in $i^{th}$ bin of
mass distribution\\
$n_d$ & total dust number density including particles of all
masses \\
$f_i$ & fractional of particles in $i^{th}$ bin of mass histogram \\
$M_{total}$ & total dust mass in a grid cell \\
$w_i$ & statistical weight of $i^{th}$ bin of mass histogram \\
$N_s$ & number of Monte Carlo particles used in settling \\
$\delta t$ & dust evolution timestep \\
$\delta t_{settle}$ & settling timestep \\
$\Delta t$ & radiative transfer timestep \\
\vspace{-0.2cm}\\
\enddata
\end{deluxetable}

During the random selection of sets q and k of potential colliders, the number of particles we draw from each mass bin is $f_i N$, unless particles of mass $m_i$ are rare enough that $f_i N < 1$. Here we introduce a weighting scheme to make sure that the rare particles are not lost from the simulation, as a few large particles may dramatically alter the particle mass distribution by sweeping up smaller grains \citep[e.g.][]{windmark12a}.  From the particle mass bins with $f_i N < 1$, a single particle is randomly selected from each bin's mass range and a weight $w_i=f_i N$ is assigned to that particle. For particles drawn from bins with $f_iN>1$, $w_i=1$. The weight can be interpreted as a fraction of the particle that truly exist in the distribution. After picking the selected number of particles from each bin, the order of particles in the two arrays q and k is randomized.

In each grid zone, for each timestep, we allow only one collision to proceed successfully. For each particle in set q, collisions with its partner in set k proceed at rate
\begin{equation}
P_{k}=n_k \sigma_{qk}\delta v_{qk},
\label{eqn:collprob}
\end{equation}
collisions per second, where $\sigma_{qk}$ is the collision cross-section $\pi(r_q+r_k)^2$.  The relative collision velocity $\delta v_{qk}$ is calculated according to the prescription outlined in section \ref{sc:relativevelocity}. The number density $n_k$, reperesented by the $k^{th}$ particle, is:
\begin{equation}
n_k=\frac{n_{d,i}}{N_i}=\frac{f_i n_d}{N_i},
\end{equation}
where $i$ is the mass bin corresponding to particle $k$.  At this point, we choose a single pair of particles from $N$ possible collision pairs from the distribution of $P_k$ obtained from equation \ref{eqn:collprob} by using a single random number drawn from a uniform distribution between $0$ and $1$. At any particular step only a single collision is allowed and the corresponding time is updated by the method explained in \S \ref{sc:timeevolution}.

For a sticking event between two particles of masses $m_q$ and $m_k$, the final mass is set as $w_q m_q+w_k m_k$ and is transferred to the bin containing mass $m_q+m_k$. Similarly, for a fragmentation event, the total mass put into the size distribution of fragments is $w_q m_q+w_k m_k$. After the collision, a new particle mass distribution is calculated.  For the next timestep, the mass distribution is again transformed into number density space using Equations \ref{eqn:massdist1}-\ref{eqn:massdist3} and the new sets q and k are selected to again make N particle pairs.

\subsection{A Two-Step Random Selection}
\label{sec:randomselection}

  The model presented here consists of repeated sampling of the size distribution to select particles and their collision partners. In the process of collisional growth, before the fragmenting threshold velocity is reached, the bins towards the higher mass end of the distribution contribute single particles to the total population of $N$ particles. It is not always guaranteed that the single particle will be picked during random sampling and hence the growth can be hindered artificially. To circumvent this problem, the random selection is done in a two-step process. First, the number of particles to be selected from each bin is calculated according to Equations \ref{eqn:massdist1} and \ref{eqn:massdist2} (See \S \ref{sec:montecarlo}). Next, after ensuring the right number of particles are drawn from each bin, the array of particles is randomized. The same process is followed for the selection of collision partners as well.

\subsection{Mass Conservation}
\label{sec:massconservation}

 One important part of our code is ensuring mass conservation locally and in each vertical column (recall that particles are not allowed to migrate radially between columns). Mass is conserved during collisions, but may be lost or gained in numerical noise when computing the particle mass histogram after each timestep. Mass loss is more likely, since the largest and rarest particles contain the bulk of the mass: at the large end of the size distribution, the difference between the maximum mass in a histogram bin and the bin center can be a significant fraction of the total particle mass in the grid zone. We conserve mass in our simulation by updating the total number density $n_d$ in each grid zone after each timestep. The number density change is
\begin{equation}
n_d \leftarrow \frac{M_{total}}{\sum_i m_i f_i},
\label{eqn:massconserve}
\end{equation}
ensuring that each mass histogram bin will contain the correct fraction of the grid zone's total mass. No bin $m_i$ can then lose mass by dropping a particle near its upper mass boundary. This allows us to conserve mass to almost the machine precision.

\subsection{Artificial Oscillation \& Partial Particle Tracking}
\label{sec:oscillationtracking}

 The imposed mass conservation can cause artificial oscillation in the number of particles at the smaller sizes in the dust-mass spectrum.  Given that the masses of the bigger particles are not necessarily equal to the respective bin center masses, using equation \ref{eqn:massconserve} would force the total number of small particles to change to accommodate a single large particle's shift to the bin center. We remove the oscillations by retaining the same individual particles between timesteps in the low-statistics bins instead of subsuming them into the new particle mass histogram calculated at the end of each timestep. This technique helps to track the dust growth in a more accurate way. For this work, we tracked particles from any bin contributing less than $10$ particles and this number is kept constant throughout the simulations.

\subsection{Calculating the Timestep Between Collisions}
\label{sc:timeevolution}

After finding the successful collision in each grid zone, the next step in our simulation is to calculate the timestep $\delta t$.  Codes that follow the kinetic Monte Carlo method set $\delta t$ by first computing the $P_{total}$, the total collision rate from {\it all} {\bf $N(N-1)/2$} possible particle pairs from sets q and k; then using a random number $r$ selected from uniform distribution, calculate
\begin{equation}
\delta t = -\frac{1}{P_{total}} \ln(r)
\label{eqn:deltat}
\end{equation}
\citep[e.g.][]{gillespie75,ormel07a,zsom08}. 

 It is evident that in our method it is not possible to calculate the time evolution following Equation \ref{eqn:deltat} as we do not calculate the rates for all $N(N-1)/2$ possible pairs while selecting the successful collision at any step. Instead, we implement a matrix partitioning method in which the particles are first binned in the mass histogram. We now assume that out of these $N$ particles, the $N_i$ ones belonging to the $i^{th}$ histogram bin represents the same mass $m_i$ corresponding to that particular bin center. Moreover, the original particles being drawn from the number density distribution $f(n)$, every particle grouped into a single bin represents the same number density in the underlying population. So, instead of calculating individual rates, we assume every pair belonging to same histogram bin contributes equally to the total rate $P_{total}$. Consider that we are calculating the rate between $i^{th}$ bin for the first array and $j^{th}$ bin for the collision pairs, we calculate the rate of collision for a single pair by $n_{d,j}\sigma_{ij}v_{rel,ij}$ (Note the change in index from $q$ and $k$ to $i$ and $j$ as particles are represented by their corresponding bins). Thus, for total rate for  all the pairs coming from the $i^{th}$ bin for the first array of particles  and $j^{th}$ bin for the collision partners can be written as:
\begin{equation}
\label{eqn:binrate}
P_{ij}=n_{d,j}\sigma_{ij}v_{rel,ij}N_i N_j
\end{equation}
The total collision rate at any point in time is then obtained by summing equation \ref{eqn:binrate} over each histogram bins and can be written as:
\begin{equation}
P_{total} = \sum_{i=1}^{N_H} \sum_{j=1}^{N_H} n_{d,j} \sigma_{ij} v_{rel,ij} N_i N_j.
\label{eqn:ptotal}
\end{equation}
 For collisions between particles of exactly equal mass, $\delta v_t = 0$ when the particle size is very small and no random velocities are excited by the  class 2 eddies for which the particle stopping time $(t_{fric})$ is less than the eddy turnover time at the Kolmogorov length scale \citep[see][]{ormel07b}. Hence, to accurately capture the turbulent velocity contribution to $P_{total}$ the masses of the $N_i$ and $N_j$ particles in equation \ref{eqn:ptotal} are chosen randomly between the bin edges instead of the mass equal to the bin center. Also, similar to \S \ref{sec:oscillationtracking} above, for the particles featuring low statistics, the exact particle mass is used to calculate the rate. Finally, we use Equation \ref{eqn:deltat} to select the timestep $\delta t$, which ranges from a few seconds to $\sim 10^3$ seconds depending on time and  location in the disk. For finely spaced particle mass histograms, Equation \ref{eqn:ptotal} is an excellent match to the kinetic Monte Carlo method. 

Appendix \ref{apn:codetest} shows that our method closely reproduces a range of results from the literature. In Appendix \ref{apn:largestparticles} we check the masses of the largest particles produced by our code against analytical estimates of the maximum mass when turbulence controls the collision speed.

\subsection{Gaining Efficiency}
\label{sec:efficiency}

 In general, Monte Carlo is an {\it O}$(n^2)$ method in which most of the computation time is spent on calculating the rates of collision between different particle pairs. For $N$ number of Monte Carlo particles used in the simulation, the {\it O}$(n^2)$ method involves $N^2$ rate calculations and the CPU time becomes proportional to $N^2$. Our method, being effectively an {\it O}$(n)$ model on the other hand, calculates only $N+N_H^2$ collision rates of which the $N_H^2$ is for computing the time evolution. As long as $N_H \ll N$, the time saved is significant. Here we use $N_H = 80$ which provides good resolution in the mass histogram while satisfying $N_H \ll N$. For example, calculating the steady-state size distribution at a single grid point takes $\sim 3 - 10$ hours to reach steady state depending on the model parameters with a single processor. The global model with MMEN surface density and $\alpha = 10^{-5}$ takes $\sim 3$ days with $48$ processors. For comparison, \citet{drazkowska13} reported their computation time for global model as a few weeks.

\subsection{Vertical Motion of Particles}
\label{sc:settling} 

Throughout most of the disk and for most grain masses, the settling timescale $t_s \approx z / v_z$ exceeds the collision timescale $t_c = 1 / (n_d \sigma \bar{v}_{rel})$, where $\bar{v}_{rel}$ is the average relative speed between a grain of mass $m$ and collision partners of all possible masses. The growth timescale of a grain of a given mass, $t_g = m / \dot{m}$ (where $\dot{m} = m / t_c$) is also short compared to $t_s$ in most of the disk.  We therefore model collisional growth and settling using an operator-splitting approach: during a time interval $\delta t_{settle} = 1$~year (where $\delta t_{settle} \gg \delta t$), we first simulate the collisional evolution of grains in each grid zone. We then connect the grid zones in each vertical column and evolve the grains' heights $z$ above the midplane during the same 1-year period, splitting the time into $\epsilon \times \delta t_{settle}$ fine timesteps; here we use $\epsilon = 0.001$. Our algorithm, which follows \citet{charnoz11}, includes grain settling, diffusion toward the midplane density maximum, and turbulent stirring (see \citet{krijt16a} for a slightly different approach). The algorithm has the following steps:

\begin{enumerate}

\item At each disk radius $R_j$, select $N_s$ Monte Carlo particles
(subscript $u$) of mass $m_i$ to represent {\it each} bin $m_i$ in the particle-mass
histogram.(We find a smooth representation of the vertical number density
distribution with $N_s = 10^5$ and adopt that value for all simulations
presented here.)

\item At each disk radius $R_j$, for each particle mass $m_i$,
distribute the Monte Carlo particles in height $z_u$ above the
midplane according to the vertical number-density distribution
$n_{d,i}(z)$ from the previous step. At $t=0$, the distribution is Gaussian, following the background gas density profile.

\item Calculate $\mu_u$ and $\sigma_u$ for each Monte Carlo particle $u$
according to equations \ref{eqn:mu} and \ref{eqn:sigma}, replacing
$\delta t$ (collision timestep) with $\epsilon \times \delta t_{settle}$
(vertical motion fine timestep).

\item Draw an array of random numbers $r_u$ of size $N_s$ from a
standard normal distribution. Update the particle heights as
\begin{equation}
z_{u,{\rm new}} = z_u + \mu_u + r_u \sigma_u.
\label{eqn:MCsettling}
\end{equation}
Repeat for $1 / \epsilon$ iterations.

\item Update number density corresponding to mass $m_i$ for each cell following the fraction of $N_s$ moved out of or received by any particular cell.

\item Repeat the process for each particle size with non-zero contribution to the total mass.

\item For each vertical cell, calculate the new particle mass
histogram before moving on to the subsequent collision routine.

\end{enumerate}

Since we assume that MRI, which is subsonic, is the main source of disk turbulence, we apply sonic cut-off in the Gaussian distribution of $\Delta z$: no particle may move a greater vertical distance than $\Delta z = c_s \delta t_{settle}$. At the disk surface we adopt an outflow boundary condition so that particles that are turbulently stirred above the top of the grid are contained in a ``ghost zone'' and do not re-enter the grid. Our results are not affected by this assumption as the amount of mass lost to the ghost zone is several orders of magnitude less than the total dust mass. For dust particles in grid zones along the disk midplane we use a reflecting boundary condition.

Appendix \ref{apn:settlindiffusion} contains results of the tests of our vertical motion algorithm.

\section{Results: Dust Growth \& Settling}
\label{sc:results}

Here we present the results of our dust growth, settling, and turbulent diffusion simulations. In \S \ref{ssc:sstime} we discuss the timescales required to reach steady state and compare them with the growth and vertical diffusion time scales. In \S \ref{sc:sizedistributions} we discuss the evolution and steady state of our grain size distributions as a function of disk mass and $\alpha$.

\subsection{Steady State Timescales}\label{ssc:sstime}

  For a single disk mass, the timescale to reach steady state increases with $v_{frag}$ and decreases as the value of $\alpha$ increases.  In all simulations the final snapshots are taken within $\sim 3\times 10^4$ years. For example, for the MMEN model, the timescales of the results shown for disk with $\alpha=10^{-3}$, $10^{-4}$ and $10^{-5}$, and $v_{frag}=50$ cm~s$^{-1}$ are $\sim 23,000$, $27,000$ and $29,000$ years, respectively. However, the growth and vertical diffusion timescales are much shorter than the timescales required to reach steady state. For example, the maximum particle size is achieved in the MMSN model at $10$~au with $\alpha=10^{-5}$ within $\sim 2500$ years. When $\alpha$ is increased, the growth process is affected in two ways. First, the relative velocity of collisions increases due to increased turbulence strength, reaching $v_{frag}$ faster and restricting the growth. Secondly, the collision timescale decreases due to increased collision velocity $(\tau_c \sim 1/n\sigma v_{rel})$. Both these effects reduce the time required to reach the maximum grain size. As an example, the growth timescale for the same MMSN model at $10$~au and $\alpha=10^{-4}$ is $\sim 2200$~years, and $\sim 1600$~years for $\alpha=10^{-3}$. For MMEN disk models the growth timescales for $\alpha = 10^{-4}$ and $10^{-5}$ at $10$~au are $\sim 2300$ and $2900$~years, respectively.

The vertical diffusion timescales generally vary between $\sim 10^4 - 10^6$~years, the longer timescales being relevant for strongly coupled (sub)$\micron$ particles in the inner disk only. However, using local dust-gas coupling by calculating local Stokes numbers results in a shorter diffusion timescales compared to the ones calculated using the midplane values \citep{mulders12}. The enhanced dust abundances in regions near the midplane generate particles slightly bigger than those estimated theoretically using equation \ref{eqn:amax} (see Figure \ref{fig:dustdist}). Hence, although the results do not change significantly beyond $\sim 15000$~years, we run our simulations until $t \sim 30000$~years to be absolutely sure that the size distributions we present here are the true steady state results. 

\subsection{Grain Size Distributions}
\label{sc:sizedistributions}

 \begin{figure*}
	\begin{subfigure}{0.33\linewidth}
		\centering
		\includegraphics[scale=0.33]{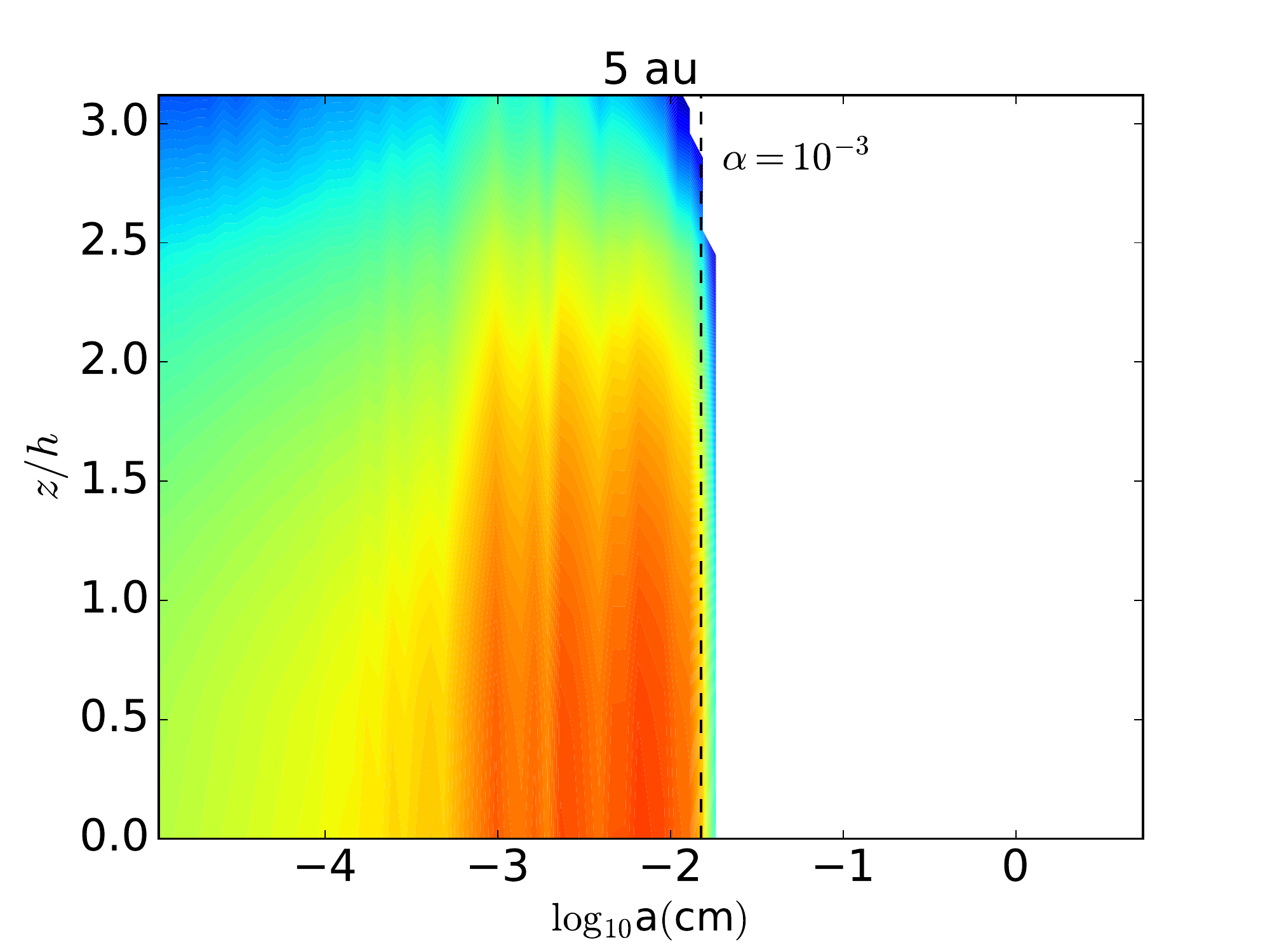}
	\end{subfigure}%
	\begin{subfigure}{0.33\linewidth}
		\centering
		\includegraphics[scale=0.33]{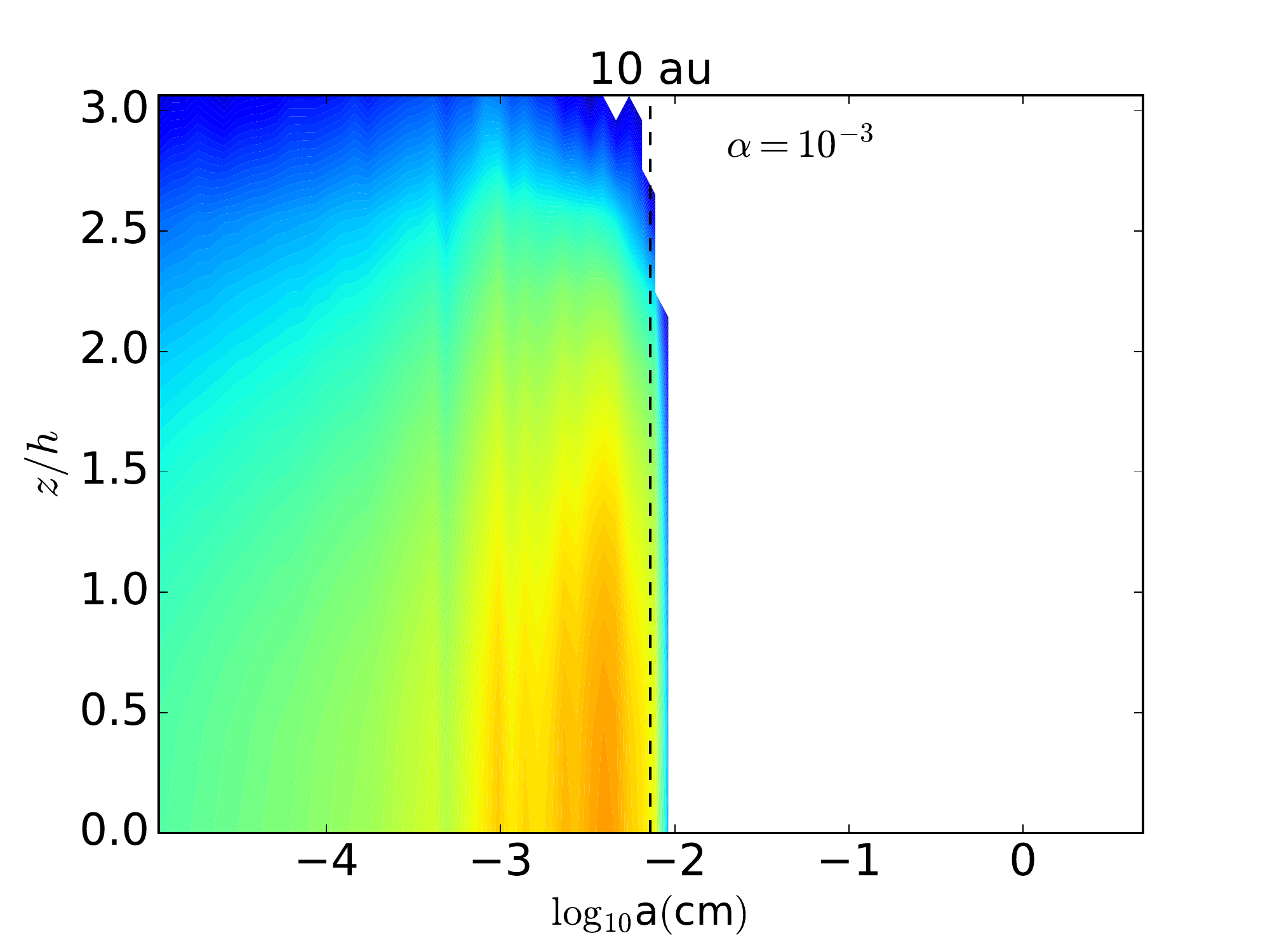}
	\end{subfigure}%
	\begin{subfigure}{0.33\linewidth}
		\centering
		\includegraphics[scale=0.335]{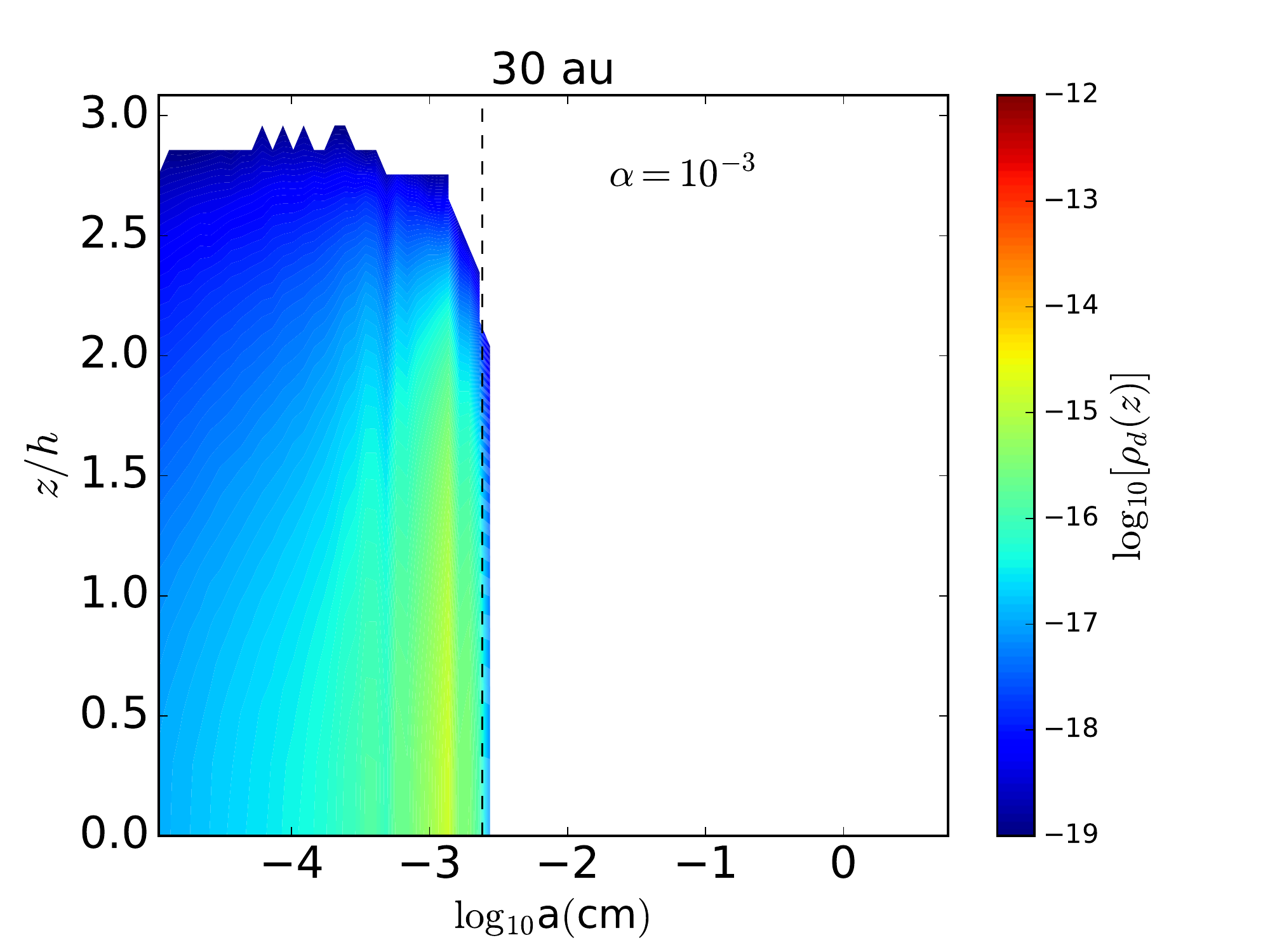}
	\end{subfigure}\\[1ex]
	
	\begin{subfigure}{0.33\linewidth}
		\centering
		\includegraphics[scale=0.33]{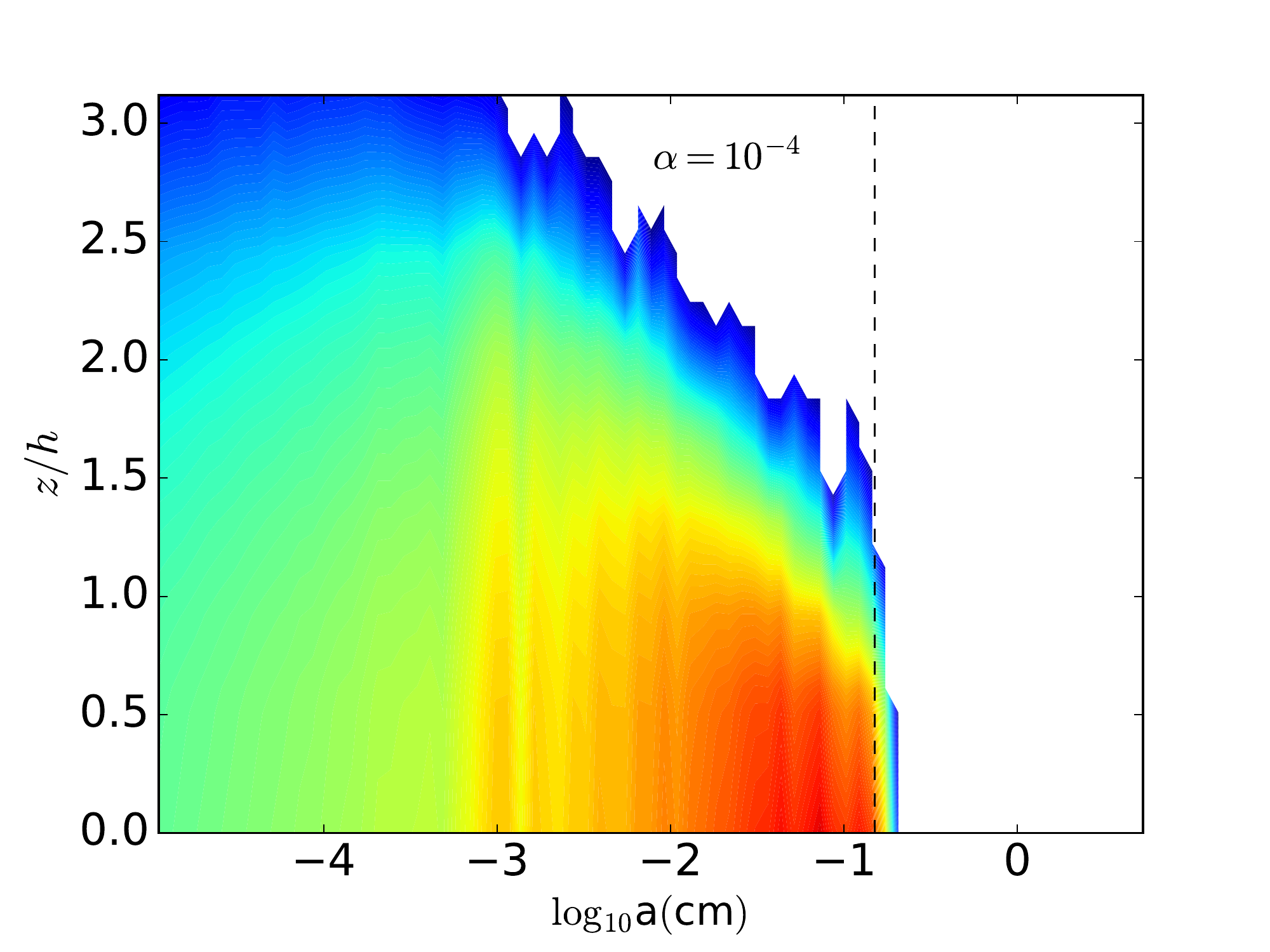}
	\end{subfigure}%
	\begin{subfigure}{0.33\linewidth}
		\centering
		\includegraphics[scale=0.33]{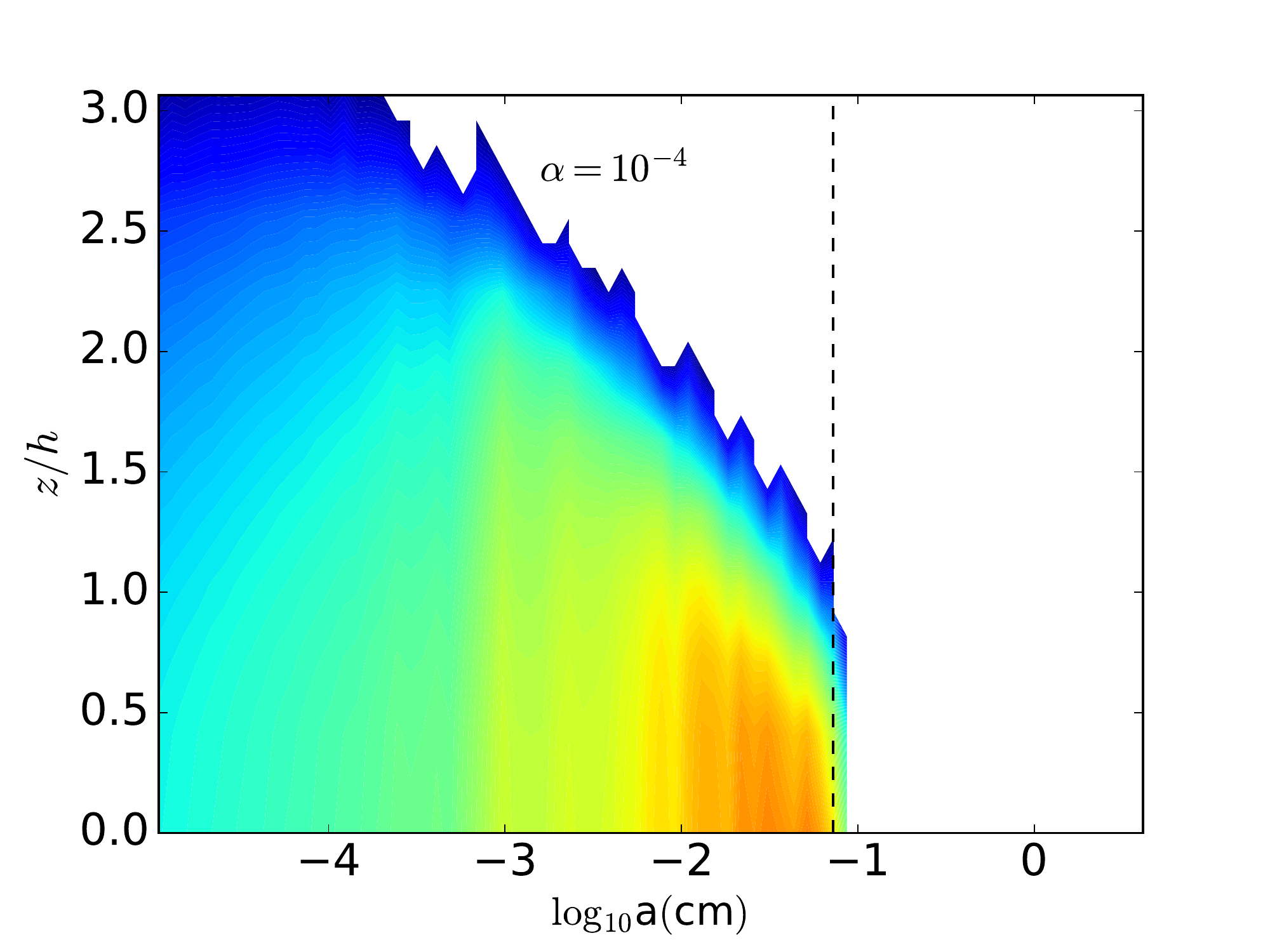}
	\end{subfigure}%
	\begin{subfigure}{0.33\linewidth}
		\centering
		\includegraphics[scale=0.335]{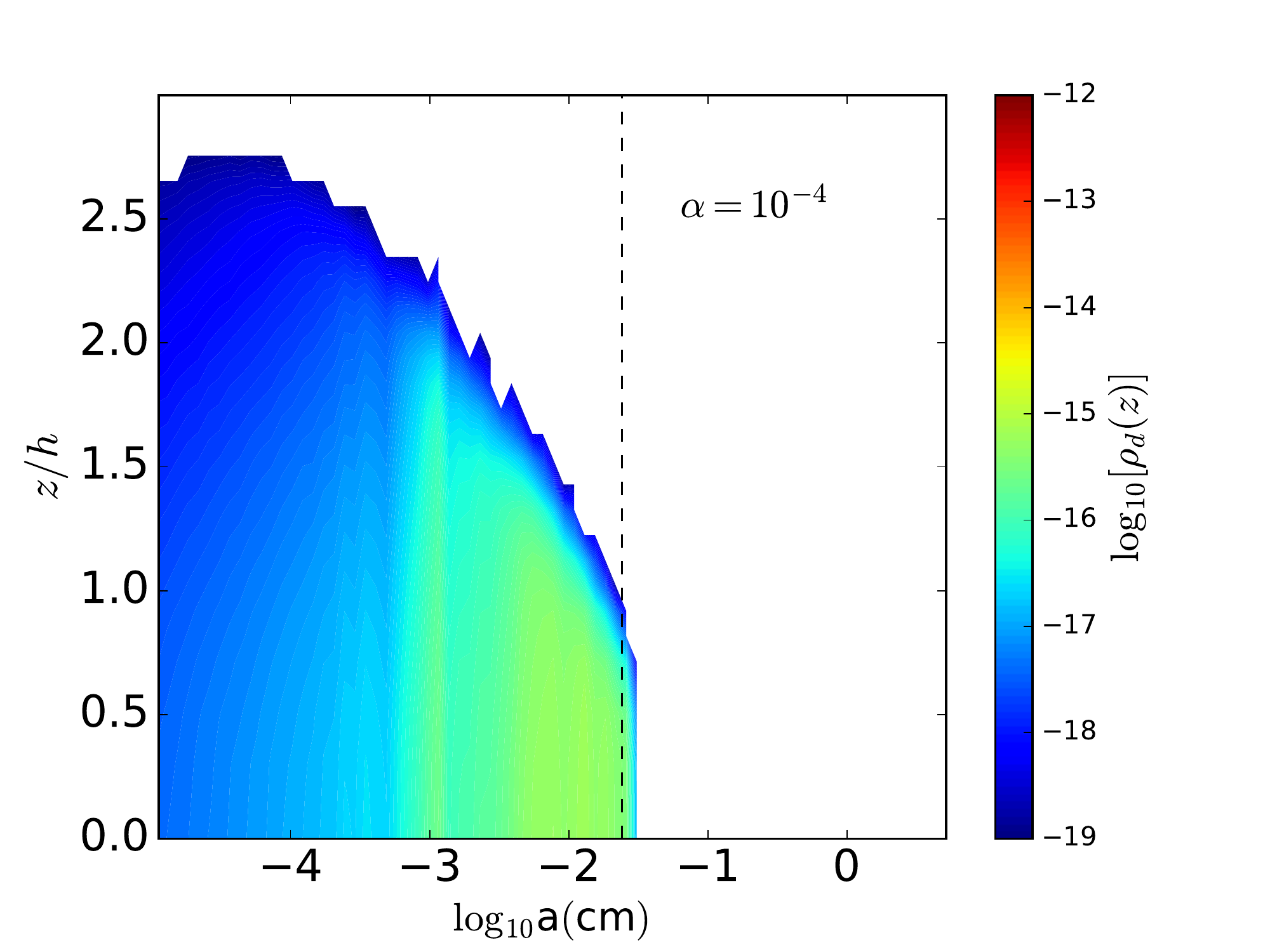}
	\end{subfigure}\\[1ex]
	
	\begin{subfigure}{0.33\linewidth}
		\centering
		\includegraphics[scale=0.33]{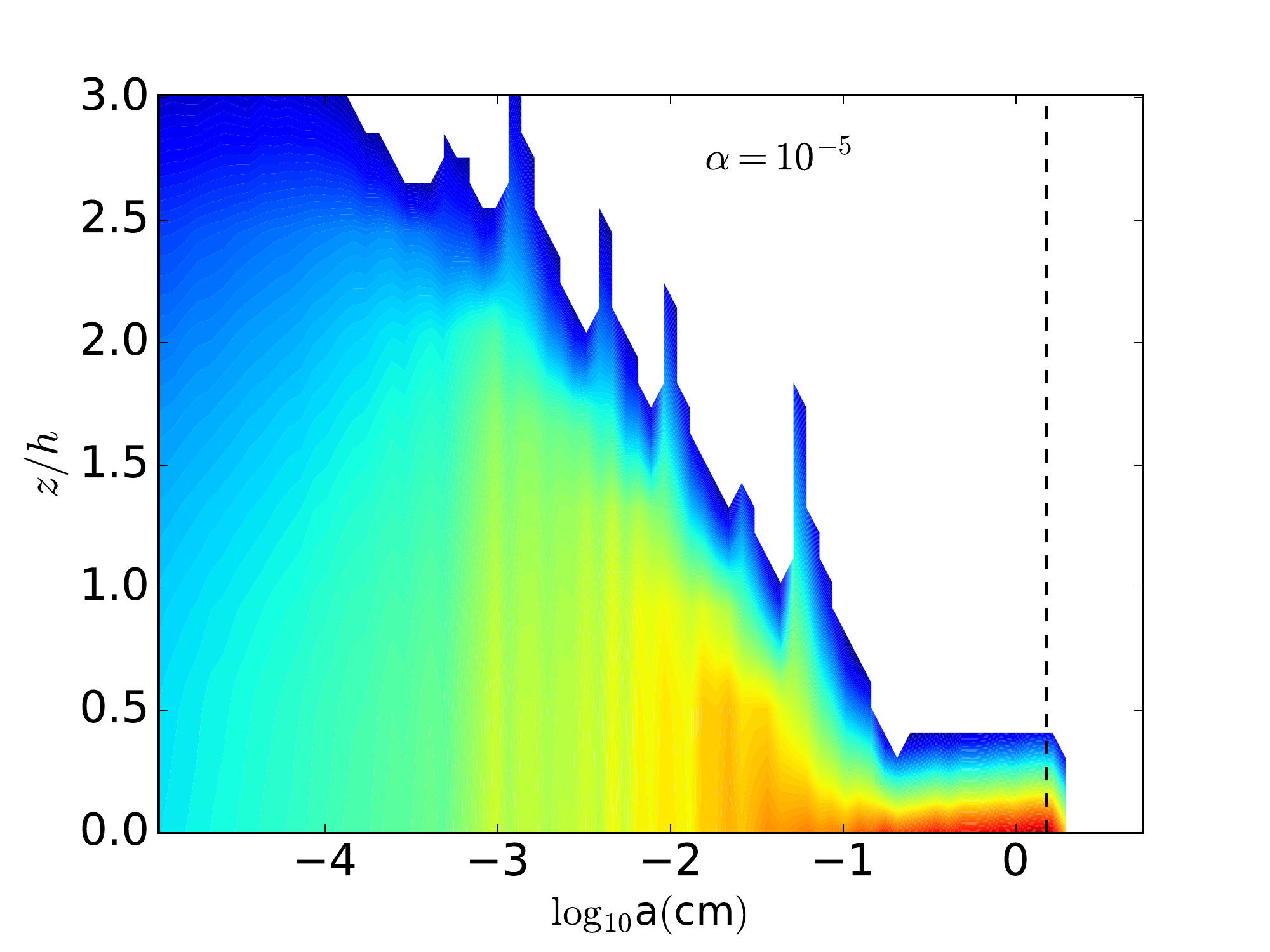}
	\end{subfigure}%
	\begin{subfigure}{0.33\linewidth}
		\centering
		\includegraphics[scale=0.33]{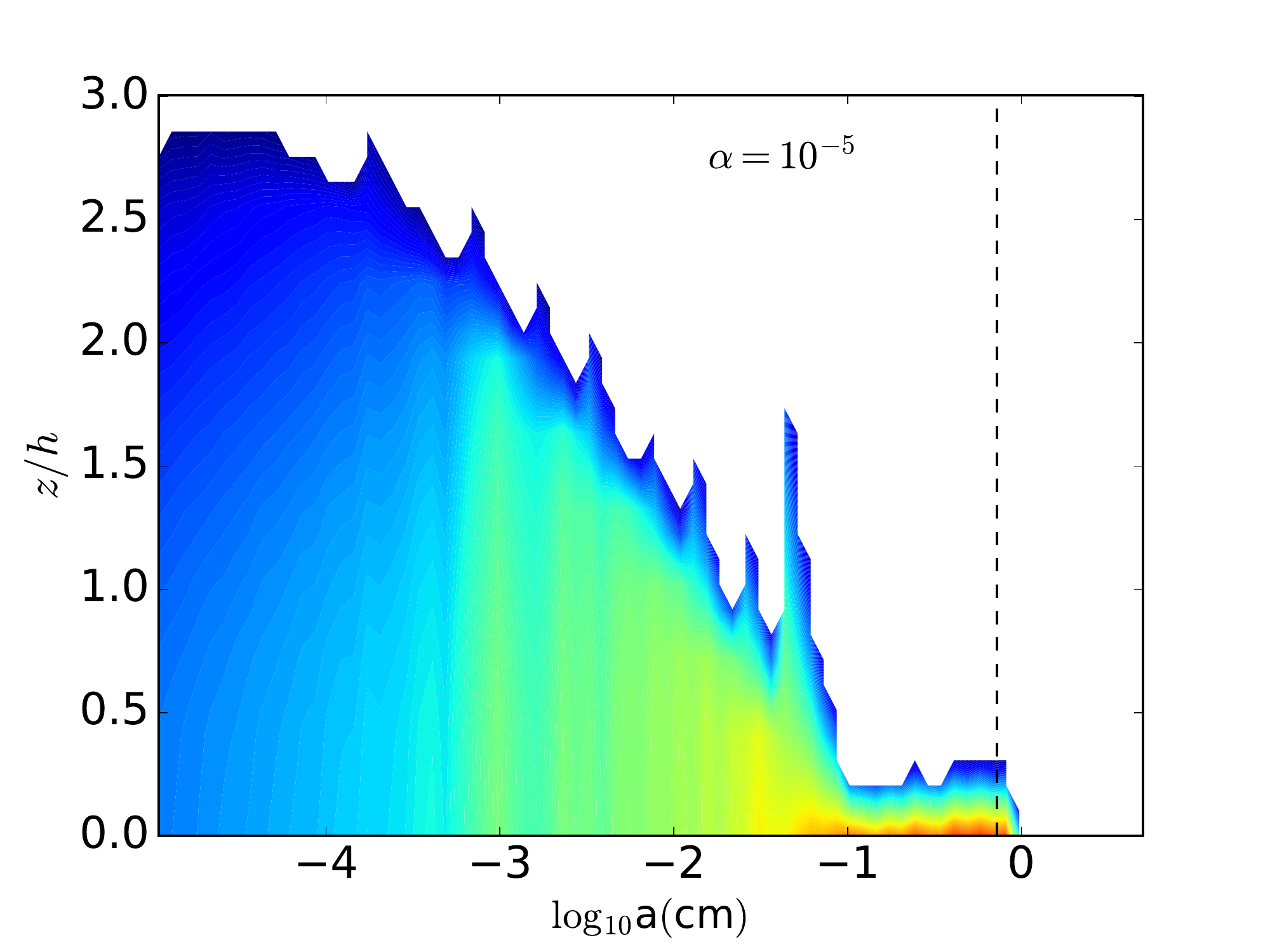}
	\end{subfigure}%
	\begin{subfigure}{0.33\linewidth}
		\centering
		\includegraphics[scale=0.335]{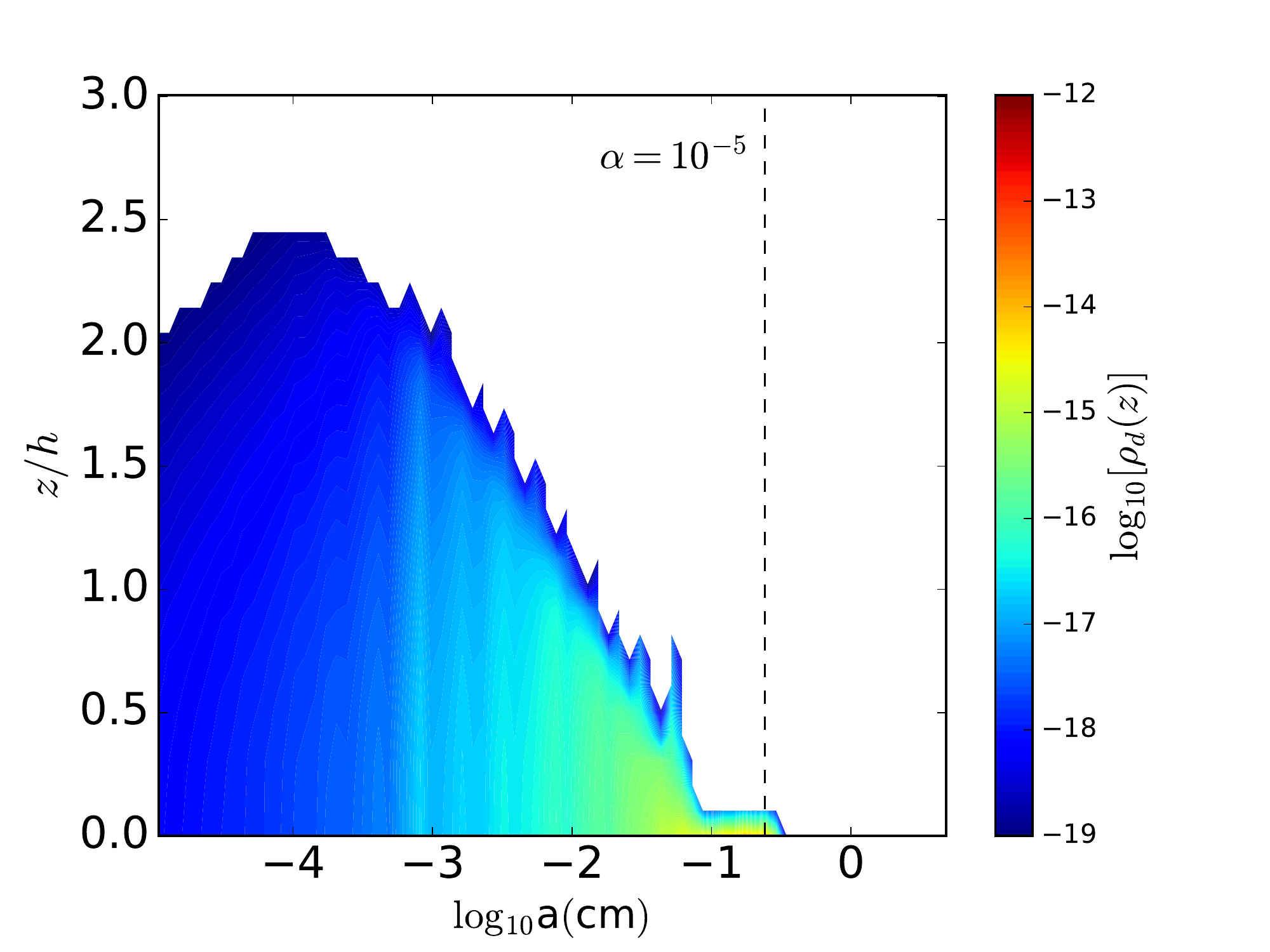}
	\end{subfigure}\\[1ex]
\caption{Steady state dust distribution for MMEN disk model with $\alpha=10^{-3}$ (F1), $10^{-4}$ (F2), and $10^{-5}$ (F3) from top to bottom at a vertical column at $5$, $10$ and $30$ au (from left to right). The colorbar in each case represents dust density (g~cm$^{-3}$ of disk volume) in $\log$ scale. As in figure \ref{fig:mmsncon}, the dotted vertical lines show the maximum dust size permissible according to equation \ref{eqn:amax}. Grain growth and settling as a function of $\alpha$ can be seen comparing figures from different rows. Also, the growth becomes less effective as we move towards the outer disk regions due to the lower gas density and higher Stokes number of dust particles. Similar sized grains attain $v_{frag}$ faster in the outer disk because of the low dust-gas coupling. The spikes in the figure are due to Monte Carlo noise.}
\label{fig:mmencon}
\end{figure*} 

Figure \ref{fig:mmencon} shows steady-state size distributions $\rho_d(a,z)$ (where $a$ is the particle radius) from our suite of constant-$\alpha(R,z)$ MMEN simulations (F1-F3) at 5, 10, and 30 AU. The figure confirms several results from the literature: 

\begin{itemize}

\item {\bf As $\alpha$ decreases, the maximum particle size increases}. In simulations including radial drift and coagulation, but not fragmentation, \citet{brauer08} find a similar trend for the most common particle size (which we also see in our results) but note that the effect is modest: only a factor of two increase in predominant particle size with a $10^2$ decrease in $\alpha$. We find that a factor-of-10 {\it decrease} in $\alpha$ yields nearly a factor-of-10 {\it increase} in maximum particle size at a given radius---true for both our test simulations of the MMSN (not pictured) and our science simulations of the MMEN. Note that this difference originates from the adopted value of the fragmentation velocity as well (see Section \ref{sc:discussion} for more discussion).  \citet{brauer08} suggested that turbulent stirring at higher values of $\alpha$ keeps number densities $n_d(R,z)$ lower, leading to less frequent collisions and frustrated growth. Our simulations have the added effect of more vigorous fragmentation at high $\alpha$ due to the higher relative velocities from stronger turbulence \citep{weid84}.

\item {\bf Particles reach larger sizes in the inner disk than the outer disk}, as seen in figures \ref{fig:mmencon} and \ref{fig:mmsncon}. Figure \ref{fig:stokesnumber} shows Stokes number as a function of $z/h_g$ at 5, 10, and 30~AU for three different grain sizes in the MMSN and MMEN models.  Particles in the outer disk have higher Stokes number at a given grain size and value of $z/h_g$ than particles in the inner disk, so decouple from the gas more easily. Small particles in the outer disk can then attain high values of $v_{rel}$ \citep[e.g.][]{ormel07b} and hit the fragmentation threshold velocity, while the same particles in the inner disk would keep growing \citep{birnstiel09, estrada16}.

\begin{figure}
	\centering
	\includegraphics[scale=0.45]{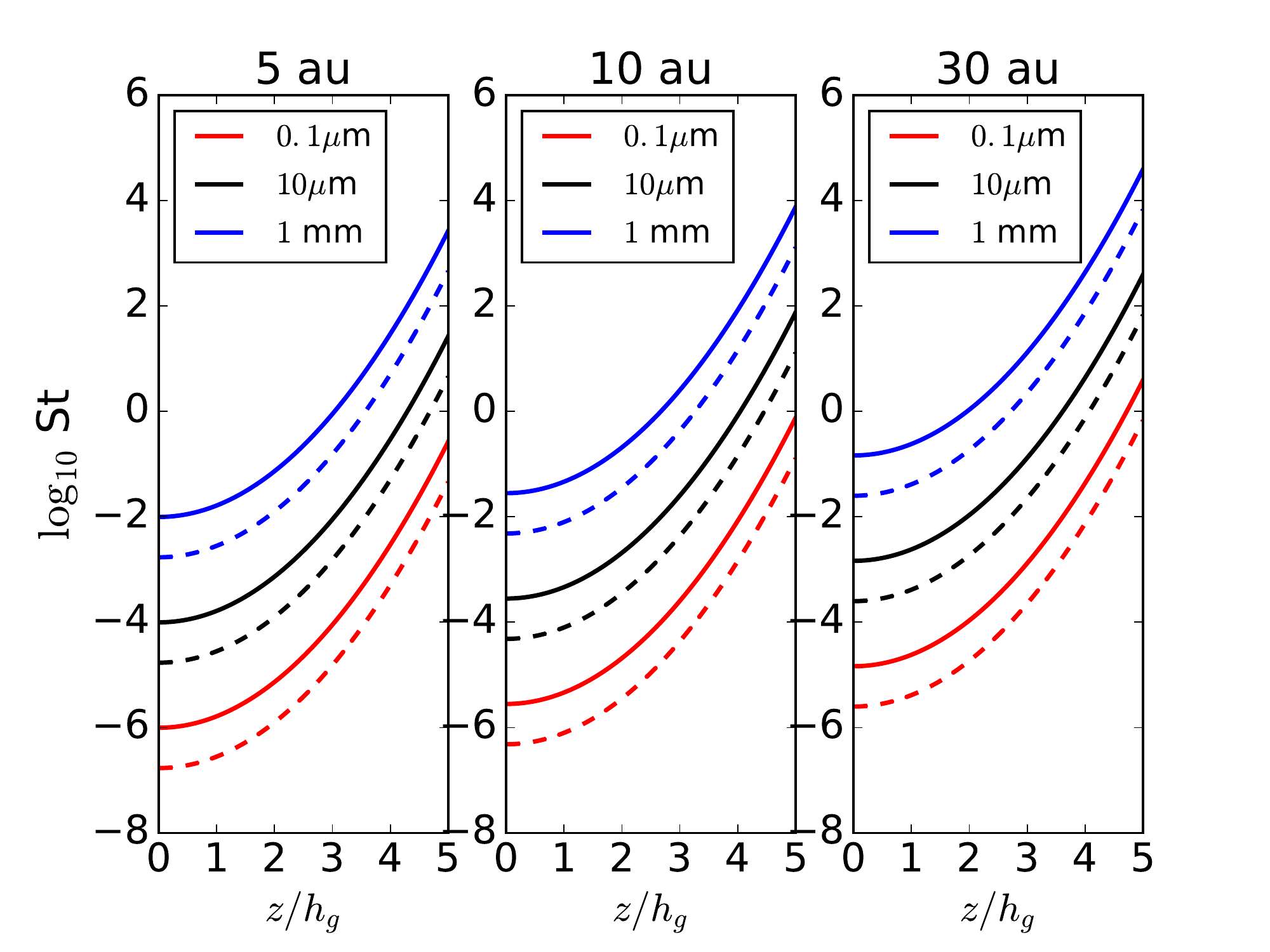}
	\caption{Stokes number as a function of height for particles of
	different size. The solid lines show the MMSN model and the
	dashed lines represent the MMEN (model F2).}
	\label{fig:stokesnumber}
\end{figure}

\item {\bf Even weak turbulence can keep particles as large as 0.1~mm stirred into the disk's upper layers} \citep[e.g.][]{dubrulle95}. Figure \ref{fig:dustdist} shows $\rho_d(a)$ at the midplane and $3 h_g$ at 5~AU for model T2 (MMSN, $\alpha = 10^{-4}$). Although $\alpha = 10^{-4}$ is near the lower limit of expected turbulent efficiency due to the likely onset of hydrodynamic instabilities where MRI is inactive \citep[e.g.][]{nelson13}, it is still possible to find 0.1-mm particles at $3 h_g$. Local, single-cell simulations without any vertical motion (solid lines) show that the maximum particle size that can grow at $3 h_g$ is only $\sim 30 \: \mu$m; turbulent diffusion introduces particles with five times larger radii that grew near the midplane. 

 In all our simulations, we have used an outflow boundary condition where particles leaving the surface of the disk are not tracked.However, as can be seen from Figure \ref{fig:mmencon}, the dust density in the upper layers of the disk at $\sim 3h_g$ is already several orders of magnitude less than that of the midplane. The same trend can be observed in our test simulations T1-T4 as well in Figure \ref{fig:mmsncon}. As a result, an insignificant grain mass is lost over the course of the simulation ($\Delta m / m \la 10^{-6}$). Also, the vertical temperature profile becomes flat at the upper layers of the disk (See Figure \ref{fig:tempvert}) which suggests that our choice of the particular boundary condition at the disk's surface does not affect the vertical temperature stratification.

 \begin{figure}
	\centering
	\includegraphics[scale=0.45]{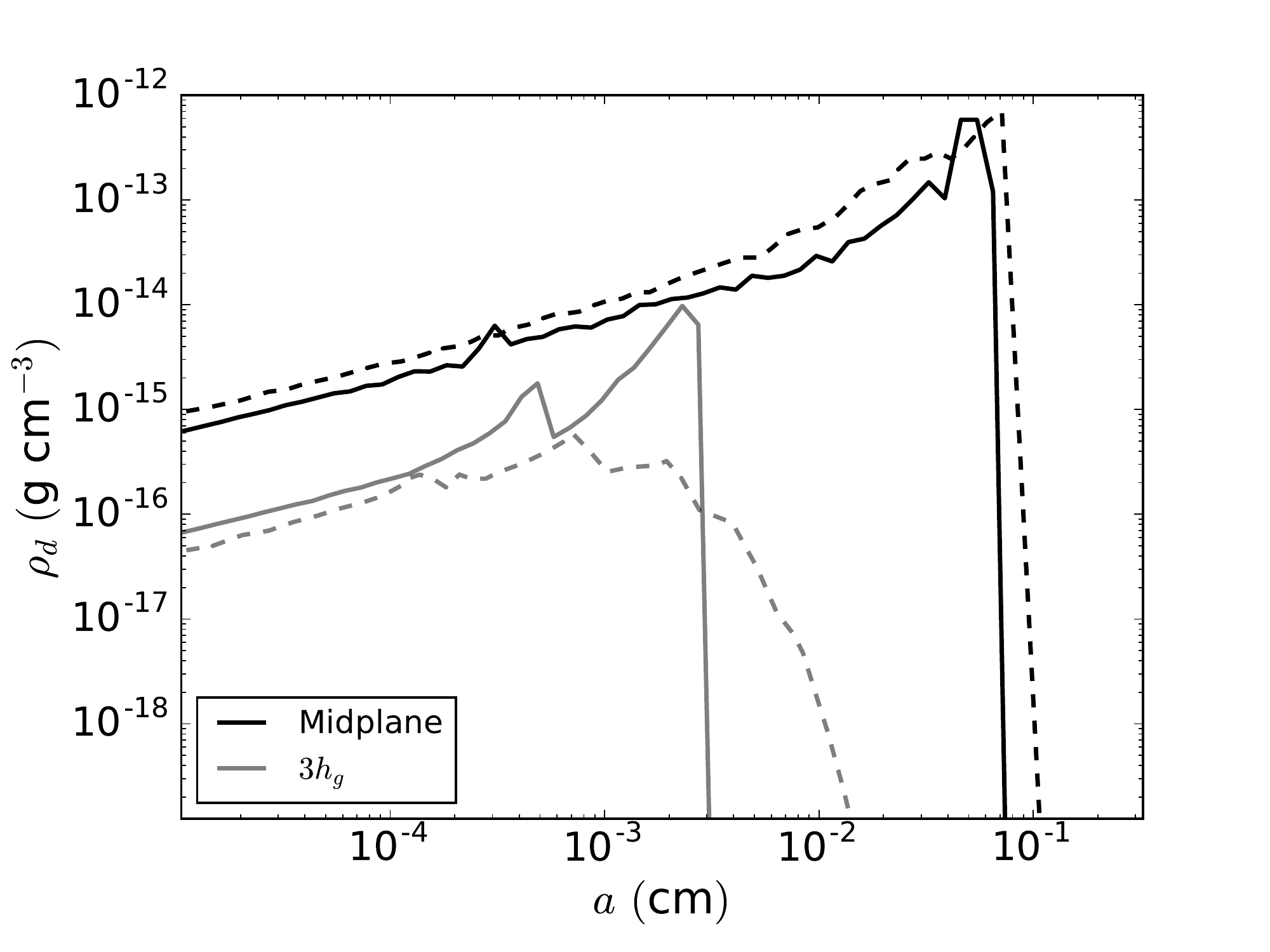}
	\caption{The steady state dust abundance at midplane and $3$ scale-heights above midplane for an MMSN disk with $\alpha=10^{-4}$ at $5$ au. The solid lines show the dust distribution that we would have achieved from a local simulation and the dotted curve show the distribution obtained from simulation with full dust dynamics in the vertical direction implemented. The extra growth at midplane takes place due to enhanced dust abundance from vertical settling. The abundance of dust grains of sizes $\sim$ a few tens of micron at $3$ scale heights is not due to the local collisional growth, rather can be attributed to the vertical turbulent stirring.}
	\label{fig:dustdist}
\end{figure}

\end{itemize}

\begin{figure*}
	\begin{subfigure}{0.33\linewidth}
		\centering
		\includegraphics[scale=0.32]{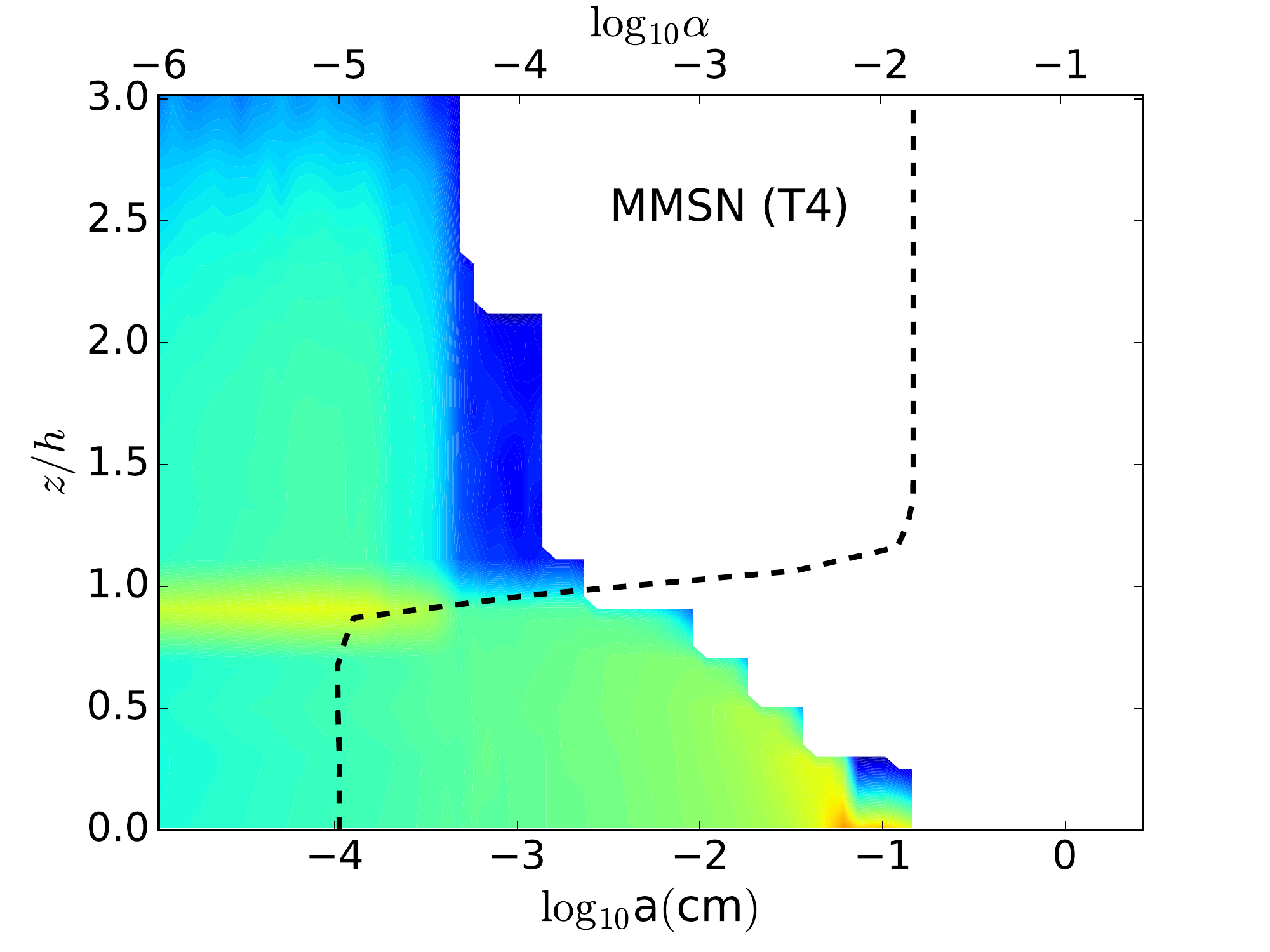}
	\end{subfigure}%
	\begin{subfigure}{0.33\linewidth}
		\centering
		\includegraphics[scale=0.32]{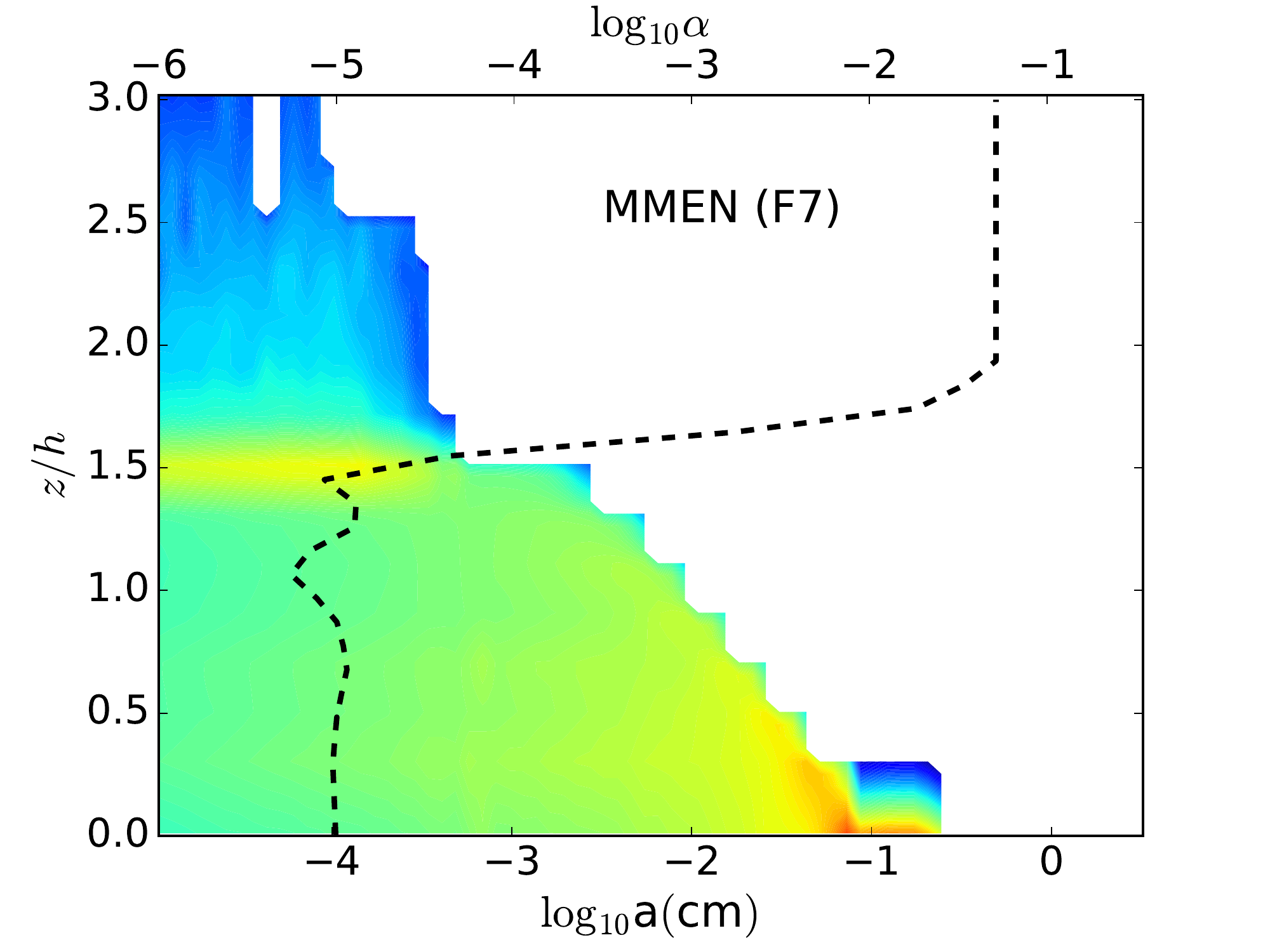}
	\end{subfigure}
	\begin{subfigure}{0.33\linewidth}
		\centering
		\includegraphics[scale=0.32]{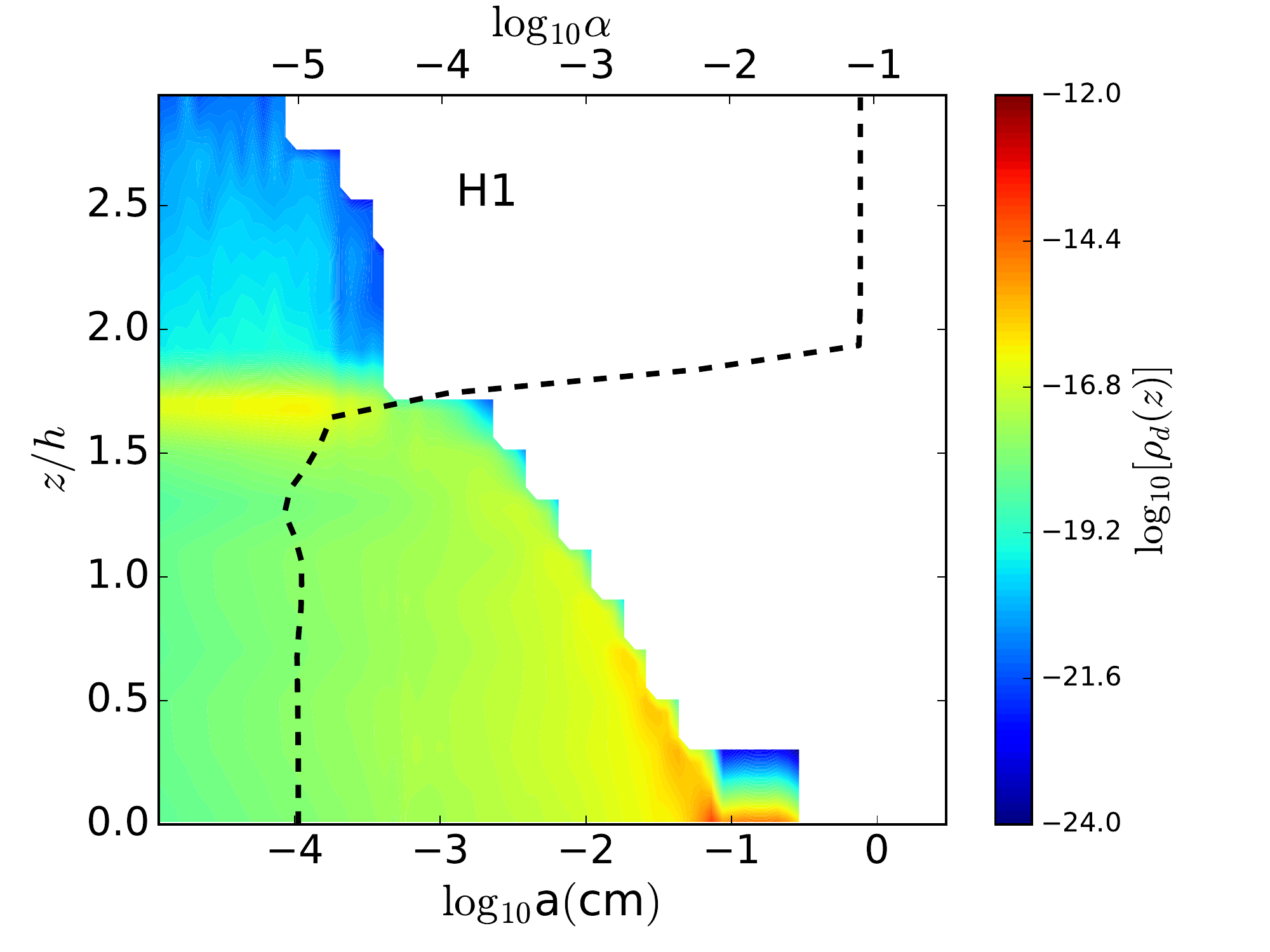}
	\end{subfigure}
	\caption{Steady-state dust density distribution with variable $\alpha(R,z)$ profile for MMSN (left), MMEN (middle) and H1 (right) disk models at $50$ au. $v_{frag}=100$~cm~s$^{-1}$ for all three cases. The colorbar represents dust density (mass per unit disk volume) in $\log$ scale. The values of $\alpha(R,z)$, obtained from the ionization-recombination chemistry model of \citet{landry13}, are shown with black dashed line with the axis on the top of each plot. The pattern of the steady-state distributions are markedly different from that for constant $\alpha$ profile shown in figure \ref{fig:mmencon}. Dust becomes sequestered in the midplane dead zone, where weak turbulence prevents grains from getting kicked upward. In all the simulations, a slightly higher concentration of smaller dust grains is obtained at heights above where $\alpha(z)$ makes a sharp transition. However, this feature may not be present for an $\alpha(R,z)$ profile  evolving in time with the evolution of gas-to-solid ratio. }
	\label{fig:vardisk}
\end{figure*}

We now turn to disk models with variable $\alpha(R,z)$.  Figure \ref{fig:vardisk} shows steady-state dust density distributions $\rho_d(a,z)$ at 50 au for simulations T4 (MMSN), F7 (MMEN) and H1. The black dashed lines show $\alpha(R,z)$ as calculated using the methods  of \citet{landry13}. In each disk, at $50$~au we can see the existence of a dead-zone: the midplane is quiescent, with $\alpha(z=0) = 10^{-5}$ due to suppression of MRI turbulence (a value that might be low enough to trigger hydrodynamical instabilities, which is the case for F8); and the surface layers have strong turbulence \citep[e.g.][]{gammie96} (though the turbulence may be confined to heights above the upper z-axis limit in Figure \ref{fig:vardisk}). Unsurprisingly, there is a strong vertical stratification in dust density that mirrors the rapid change in $\alpha(z)$. Any particle that dips below $z / h_g = 1.5$--2 is unlikely to be kicked upward again due to the weak turbulence, so grains stay sequestered near the midplane. Also, we can see a local accumulation of small dust grains with $a \la 10 \micron$ at a height where $\alpha$ suffers a sharp transition. While disks with constant $\alpha(R,z)$ have dust density profiles that are vertically Gaussian (Figure \ref{fig:verttest}), disks with variable $\alpha(R,z)$ have vertical dust density profiles that are strongly non-Gaussian, having a sharp cutoff at some height $z$.


The fact that different disk models used in this paper have different size distributions $\rho_d(a,R,z)$ means that they will have different vertical optical depths, angles at which starlight is absorbed, and temperature structures. We explore the opacity, optical depth, and gravitational stability of our model disks in the next section.

 \section{Opacity Model and Thermal Evolution}\label{sec:opacity}

After computing the dust number density $n(a,R,z)$, we require an opacity prescription to find the disk temperature $T(r,z)$. For the majority of the disk mass, which lies near the midplane, it is reasonable to assume that the gas temperature and dust temperature are equal. For temperatures less than $\sim 2000$ K, dust is the dominant opacity source \citep[e.g.][]{kama09}, so we neglect opacity contributions from gas. We adopt the ``Utilitarian opacity model'' from \citet{cuzzi14} (C14 hereafter) to calculate the extinction efficiencies $Q(\lambda,a)$ as a function of wavelength and dust size. Following the calculation of extinction efficiency, the opacity per gram of dust is calculated as
\begin{equation}\label{eqn:opac}
\kappa_{\lambda}(a) = \frac{3}{4}Q(\lambda ,a)\frac{1}{a \rho_m}.
\end{equation}

\begin{figure}
\centering
\includegraphics[scale=0.45]{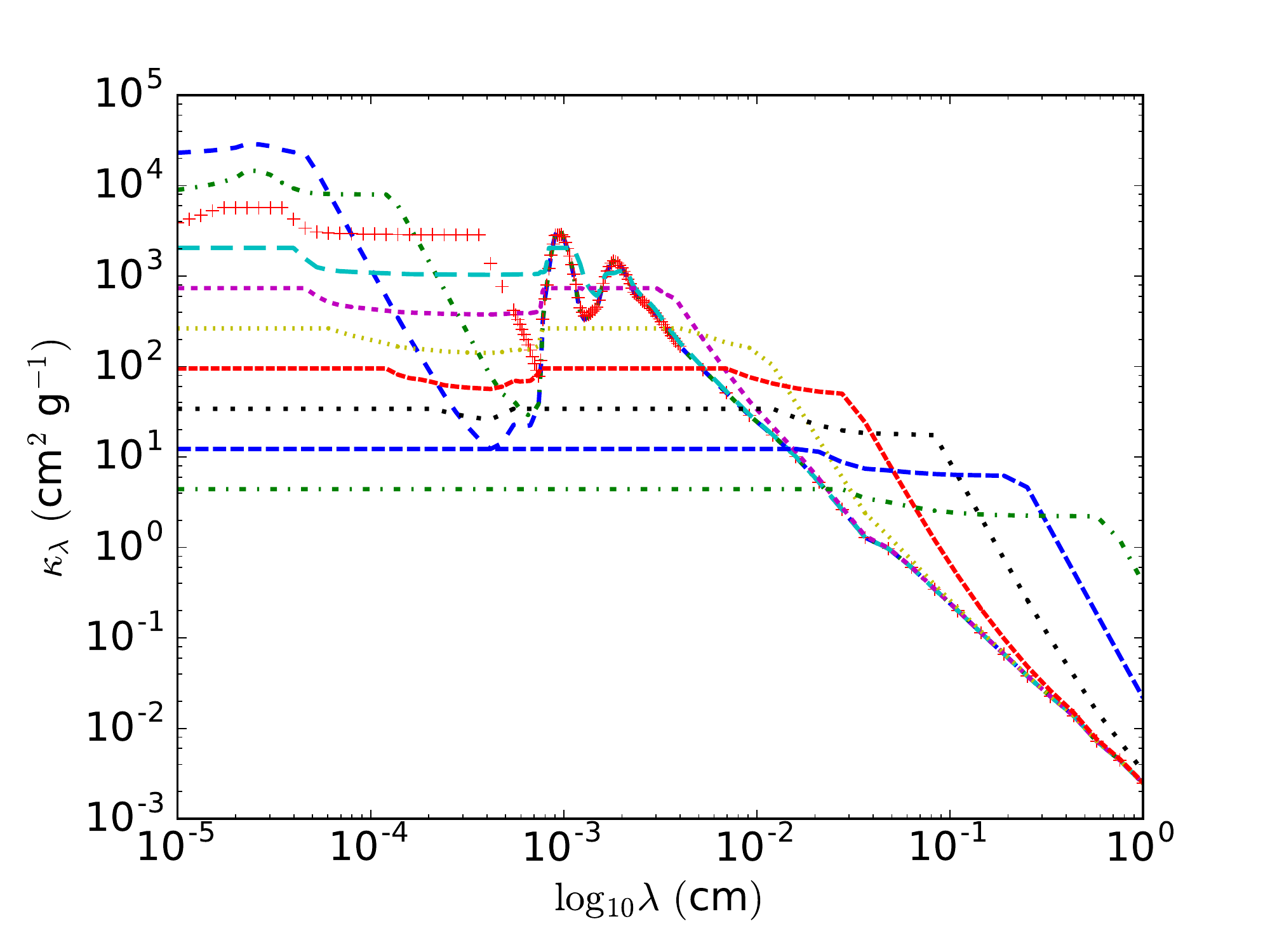}
\caption{Opacity as a function of wavelength for $100\%$ silicate grains. The opacities shown are for dust sizes between $0.1~\upmu$m to $1$ mm, from top to bottom, equi-spaced in $\log$ scale. The ratio of the particle diameters between any two successive lines in the figure is 2.78.}
\label{fig:dustopacity}
\end{figure}

We assume that the composition of dust particles is $100\%$ astronomical silicate, [Fe$_x$Mg$_{1-x}$]SiO$_3$ with $x=0.3$, and has a material density $\rho_m=3.4$ g~cm$^{-3}$. The real and imaginary refractive indices are taken directly from the MPIA website\footnote{\url{https://www2.mpia-hd.mpg.de/home/henning/Dust_opacities/Opacities/opacities.html}}. The reader is advised to look into C14 for further details of the model. See figure \ref{fig:dustopacity} for the dust opacities calculated using C14 and used in radiative transfer calculations.

 We compute the temperature profile of the disk using RADMC-2D \citep{dullemond04} which performs Monte-Carlo dust continuum radiative transfer based on the method of \citet{bjorkman01} with modifications to produce smoother results with a reasonable number of photons. The working principle of this code involves dividing the luminosity of the source into a finite but large enough number of photon packets, each with the same amount of energy. However, the number of physical photons, actually contained by each packet, depends on the frequency. After a photon packet is injected into the disk with an assigned random frequency chosen from the spectral energy distribution of the central star, the code follows the packet through absorption and scattering by dust grains. The photons once absorbed by the dust get re-emitted immediately with frequencies randomly chosen from the difference between the thermal spectra before and after the packet is absorbed. This process continues until the photon escapes the disk through its physical boundary. The increase in temperature of the cell, where absorption/re-emission or scattering takes place,  is computed after each event. The frequency of the incident photon determines the dust opacity which is used for temperature re-calculation. 

To use the code, we treat dust of each size of our histogram as separate species and provide RADMC the monochromatic absorption and scattering opacity {\it per gram of dust} calculated using equation \ref{eqn:opac}.  Based on convergence tests, we find that we achieve an accurate temperature profile using $10^6$ photon packets. 

Note that the vertical temperature structure at a particular column at $t=0$ obtained from RADMC is different than the canonical power-law temperature profile given by Equation \ref{eqn:tempt0}, which assumes the vertical column to be isothermal. We use isothermal prescription to define the initial gas scale height $h_g$ which remains the same throughout the simulation as the dust physics is implemented on a fixed gas background.This implies that our steady state solutions are not in hydrostatic equilibrium. The disk interior is cooler that the initial stage, so restoring vertical force balance would make the disk even thinner.

\subsection{Results: Opacity, Temperature and Gravitational Stability}
\label{sc:stability} 

\begin{figure*}
	
	\begin{subfigure}{0.49\linewidth}
	\centering
	\includegraphics[scale=0.47]{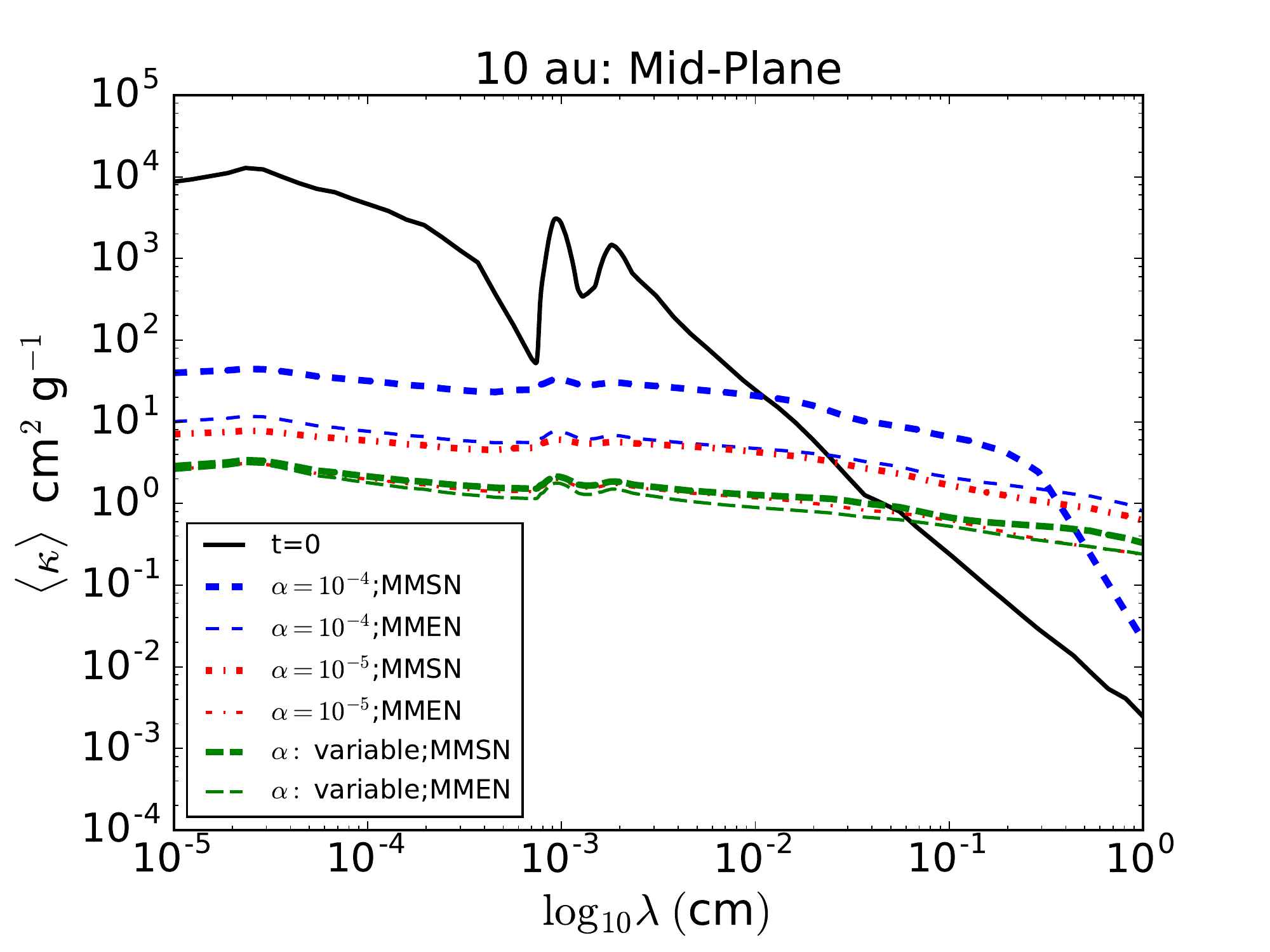}
	\caption{}
	\end{subfigure}	
	\begin{subfigure}{0.49\linewidth}
	\centering
	\includegraphics[scale=0.47]{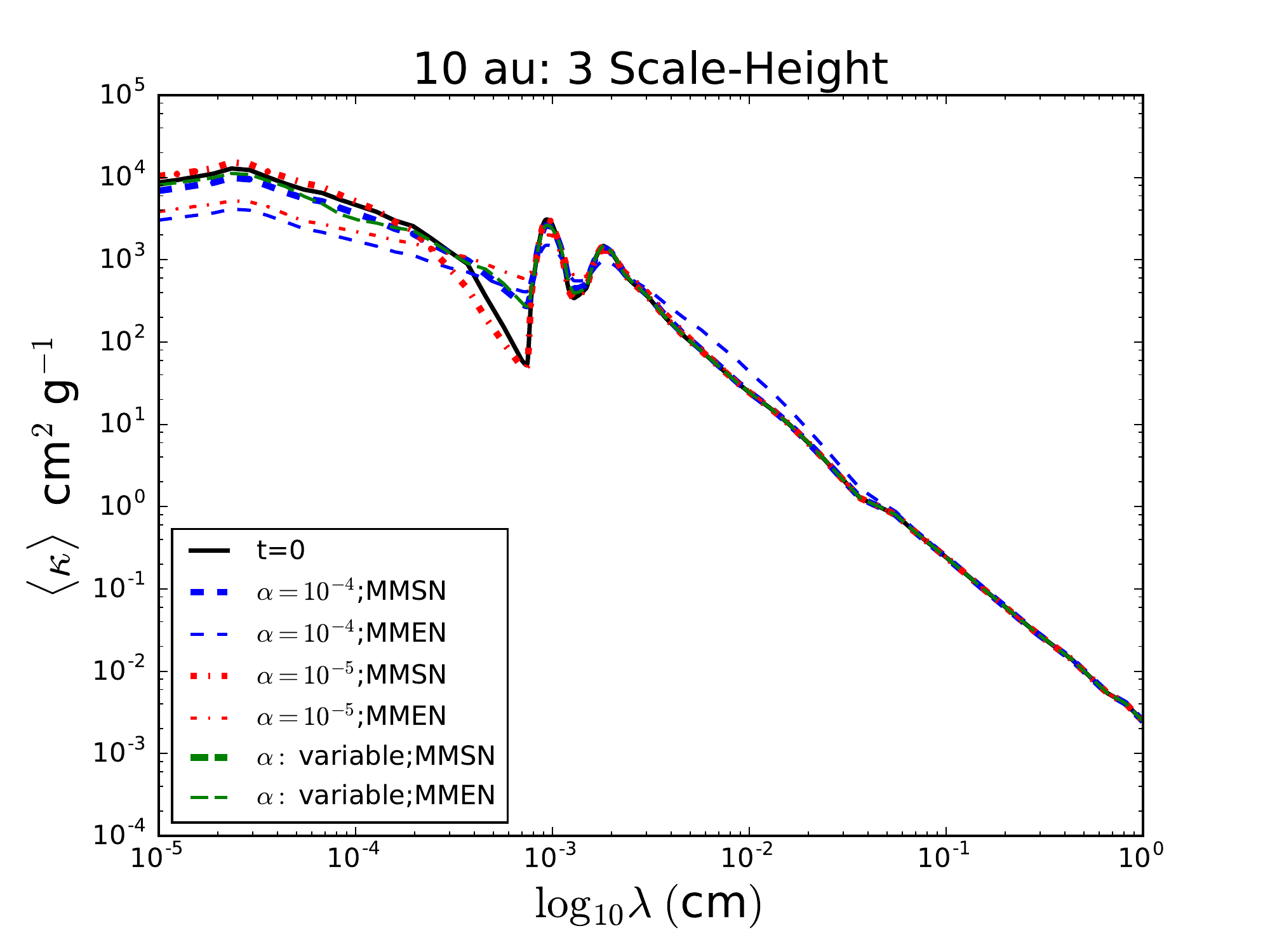}
	\caption{}
	\end{subfigure}\\[1ex]
	
	\begin{subfigure}{0.49\linewidth}
	\centering
	\includegraphics[scale=0.47]{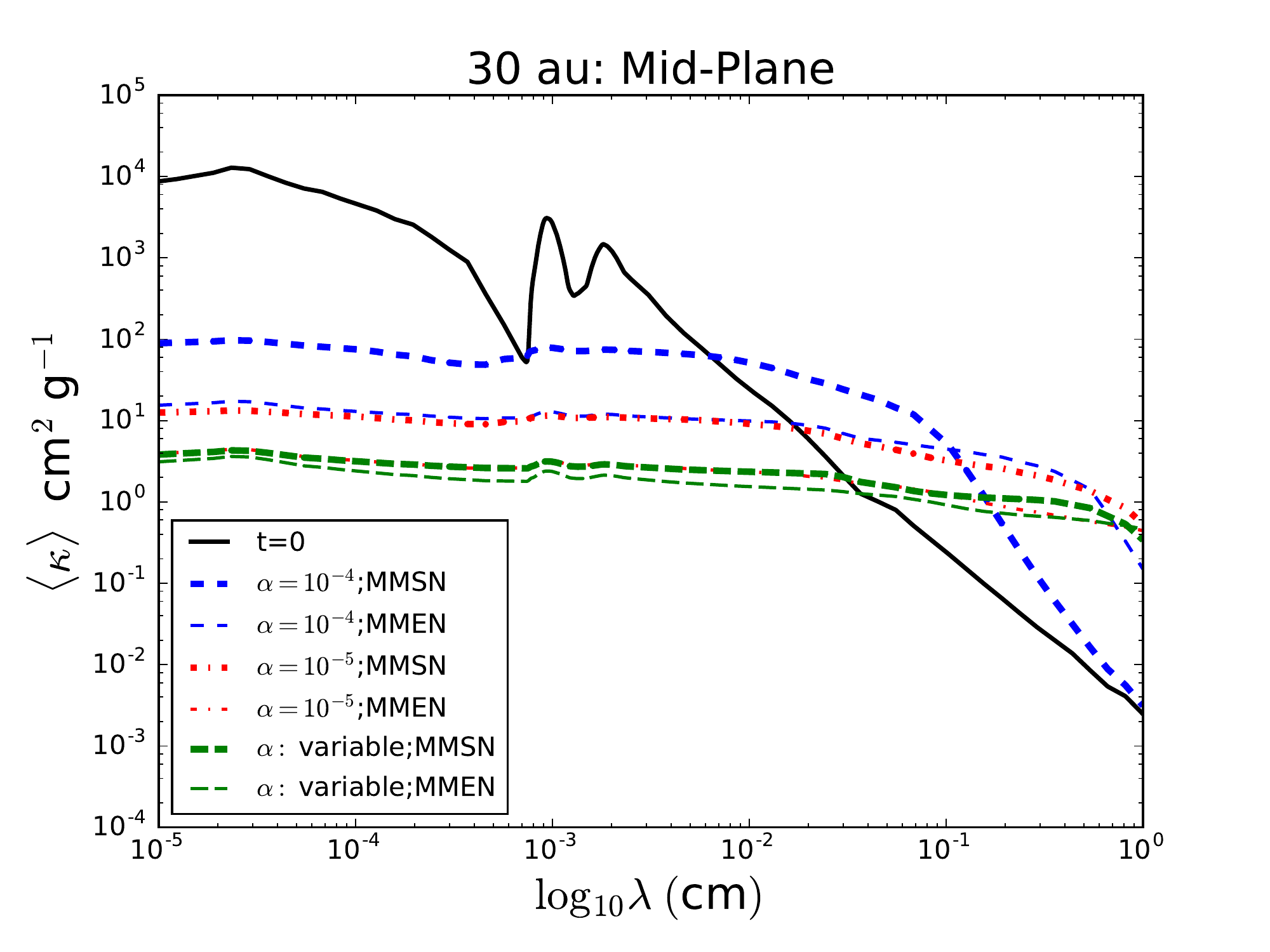}
	\end{subfigure}
	\begin{subfigure}{0.49\linewidth}
	\centering
	\includegraphics[scale=0.47]{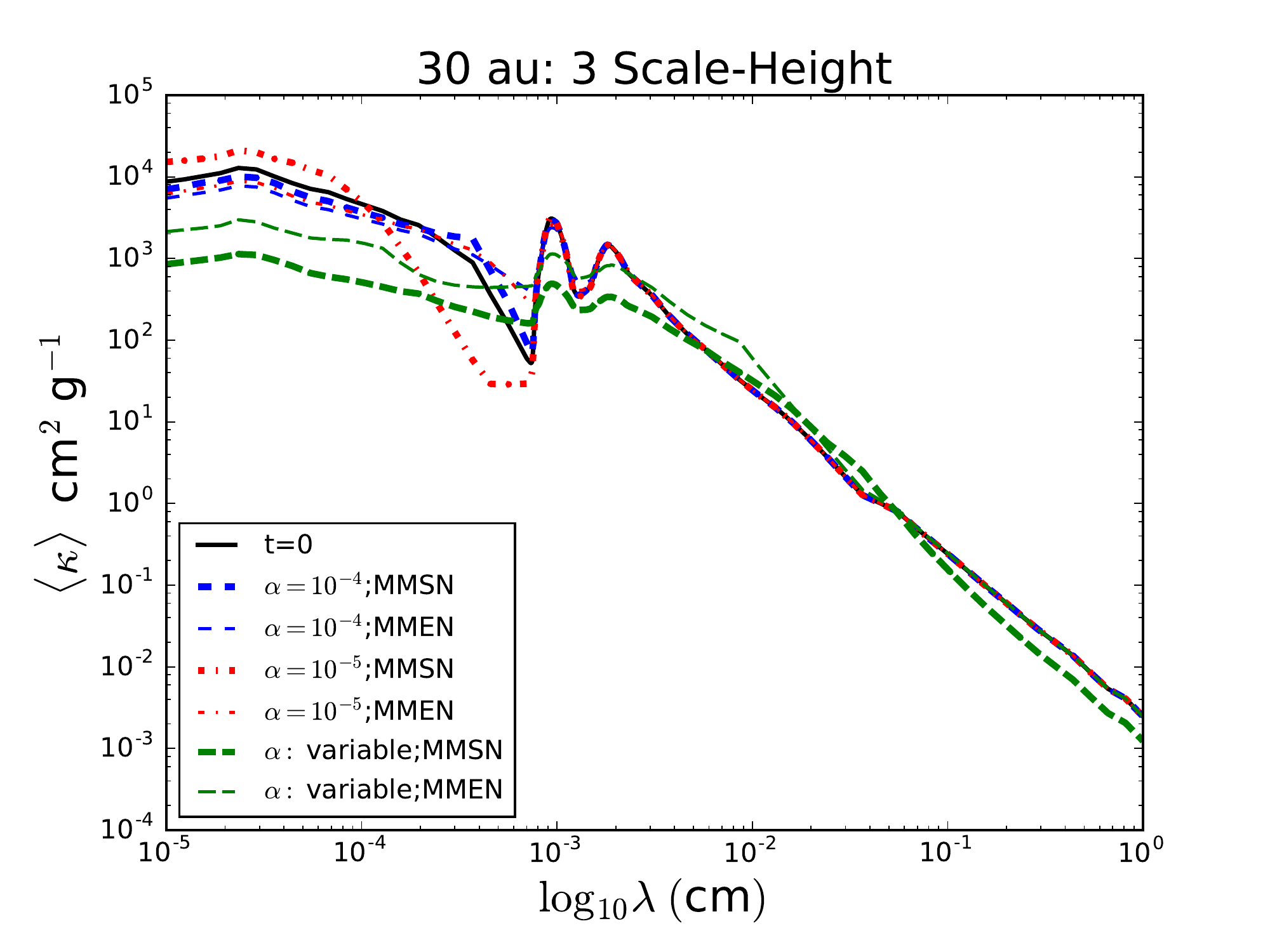}
	\end{subfigure}
\caption{Opacities $\langle\kappa(\lambda)\rangle_{\rho_d}$ from the $t = 0$ disk (solid line) and the steady-state size distributions (dashed lines) for the MMSN and MMEN disk models. Grain growth significantly reduces the short-wavelength opacity and increases the long-wavelength opacity at the midplane, while having much weaker effects at three scale-heights. Depending on the position in the disk and strength of the turbulence, the quantity $\langle\kappa\rangle_{\rho_d}$ can differ by more than an order of magnitude.}
\label{fig:densityweightedopacity}
\end{figure*}

In figure \ref{fig:densityweightedopacity} we show opacity as a function of wavelength for $t = 0$ and steady-state size distributions in the MMSN (T2-T4) and MMEN (F5-F7) models, all with $v_{frag}=100$~cm~s$^{-1}$. We define the mean opacity of a grain size distribution $\langle\kappa\rangle_{\rho_d}$ as
\begin{equation}
\langle\kappa(\lambda)\rangle_{\rho_d}=\frac{\int
\rho_d(a)\kappa_{\lambda}(a)\,da}{\int \rho_d(a)\,da}.
\label{eqn:kappaavg}
\end{equation}
Grains absorb and emit light most efficiently at wavelengths  shorter than $2 \pi a$ at which point the profile of opacity of dust starts to drop. We see that in steady state, the opacity contribution from small grains at the disk midplane has decreased by 2--3 orders of magnitude from $t = 0$ due to grain growth. Meanwhile, the opacity contribution from particles with $a \ga 30 \micron$ has increased. At height $3 h_g$, the mean opacity across the size distribution does not evolve as much between $t = 0$ and steady state, though an opacity deficit develops from 1--5~$\micron$ as the 0.1~$\micron$ monomers are left behind {\bf due to selective grain settling. This reduction in opacity is also prominent in the top-most curve of figure \ref{fig:dustopacity}.  } The silicate resonance features at 10--20~$\micron$, which are produced by warm grains of $1 \la a \la 10 \micron$, also weaken in the midplane, nearly disappearing for the disks with $\alpha = 10^{-5}$.  The decrease in opacity at short wavelengths can be attributed to the collisional growth of dust which reduces the abundance of particles with sizes $2\pi a \la \lambda$, for which the opacity curve is wavelength independent. Larger dust particles, due to their sizes exceeding short wavelengths, gain no extinction efficiency but decrease in physical area per unit mass by a factor of the radius. This is also the reason why the opacity increases at longer wavelengths as dust particles reach those sizes due to collisional growth.

 \begin{figure*}
\begin{subfigure}{0.49\linewidth}
\centering
\includegraphics[scale=0.47]{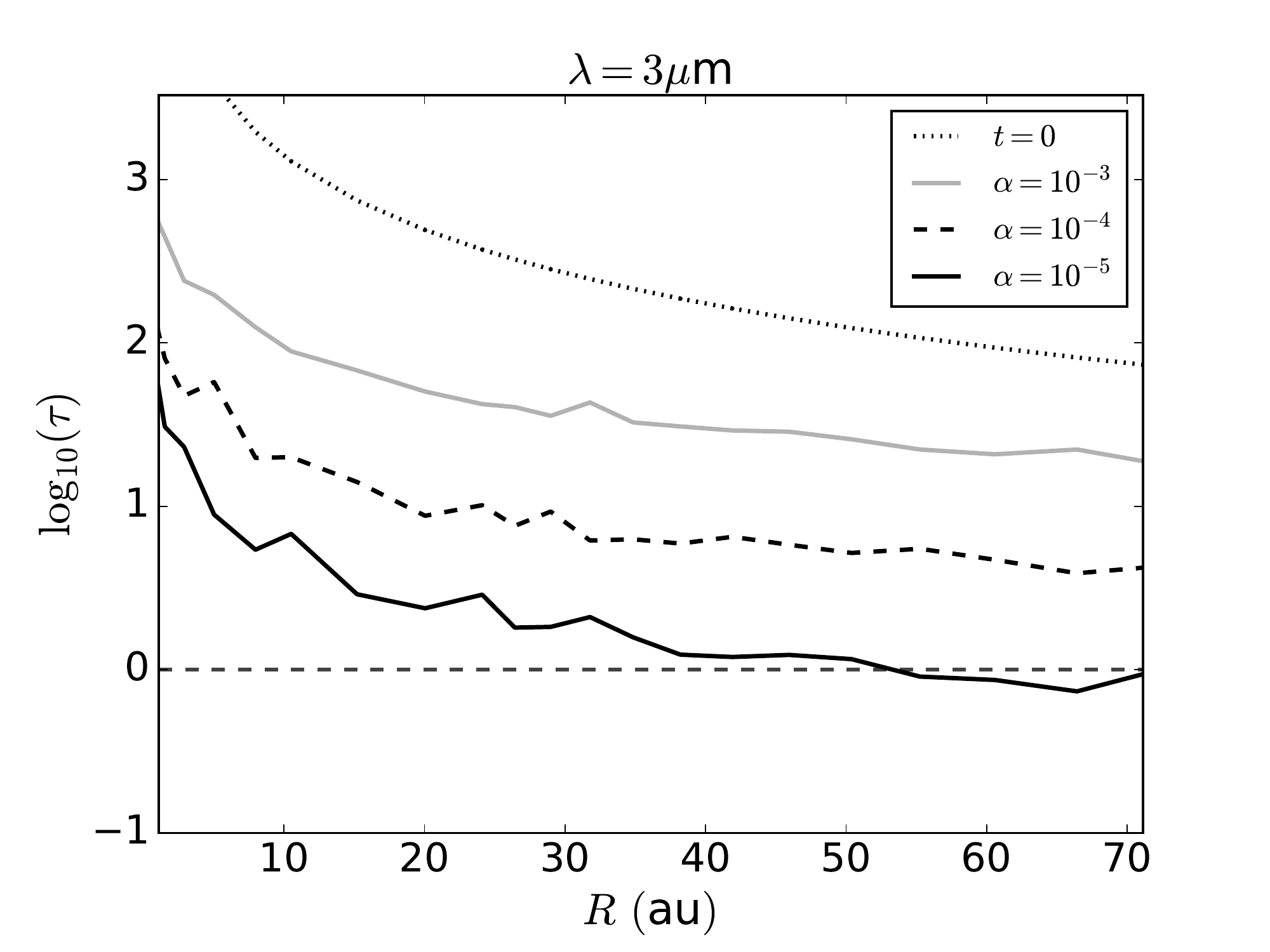}
\end{subfigure}
\begin{subfigure}{0.49\linewidth}
\centering
\includegraphics[scale=0.47]{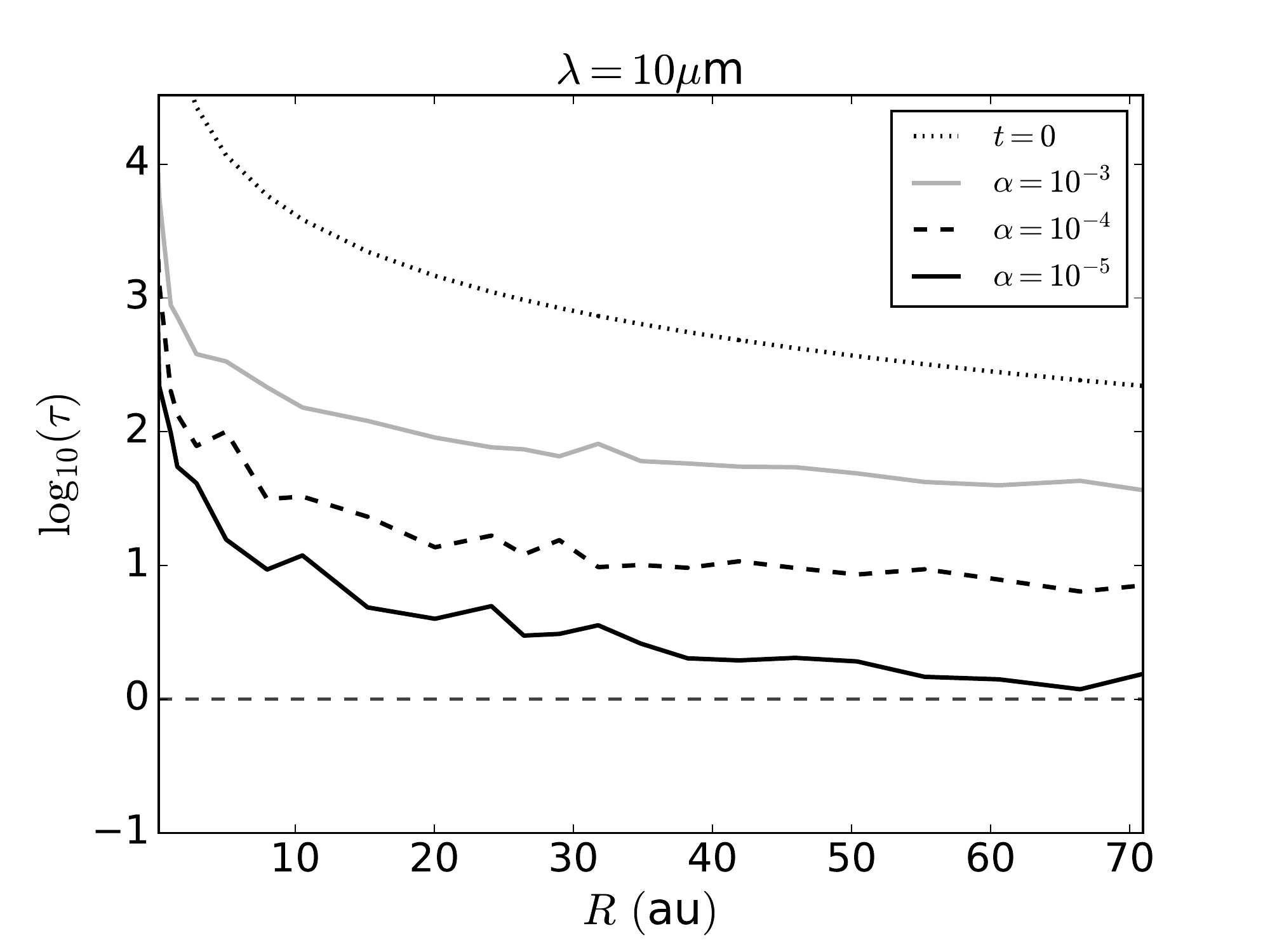}
\end{subfigure}\\[1ex]
\begin{subfigure}{0.49\linewidth}
\centering
\includegraphics[scale=0.47]{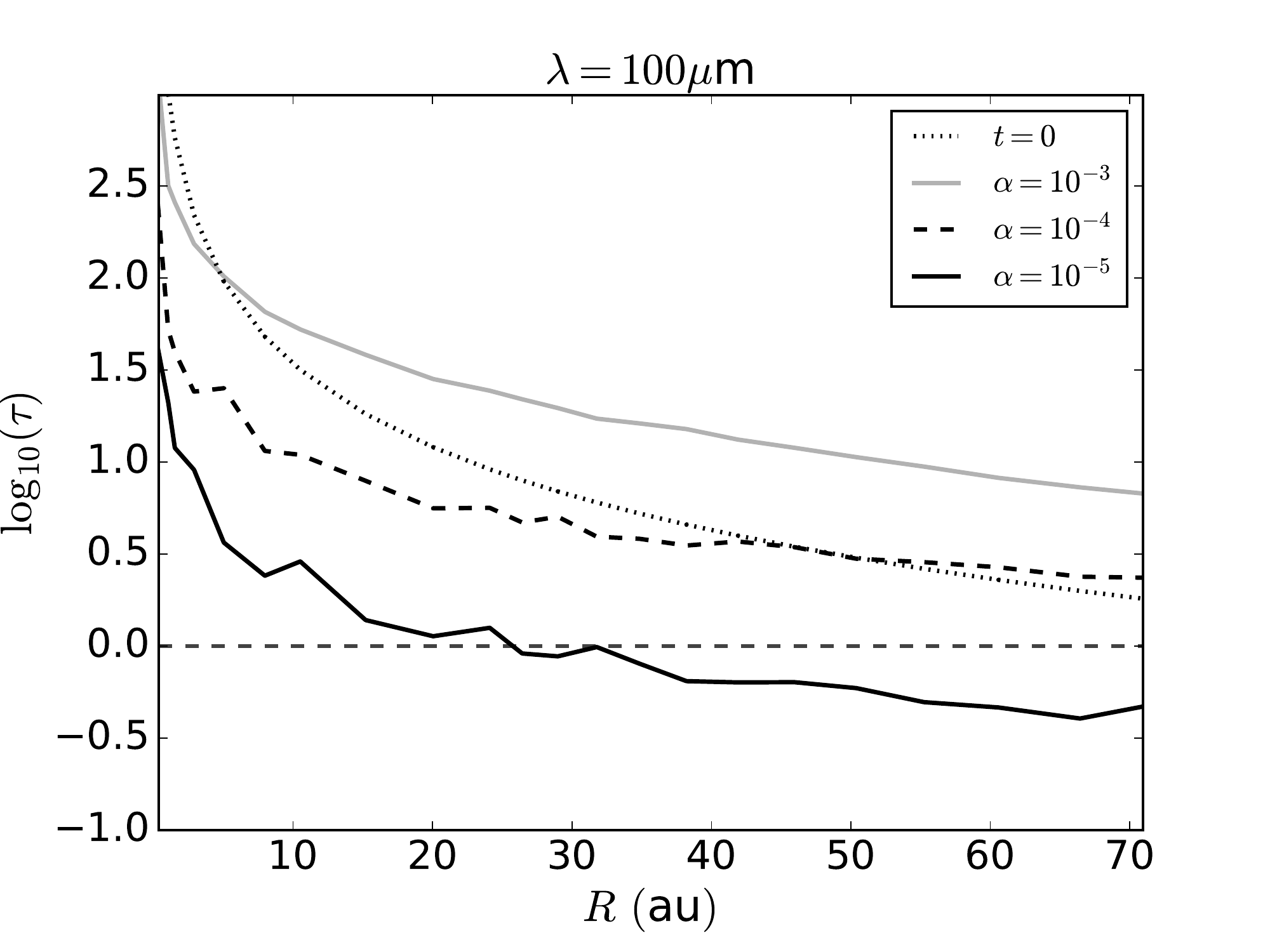}
\end{subfigure}
\begin{subfigure}{0.49\linewidth}
\centering
\includegraphics[scale=0.47]{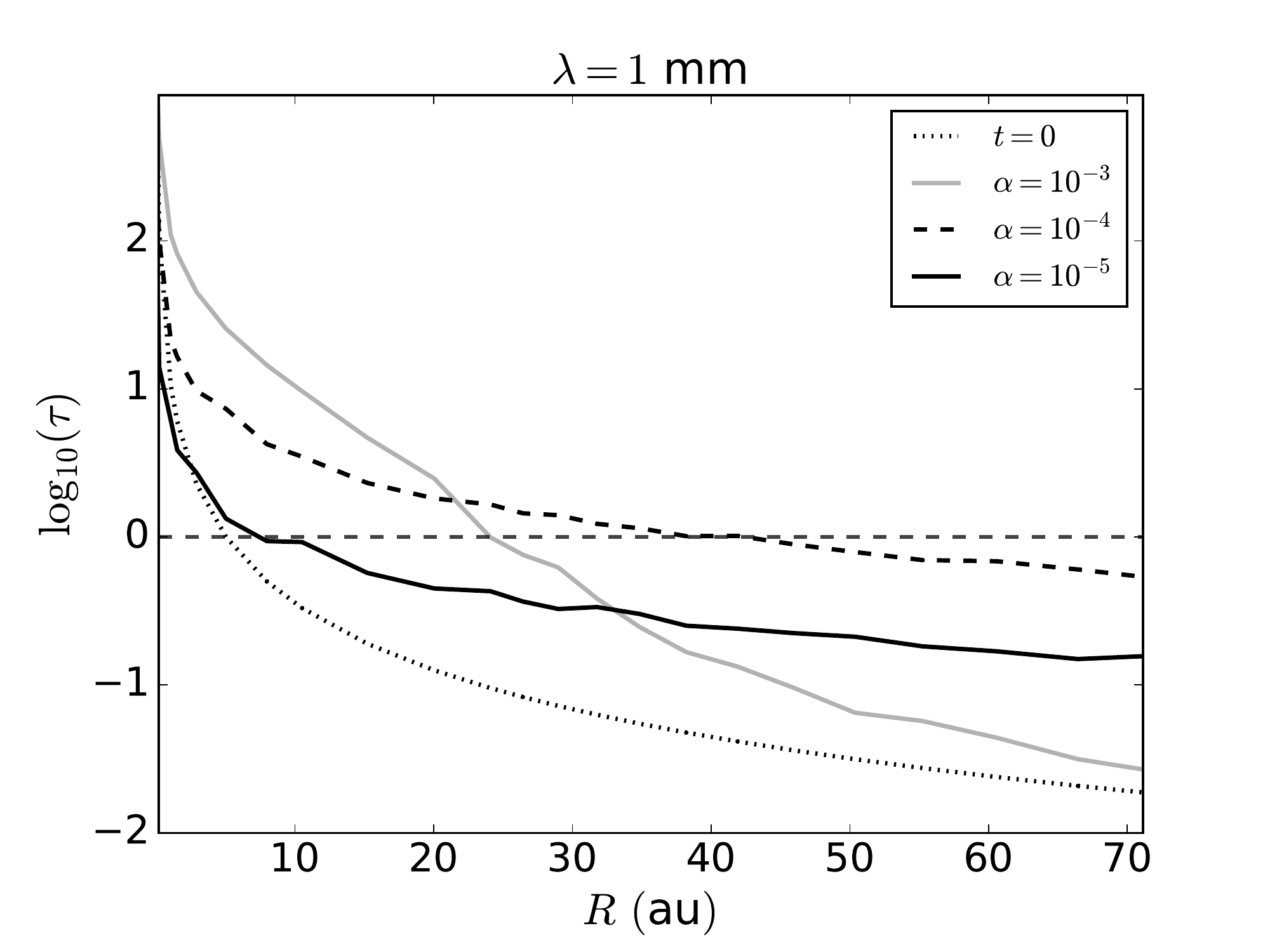}
\end{subfigure}
\caption{Optical depth $\tau(R)$ integrated from the disk surface to the midplane for the MMEN models F4 - F6 with constant $\alpha$.  The dashed horizontal line in each figure shows $\tau=1$. At $\lambda = 3 \micron$ and $\lambda = 10 \micron$ the optical depth, which is provided by the smallest grains, drops as grain growth becomes more efficient (decreasing $\alpha$). At $\lambda=100 \micron$ the disk with the highest optical depth at $R > 30$~AU has $\alpha = 10^{-4}$. Finally, while the disk starts out optically thin at $\lambda = 1$~mm outside 3~AU, its optical depth {\it increases} once grains begin to grow. For $\alpha = 10^{-4}$, the disk even becomes optically thick out to $R = 12$~AU once the dust size distribution reaches steady state.}
\label{fig:optdepth}
\end{figure*}

The optical depth, defined by
\begin{equation}
\label{eqn:taudef}
\tau=\int_0^{\infty}\kappa\rho\,dz
\end{equation}
is also affected by the growth and settling of dust grains. Figure \ref{fig:optdepth} shows the optical depth from surface to midplane of models F4-F6 [MMEN, constant $\alpha(R,z)$] plus $t = 0$ at four different wavelengths. Grain growth depletes the small grains and causes the optical depth at $\lambda = 3 \micron$ and $\lambda = 10 \micron$ to decrease as the disk reaches steady state.  At $\lambda = 100 \micron$, all disks with steady-state size distributions are still more optically thin than the $t = 0$ disk.  Finally, at $\lambda = 1$~mm, the steady-state disks with $\alpha(R,z) = 10^{-5}$ and $10^{-4}$ have {\it increased} their optical depth since $t = 0$ (at least within 35~AU of the star). For $\alpha(R,z) = 10^{-4}$, the disk becomes optically thick at 1~mm inside 10~AU. Figure \ref{fig:optdepth} provides a caution that calculating the surface density of grains available for planet formation in the inner disk from (sub)millimeter observations \citep[e.g.][]{andrews13} might not work, as the disk emission may be optically thick as has already been suggested by ALMA observations (e.g., HL Tau disk).

\begin{figure}
	\centering
	\includegraphics[scale=0.45]{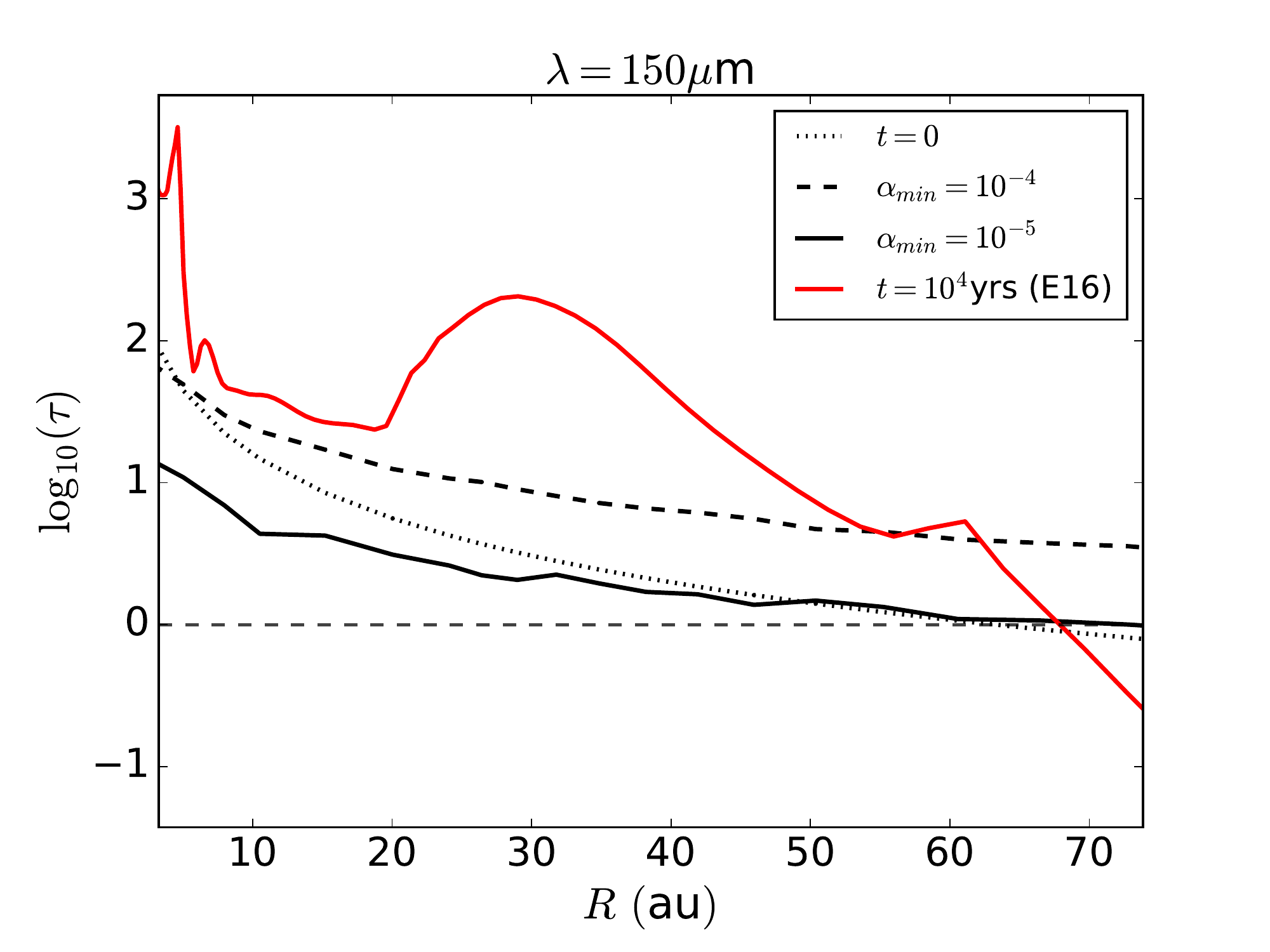}
	\includegraphics[scale=0.56]{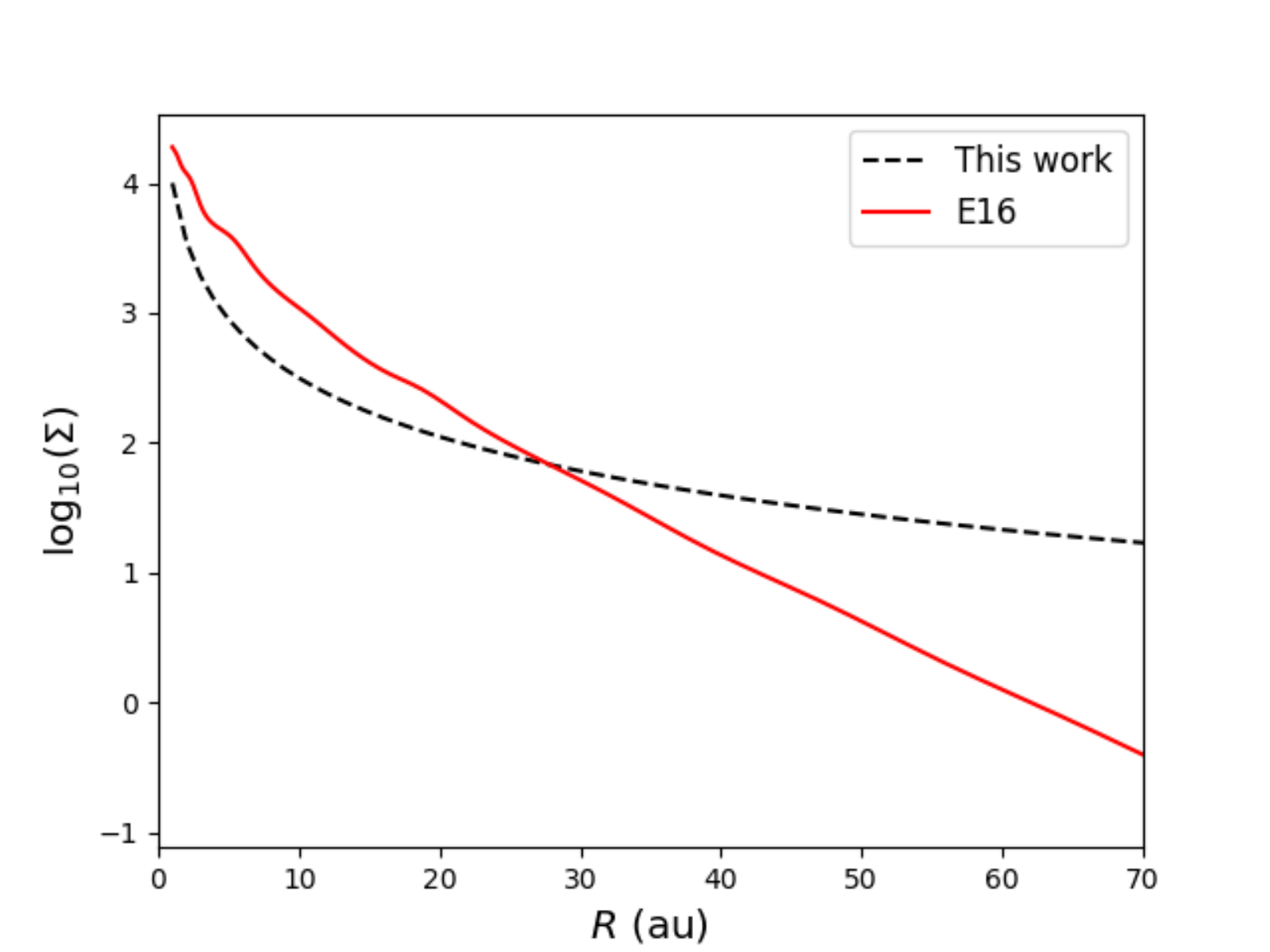}
	\caption{{\bf Top:} Steady state optical depth $\tau$ at $150 \micron$ as a function of radial distance for models F7 (solid black) and F8 (dashed black) with variable $\alpha$ profile with $\alpha_{min}$ at midplane of $10^{-5}$ and $10^{-4}$, respectively. $\alpha_{min}=10^{-4}$ at midplane is more consistent with a turbulence model where hydrodynamic processes contribute to angular momentum transport \citep{nelson13,stoll14,estrada16,neal14}. The red solid line is the  Rosseland mean optical depth from \citet{estrada16}, who find peak disk emission at $\lambda \sim 150 \micron$, with $\alpha=4\times 10^{-4}$. Gas advection in the E16 model causes the optical depth bump at $\sim 30$~au; since we hold the gas surface density fixed in our simulations, we are not able to assess whether grains should pile up anywhere in our model disks. Other differences between our optical depths and those of Estrada et al. (2016) are likely caused by grain composition (silicate vs.\ ice) and radial drift, which removes most of the grains from $R > 60$~AU. {\bf Bottom:} The surface densities for the MMEN disk model and that from the model of E16.}
	\label{fig:opdnew}
\end{figure}

 In Figure \ref{fig:opdnew} we show the optical depth $(\tau)$ for simulations F7 and F8 at $\lambda=150\upmu$m with variable $\alpha$ profile and $\alpha_{min}=10^{-4}$ and $10^{-5}$, respectively. The optical depths at the outer radii  are much lower for $\alpha_{min} = 10^{-5}$ than for $\alpha_{min} = 10^{-4}$. However, \citet{nelson13,klahr14,lyra14,marcus15} have suggested that hydrodynamic instabilities capable of sustaining angular momentum transport can operate in magnetically dead zones, making $\alpha_{min} = 10^{-4}$ a more physically realistic value. 

The opacities in our models can be directly compared to those of \citet{estrada16}. The solid red line in Figure \ref{fig:opdnew} is the          optical depth $\tau = \kappa_R \Sigma / 2$ based on the Rosseland mean opacity  $\kappa_R$. The data have been electronically extracted from two separate subfigures of Figures $3$ and $4$ of E16 ($\kappa$ from the top row of Figure $3$ and $\Sigma$ from top row of Figure $4$) and interpolated onto the same radial gridpoints as in our models. The Rosseland mean optical depth is roughly equivalent to optical depth at the wavelength where the Planck function peaks, which is $\sim150\upmu$m in the typical temperature ranges in the E16. The optical depth from from E16 is an order of magnitude more that our values at $\sim 30$ au, followed by a sharp decrease in the outer nebula.

The optical depth differences between our model and E16 are likely due to advection by gas: the E16 disk has a maximum {\it outward} gas mass flux at 20~au (see their Figure 4), with outward gas motion everywhere outside 7~au. We believe the gas flow is carrying grains outward so that they pile up at 30~au, causing the large bump in optical depth. The E16 grain pileup is probably also sourced by inward radial drift from the outer edges of the disk, correlated with the sharp drop in optical depth beyond 60~au. Other differences between our model and E16 are grain composition (they use ice opacities where $T < 160$~K where we assume silicates throughout the disk for consistency with our collision model), $\alpha = 4 \times 10^{-4}$ throughout the disk (Figure \ref{fig:opdnew} is from our models with variable $\alpha$), and surface density (bottom of Figure \ref{fig:opdnew}). The comparison with E16 highlights the importance of gas velocity: in our work, we treat the gas only as a fixed background against which particles evolve. We justify this assumption by the short timescale over which the grain size distribution reaches steady state, but note that even if the gas mass distribution does not significantly evolve over the course of a simulation, the gas velocities may be important when computing the radial distribution of solids.



 \begin{figure*}
	\begin{subfigure}{0.45\linewidth}
	\centering
	\includegraphics[scale=0.4]{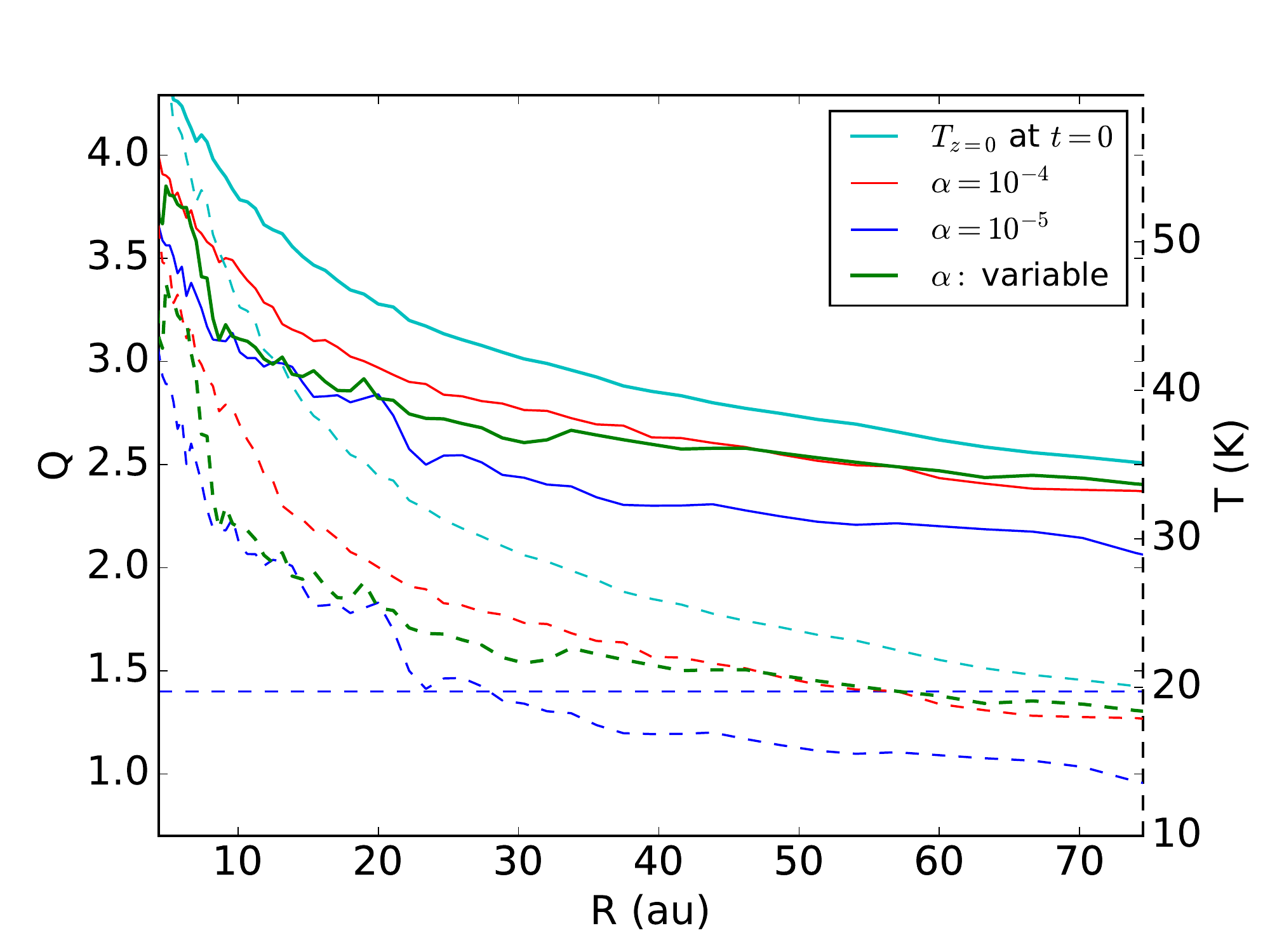}
	\caption{}
	\end{subfigure}	
	\begin{subfigure}{0.45\linewidth}
	\centering
	\includegraphics[scale=0.4]{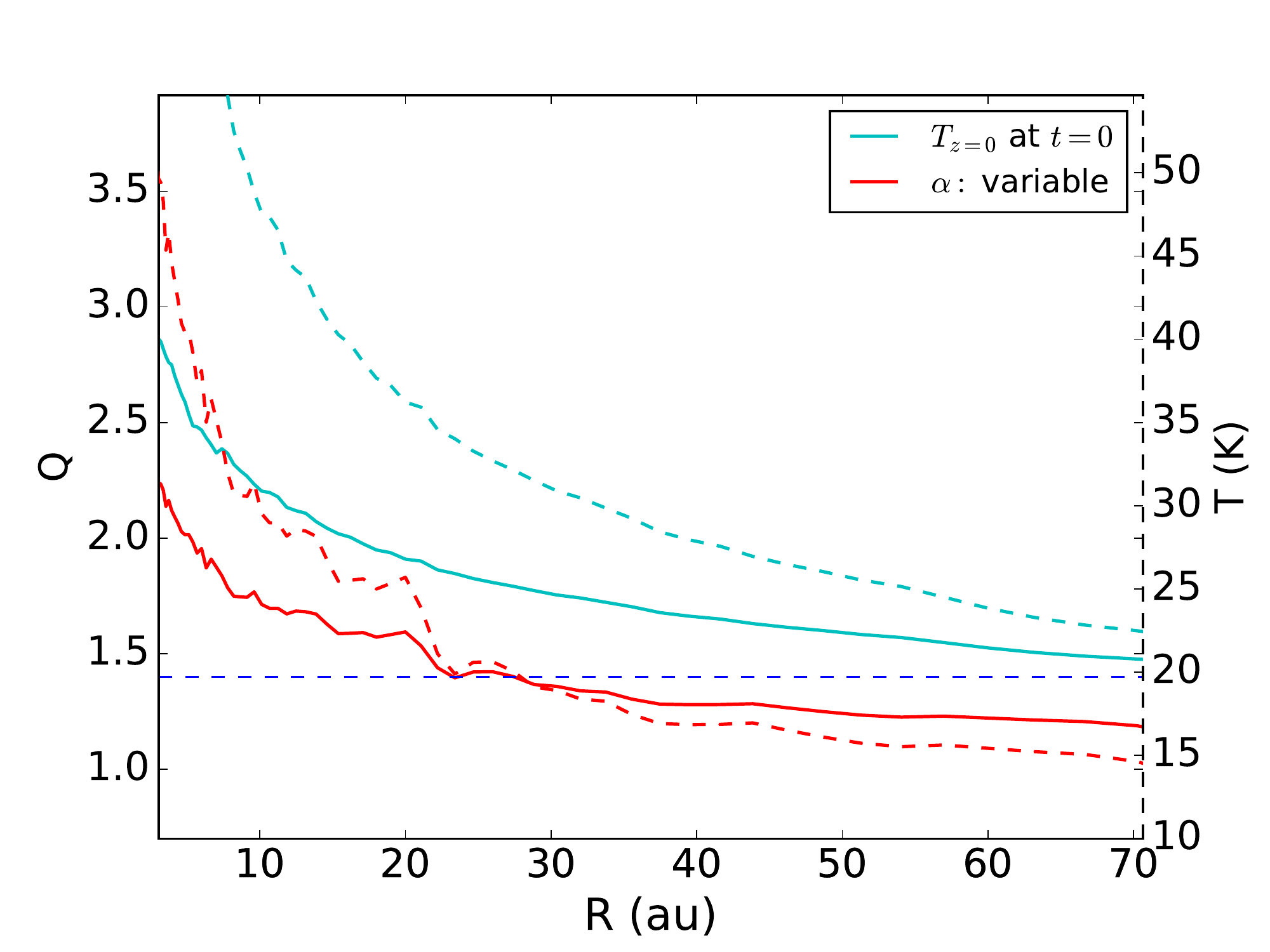}
	\caption{}
	\end{subfigure}	
\caption{Toomre Q parameter and midplane temperature as a function of radius for  MMEN (F5 - F7, $v_{frag}=100$~cm~s$^{-1}$, left) and H1 (right) disk model. Solid lines show $Q(R)$ referenced to the left axis and dashed lines show midplane temperature referenced to the dashed right axis.  The dotted horizontal line denotes $Q=1.4$, a value where the disk might become unstable to non-axisymmetric perturbations (e.g.\ spiral modes) \citep{papa91,nelson98,mayer02,johnson03,pickett03}. For both disks $Q(R)$ can drop by $0.3-0.4$ from its initial value, with the biggest drops in $Q$ and $T$ associated with disks with the weakest turbulence. The spiky features in the temperature profile and hence in $Q$ profile for the inner disk regions arise due to Monte Carlo noise from RADMC calculations.}
\label{fig:toomreprofile}
\end{figure*}

 We expect the changes in mean thermal opacity as a function of wavelength to affect the temperature of the disk interior. Stellar photons are absorbed high in the disk where from they are re-emitted towards the midplane, heating the disk interior. This energy is then re-radiated and escapes vertically to space resulting into cooling. A vertical column with higher optical depth will absorb more photons only to re-emit them towards the midplane and disk surface, making it harder for the photons to escape the disk vertically at the same time. Moreover, the grazing angle at which starlight penetrates the disk becomes smaller as dust settling proceeds \citep{chiang97,hasegawa11} due to the lack of dust particles high up in the disk, which decreases the photon absorption as well. Clearly, as the optical depth decreases through the process of grain growth and settling, cooling becomes more efficient and the interior disk temperature decreases, ultimately lowering the value of Q parameter. 

\begin{figure}
	\centering
	\includegraphics[scale=0.45]{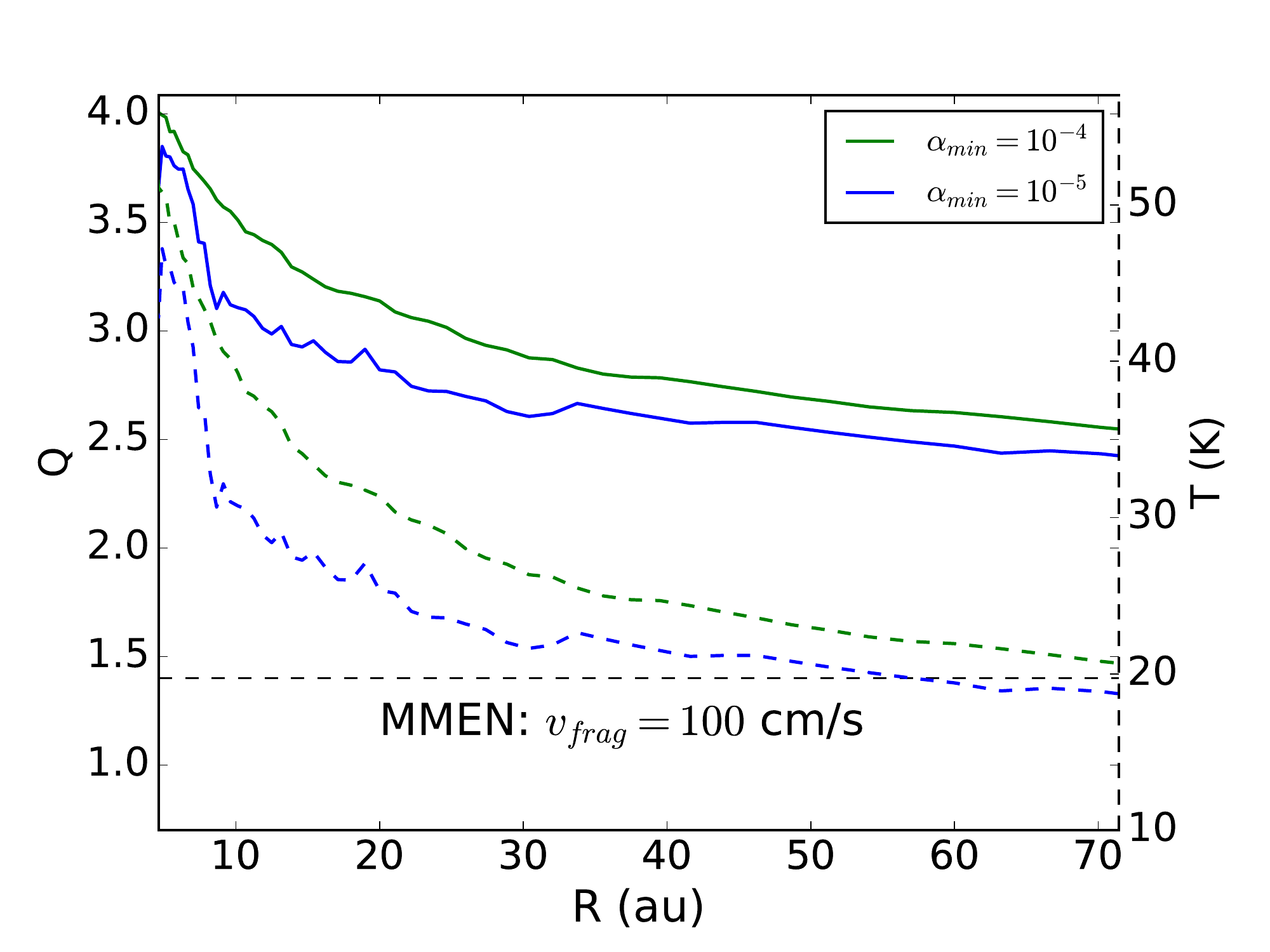}
	\caption{A figure similar to figure \ref{fig:toomreprofile} for the models F7 and F8 where $v_{frag}=100$ cm~s$^{-1}$ is used and the $\alpha$ profile is variable with $\alpha_{min}$ at midplane is taken as $10^{-5}$ (green) and $10^{-4}$ (blue) respectively.  The temperatures are also shown with the dashed curve and with an axis placed on the right-hand-side. The overall temperature difference is not highly sensitive to the minimum value of $\alpha$ chosen for the outer disk. The dashed black horizontal line corresponds to $Q=1.4$.}
	\label{fig:mmen100}
\end{figure}

 In figure \ref{fig:toomreprofile} we plot the Q (Equation \ref{eqn:tq}; left axis, solid lines) and midplane temperature (right axis, dashed lines) as a function of $R$ for MMEN models F2-F4 (left) and disk H1. Both disks show a midplane temperature decrease and corresponding drop in $Q$ as the disk evolves from $t = 0$ to steady state. The disk with the least efficient turbulence at the midplane [$\alpha(R,z) = 10^{-5}$] becomes the coldest and least stable to axisymmetric perturbations. For model H1 (variable $\alpha$), the drop in $Q(R)$ caused by grain growth pulls the disk below the $Q=1.4$ threshold \citep[e.g.][]{papa91,nelson98,mayer02,johnson03,pickett03}---at which non-axisymmetric modes may begin to grow exponentially beyond $\sim 20$ au.  In Figure \ref{fig:mmen100}, we present a similar plot is presented for the models F7 and F8 where $v_{frag}$ is taken as $100$ cm~s$^{-1}$. Two different values for $\alpha_{min}$ are used for the variable $\alpha$ profile: $10^{-4}$ and $10^{-5}$ for the midplane. As expected, the temperature at the midplane is higher for $\alpha_{min}=10^{-4}$ compared to $\alpha_{min}=10^{-5}$ by $\sim 5$ K inside $20$ au. Beyond $20$ au the temperature difference is $\sim 2 - 3$ K. Overall, the radial midplane temperature profile is not very sensitive to the choice of $\alpha_{min}$, specially at the outer radii. However, the inclusion of viscous heating may result in a bigger temperature difference.

In the next section we discuss the significance of our results and the caveats about our models' physical realism.

\section{Discussion and Model Limitations}
\label{sc:discussion}

Here we have presented a proof-of-concept experiment showing that grain growth alone, with no triggers such as infall or vortices, may be able to drive a massive protoplanetary disk to gravitational instability. Yet instability does not necessarily lead to companion formation, and only if our adopted assumptions are fulfilled, our conclusions become fully applicable to observed protoplanetary disks. Here we discuss the limitations of our model and the robustness of our conclusions.

\subsection{Only Sticking and Fragmentation (SF) collision outcomes}
Out of many collision outcomes---up to nine possibilities presented by \citet{guttler10} but notably erosion, mass transfer, and bouncing \citep{windmark12a}---we have restricted our simulations to just two. Any outcome that tends to keep particles small, such as bouncing or erosion, would work against grain settling and disk instability. Likewise, we have not included planetesimal formation or planet growth, though large bodies increase the velocity distributions of nearby objects, leading to more destructive collisions \citep[e.g.][]{dobinson16}. Our simulations only apply to young disks at the very beginning of disk evolution. However, it is important to remark that larger grains may already be present in young stellar objects \citep{jorgensen07, steinacker10, ricci10, cox15}.

\subsection{Viscous Heating}

Although we assume our disks are MRI-turbulent, we do not include viscous heating and assume that the disk is heated only by the stellar photons. The importance of accretion heating depends on the disk accretion rate. For disks with a higher accretion rate, the region where accretion heating dominates expands towards the outer disk. For a classical T-Tauri star with an accretion rate of $\dot{M} \sim 10^{-8} M_{\odot}$~yr$^{-1}$, the iceline is located around $2$~au \citep{hasegawa11, min11}. The heating due to the central star varies as $R^{-1/2}$ while the heating due to the accretion process is much steeper with an $R^{-3/4}$ variation \citep[see][for a detailed review]{dullemond07}. Hence, for classical T-Tauri stars, with an $\dot{M} \sim 10^{-8} M_{\odot}$~year$^{-1}$ the viscous heating dominates only within $1-2$AU \citep{jang04, yu16} (Also see our figure \ref{fig:sigmaprof}).  \citet{landry13} argue that outside the disk region where $\Sigma\sim 20$ g~cm$^{-2}$ the disk can be assumed to be fully MRI-active. Our MMEN and H1 disk models  are substantially heavier that those used in \citet{landry13} (See table \ref{tbl:simulations}), which extends the dead zone to beyond 65 au for heavy disk models.  In figure \ref{fig:sigmaprof}, we have shown the surface density profile for our MMEN and H1 disk model, where the surface density is more than the $20$~g~cm$^{-2}$ threshold throughout the radial range of our simulations.  However, we expect that the disks with $\alpha(R,z) = 10^{-3}$ might be significantly warmer than what our RADMC simulations of passive heating predict, and so do not include these disks in Figure \ref{fig:toomreprofile} or make predictions about their gravitational stability.

Apart from that, the disk angular momentum can be removed by magnetically induced disk winds when vertical magnetic flux is relatively strong. For such cases, accretion heating can be neglected even in the inner part of the disk since the value of $\alpha$ due to disk turbulence should be relatively small \citep{bai16, bai17, simon17}.  The existence of a disk wind and its fractional contribution in angular momentum transport is a matter under debate. In our model, a disk wind would add an additional advection term for small, fully coupled dust particles. Disk winds are outside the scope of this work and merit separate investigation. However, although not consistent in the upper layers of the disk, our models with low $\alpha$ can mimic the disk wind effect at the midplane.

\subsection{Grain Composition: Silicate Particles}

 Literature on collision outcomes is far more extensive for silicates than for any other protoplanetary disk constituent, which led us to restrict our study to silicate particles. However, our model disks are cold enough that particles almost everywhere should be ice-coated, which would change both their opacity and their sticking efficiency. \citet{estrada16} and \citet{krijt16b} have already explored collisions of icy grains. Modeling volatiles also demands the addition of evaporation fronts where solid growth is enhanced.  In certain cases the dust-to-gas abundance ratio can be increased by an order of magnitude \citep[see figure $20$ of][]{estrada16}. In the context of disk opacities, porous icy grains would allow particles to grow further due to higher fragmenting threshold velocity, reducing the abundance of small particles. This effect will lower the opacities at the small wavelengths while increasing the opacities at longer wavelengths. Once experimental data on collisions of icy bodies \citep[e.g.][]{shimaki12,yasui14,deckers16} becomes more complete it would be worth repeating our experiment with collision outcomes, velocity thresholds, and opacities appropriate to porous ice.

\subsection{Radial Drift}

For this initial experiment we have not included radial drift in our simulations, though we plan to add it in future work. According to \citet{birnstiel10, birnstiel11, drazkowska13, estrada16}, the outer disk beyond 20--25~au should be drift-dominated, with the particle size spectrum significantly altered.  Consequentially, radial drift might have the effect of cooling the outer disk more than what is predicted in this work, by lowering its opacity to (sub)-mm radiation, while increasing the (sub)-mm opacity in the inner disk. However, the drift timescale is longer than the vertical settling/diffusion timescale for dust particles \citep{birnstiel10}. 

With our disk setup, the width of the annuli at $40$ and $70$~au are $7.5$ and $14$~au respectively, whereas the particles of maximum sizes at those positions travel $\sim 6.5$ and $12$~au respectively in a timescale of $\sim 10^4$~years for an MMEN disk model. Similarly in the inner disk, the width of the column at $5$~au is $\sim 1.2$~au with the maximum drift in the same timescale is $\sim 0.9$~au. These comparisons suggest that inward radial drift is an important but not dominant effect over the simulation period. 

Also, \citet{estrada16} showed that radial drift  becomes important in the outer disk in limiting the particle size to $St \sim 0.1$. So, inclusion of radial drift might have some effect on our growth model as well even in somewhat shorter timescales. However, we note that \citet{estrada16} included bouncing in their model, which slows the growth process, possibly making the growth timescale comparable to the radial drift timescale.


\begin{figure*}
	\begin{subfigure}{0.49\linewidth}
	\centering
	\includegraphics[scale=0.45]{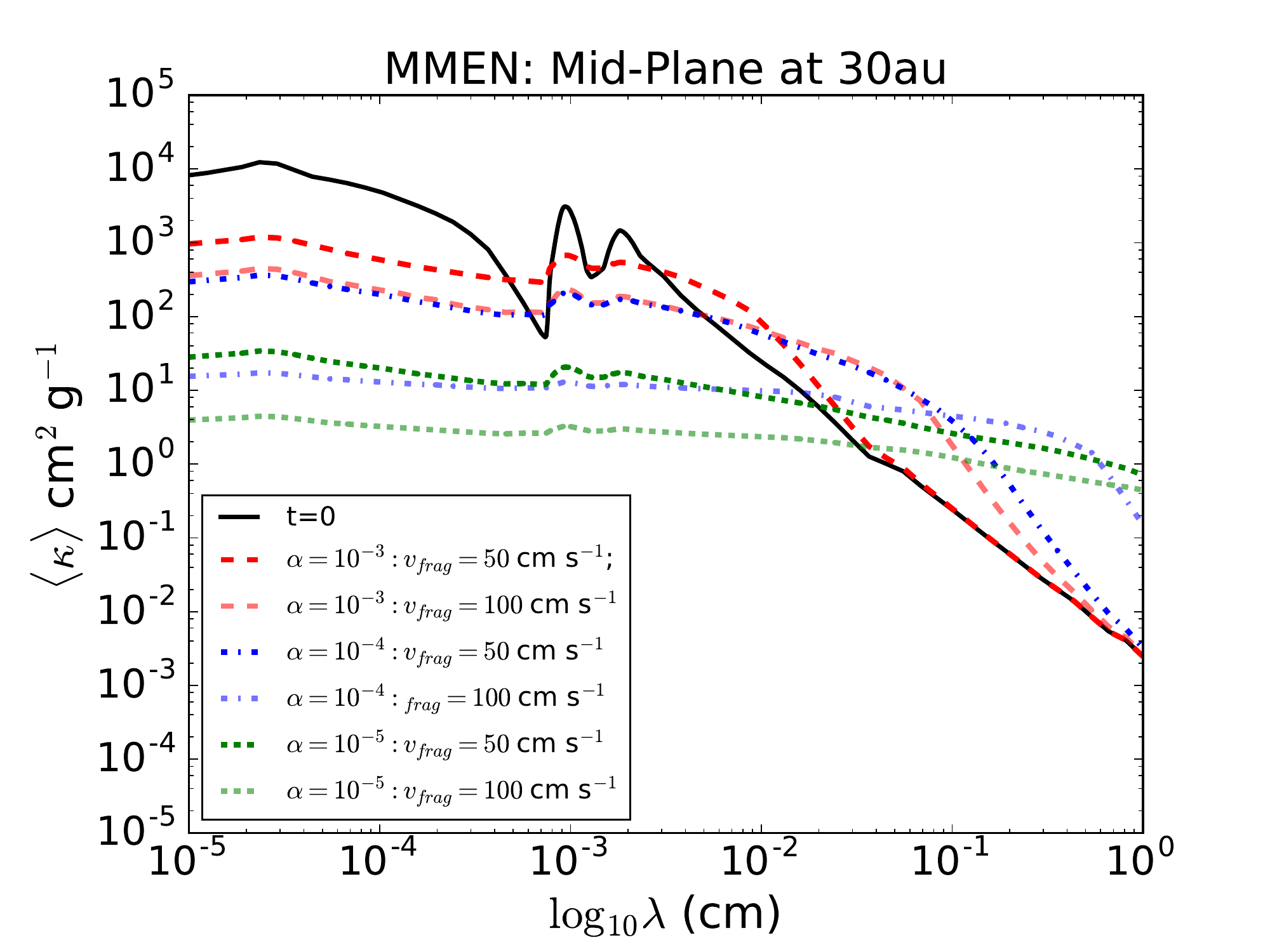}
	\caption{}
	\end{subfigure}	
	\begin{subfigure}{0.45\linewidth}
	\centering
	\includegraphics[scale=0.49]{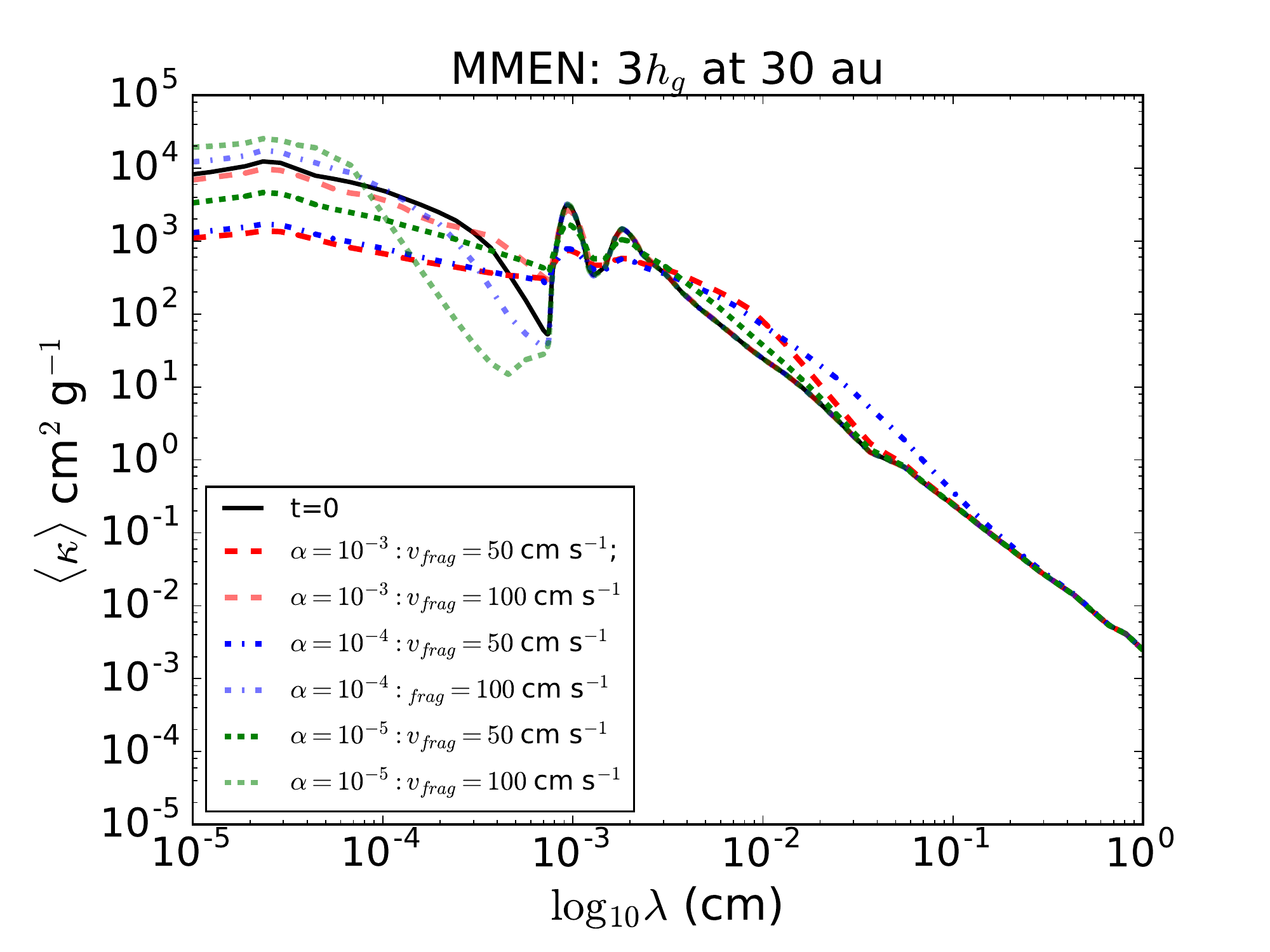}
	\caption{}
	\end{subfigure}	
\caption{Spectral opacities $\langle\kappa(\lambda)\rangle_{\rho_d}$ at mid-plane (left) and $3$ scale-heights above (right) for simulations with MMEN disk models with $v_{frag}=50$ (F1, F2 and F3) and $100$~cm~s$^{-1}$ (F4, F5 and F6) at a radial distance of $30$~au from the central star. For $\lambda < 100 \micron$, lower $v_{frag}$ leads to higher opacity, while higher $v_{\rm frag}$ allows larger particles to stick, decreasing the opacity. The relationship between $v_{\rm frag}$ and opacity is especially strong for low values of $\alpha$.  This trend, however, reverses for $\lambda \sim 100\upmu$m and larger,  due to the smaller maximum size attained in the lower $v_{frag}$ case (see equation \ref{eqn:amax}). At the disk surface (right plot), the behavior is same, although the differences in opacities are small due to restricted grain growth arising due to lower gas density and weak coupling between gas and dust. }
\label{fig:opac10050}
\end{figure*}

\subsection{Choice of $v_{frag}$}

 For this work we set $v_{frag} = 100$~cm~s$^{-1}$ for all our disk models except F1 - F3, for which $v_{frag} = 50$~cm~s$^{-1}$, allowing us to explore our results' sensitivity to fragmenting threshold velocity. Literature values include 100~cm~s$^{-1}$ \citep[experiment F1 by][]{guttler10}, 80~cm~s$^{-1}$ \citep[Monte Carlo models of][]{drazkowska13}, and 50~cm~s$^{-1}$ \citep[further work by][]{drazkowska14}. Though there is uncertainty on the appropriate value of $v_{frag}$ for silicate particles, especially when considering variations such as porosity or aggregate type, our choice of relatively low $v_{frag}$ helps keep our maximum particle sizes low, thereby minimizing Stokes numbers and keeping our neglect of radial drift appropriate. Our conclusion that grain growth and settling can trigger non-axisymmetric instability might not apply to disks with stronger particles that better resist fragmentation, where drift can alter the size spectrum.

\begin{figure}
\centering
\includegraphics[scale=0.45]{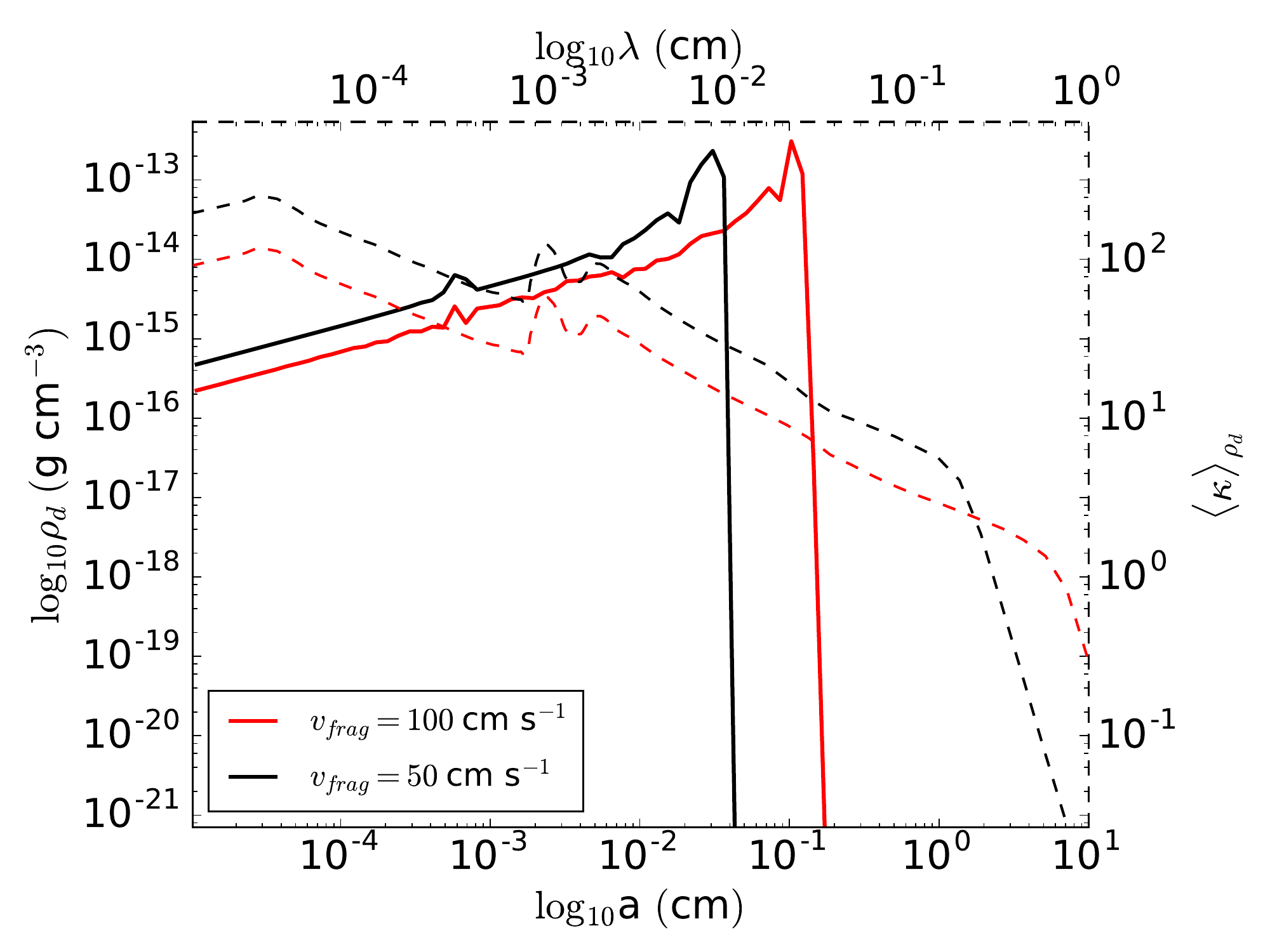}
\caption{The effect of fragmenting threshold velocity on grain size and opacity. The solid curves represent the steady-state dust distribution for $v_{frag}=100$~cm~s$^{-1}$ (solid red) and $50$~cm~s$^{-1}$ (solid black) for the same location in the disk. The dashed curves show the corresponding mean opacity with the axes placed on right and top. $\langle\kappa\rangle_{\rho_d}$ differs by a factor of $\sim 2$ between the models.}
\label{fig:vfragopacity}
\end{figure}

 The dependence of disk opacity on the choice of $v_{frag}$ can be estimated from figure \ref{fig:opac10050} where the spectral opacities for simulations F1 - F3 and F4 - F6 are plotted for both mid-plane and $3$ scale-heights at $30$~au. For $\lambda \la 100 \micron$, opacity is higher for $v_{frag}=50$~cm~s$^{-1}$ compared to $100$~cm~s$^{-1}$. This difference is amplified for lower value of $\alpha$ as well. A higher $v_{frag}$ with a lower turbulence efficiency for a similar surface density puts more mass in the larger particles leaving a small fraction of the total mass for the smaller grains, which are mostly responsible for photon absorption. This effect lowers the opacity of the disk and changes the temperature profile. This trend, however, reverses for $\lambda \sim 100\upmu$m and larger, due to the smaller maximum size attained in lower $v_{frag}$ case. Apart from that, increasing $v_{frag}$ could also intensify the tendency of dust evolution to trigger gravitational instability: Figure \ref{fig:vfragopacity} shows the particle size spectrum (solid lines, solid axes) and wavelength-dependent opacity (dashed lines, dashed axes) for the same region of the disk but with two different values of $v_{frag}$. Higher $v_{frag}$ decreases the opacity at $\lambda \la 1$~mm but increases it for longer wavelengths, the very effect that helps decrease Q (\S \ref{sc:stability}, Figures \ref{fig:densityweightedopacity} and \ref{fig:optdepth}). The experiments presented here do not cover a wide enough parameter space in collision outcomes for us to be sure that there is a {\it general} tendency for grain growth to reduce disks' gravitational stability.

\section{Conclusions}
\label{sc:conclusion}

In this paper, we have developed a new weighted Monte Carlo model of collisional sticking and fragmentation along with a Monte Carlo Lagrangian prescription for  settling, and turbulent stirring, combined with wavelength dependent opacity calculations and radiative transfer. We have used three disk models with different surface densities and have employed both spatially constant and variable turbulence efficiency $(\alpha)$ prescriptions. Our main findings are:
\begin{itemize}
\item The collisional growth of dust grains through sticking and fragmentation transfers most of the solid mass to larger particles, leaving a small portion of the total dust mass in the $\micron$ and sub-$\micron$ dust grains which provide most of the surface area for photon absorption. This results in a reduction in midplane opacities at smaller wavelengths by $3-4$ orders of magnitude compared to the initial values. At the disk surface, however, the opacities decreases mainly due to depletion of dust grains by settling and inefficient growth of the dust particles due to weak coupling between dust and gas.

\item Grain growth and settling tend to decrease the optical depths $(\tau)$ from disk's surface to the midplane at short wavelengths $(\lambda \la 10\micron)$ by a couple of orders of magnitude, while increasing $(\tau)$ at mm and sub-mm wavelengths. For a typical value of $\alpha=10^{-4}$, the optical depths at $1$~mm inside $30$ au exceed unity, which may be problematical for disk mass calculations from (sub)millimeter observations.

\item In spite of the depletion of solids in the upper layers of the disk, grains of (sub)micron sizes are stirred high up in the inner disk even when the turbulence strength is small. This effect becomes more prominent when a strong turbulence in the disk surface is considered. Because of strong coupling, these dust particles would follow the gas motion in case a disk wind is present, altering the opacities in the disk atmosphere, an essential physical process requiring an in-depth investigation.

\item The optical and thermal profiles of the disk is sensitive to the fragmenting threshold velocity $(v_{frag})$, chosen for modeling the collisional dust growth. We found the opacities at short wavelengths to be $5-10$ times smaller for $v_{frag}=100$~cm~s$^{-1}$ compared to $50$~cm~s$^{-1}$. An even higher value value of $v_{frag}$, traditionally chosen for porous icy aggregates would alter the outcomes significantly.

\item Grain growth and settling can bring an initially marginally stable protoplanetary disk down below a Toomre $Q=1.4$ threshold at which non-axisymmetric gravitational instabilities may grow. We find that the disk interior cools as the disk's surface layers are heavily depleted of small grains once the size distribution reaches steady state, decreasing its stability to gravitational perturbations. As disks with low turbulent efficiency $\alpha(R,z)$ have lower collision speeds, and allow grains to grow and settle more efficiently than disks with active turbulence, we expect to find grain-triggered instability primarily in weakly turbulent disks. The model in which we find $Q < 1.4$ throughout most of the disk is extremely massive, with almost ten times the surface density of the minimum-mass solar nebula. Interestingly, this massive disk is consistent with what theorists propose is necessary for giant planet formation \citep[e.g.][]{lissauer09}, but is much larger than typical values inferred from disk observations \citep[e.g.]{andrews13, ansdell16, pascucci16}. However, given the evidence that disk masses are systematically underestimated \citep[e.g.]{mcclure16,yu17}, our model H1 ``heavy'' disk mass may be physically plausible.

\item Finally, we note that disk instability may not necessarily lead to brown dwarf or star formation, though companions can form in overdense spiral arms \citep[e.g.][]{kratter16}. Disks that become gravitationally unstable may transport angular momentum by gravitoturbulence \citep[e.g.]{gammie01, shi14}, or growing spiral modes may saturate \citep{cossins09}, keeping the disk marginally stable. Further work would be necessary to track the eventual dynamical outcome of the grain growth and settling studied here.

\end{itemize}

\section{ACKNOWLEDGEMENTS}

We thank Til Birnstiel, Xue-Ning Bai, Wladimir Lyra and James Owen for useful discussions. We also thank Bennett Maruca for useful discussion on code design and James MacDonald for a careful review of the manuscript. We are especially grateful to Joanna Dr{\c a}{\.z}kowska for consulting on dust modeling. DS and SDR were supported by NSF grant 1520101 and the UNIDEL foundation. YS and NJT were supported in part by the JPL Research \& Technology Development Program. This research was carried out in part at the Jet Propulsion Laboratory, operated by the California Institute of Technology under contract with the National Aeronautics and Space Administration.

 \appendix
\section{Code Test Results}

\subsection{Size Distribution}
\label{apn:codetest}

\begin{figure*}\label{fig:codetest}
\centering
	\begin{subfigure}{.45\textwidth}
	\centering
	\includegraphics[scale=0.45]{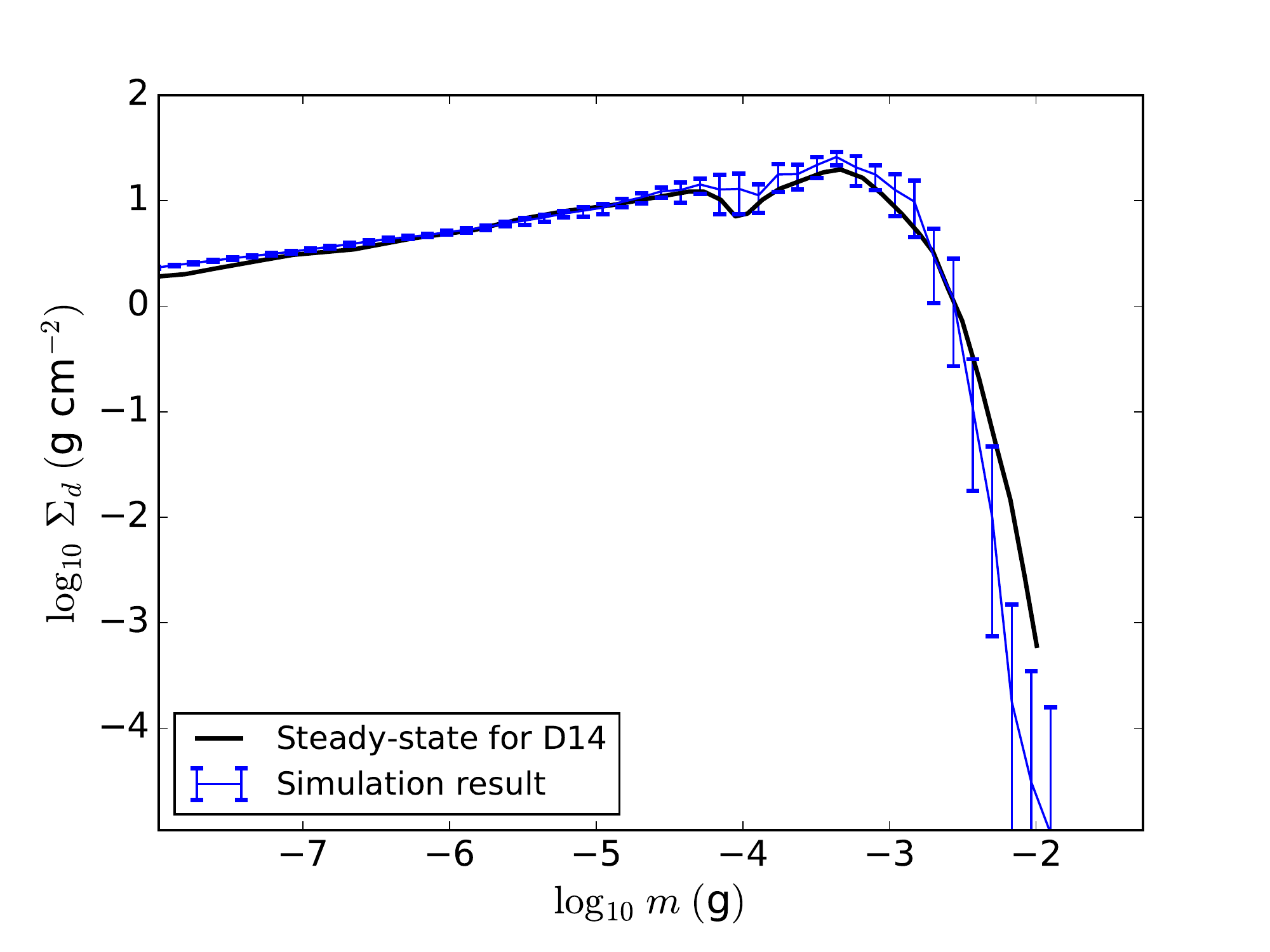}
	\caption{ }
	\label{fig:codetestd14}
	\end{subfigure}
\hfill
	\begin{subfigure}{.45\textwidth}
	\centering
	\includegraphics[scale=0.45]{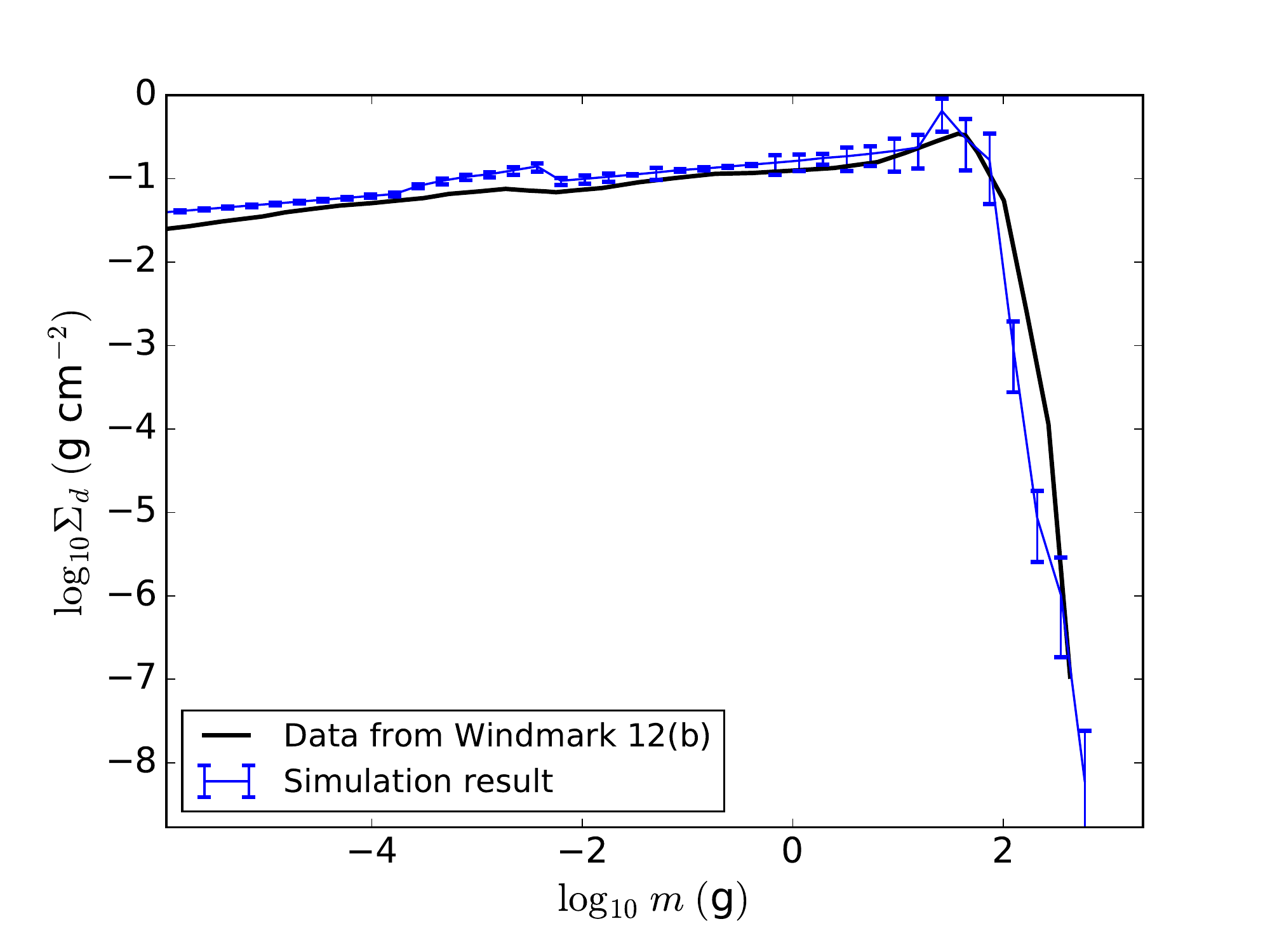}
	\caption{}
	\label{fig:codetestw12}
	\end{subfigure}
	\caption{{\bf {a:}} The steady state dust distribution for MMEN model. The plot shows the vertically averaged dust surface density for a vertical column at $1$AU with $\Sigma=9900$ g~cm$^{-2}$, $\alpha=0.01$, $\rho_m=1$ g~cm$^{-3}$, $T=280$K and a standard dust to gas mass ratio $0.01$. $v_{frag}$ is taken to be $50$ cm~s$^{-1}$. The solid black line is the data electronically extracted from \citet{drazkowska14} and the line with error bars shows results from our simulations where the average from $10$ runs with $80000$ particles each is presented. {\bf {b:}} Local dust distribution steady state comparison with \citet{windmark12a}. Simulations are carried out at $1$ AU of an MMSN disk with $\Sigma=1700$ g~cm$^{-2}$, $\alpha=10^{-4}$, $\rho_m=1$ g~cm$^{-3}$, $T=280$K and dust to gas mass ratio $0.01$. $v_{frag}$ is taken as $100$ cm~s$^{-1}$. The solid black line is the data electronically extracted from \citet{windmark12a} and the line with error bars shows results from our simulation. The average of $10$ simulations with $80000$ particles each is plotted.}
\end{figure*}

In this section we present size distributions computed by our model and compare them with models already existing in literature. In figure \ref{fig:codetestd14} we plot our steady state on top of \citet{drazkowska14} results extracted electronically from their paper for a 1-D vertical column with the same conditions.  Our model includes sticking and fragmentation only, unlike the \citet{drazkowska14} model, which includes mass transfer as well. However, \citet{drazkowska14} mentioned that panel 3 of their Figure 1 represents vertically averaged steady state size distribution they would achieve without mass transfer; it is this steady state that we plot in Figure \ref{fig:codetestd14}. Once the distribution hits the fragmentation barrier growth is stalled unless mass transfer is included.

We also make a comparison test with the results from \citet{windmark12b} where a local simulation is performed with SF model without any velocity distribution. In both the cases our model shows an excellent match in the smaller mass range of the distribution and it deviates slightly in the higher end. In figure \ref{fig:codetestw12} the points in the extreme right miss the reference plot beyond the error bars which can be attributed to the bigger dynamical range obtained by introducing the weighing method.

\subsection{Largest Particles Produced}
\label{apn:largestparticles}

As a further code test, we compare the maximum particle radius $a_{max}$ that our code produces with analytical estimates of $a_{max}$ for relative velocities dominated by turbulence. Initially, for tiny dust grains of micron and sub-micron sizes, the particle relative velocities are dominated by Brownian motion (figure \ref{fig:relativevel}) and collisional growth is efficient. When particle size exceeds $\sim 100\upmu$m gas turbulence starts playing the dominant role in setting $v_{rel}$ until collisions between the largest particles reach the fragmenting threshold velocity $v_{frag}$. Given that our collision model includes only sticking and fragmentation (``SF''), grain growth does not continue (see but results from \citet{drazkowska14} on continued growth when mass transfer is included).


The largest eddy turnover time is $t_L \sim L/U_L$, where $L$ is the largest scale of the inertial range and $U_L$ is the characteristic velocity $\sqrt{\alpha}c_s$. Taking $L\sim \sqrt{\alpha} h_g$ \citep{schrapler04}, the largest eddy turnover time becomes $t_L \sim 1/\Omega$. Hence, for particles with stopping time of the same order as $t_L$ the Stokes number is $St = t_L\Omega \sim 1$. On the other hand, the smallest eddy turnover time at the dissipation scale, $t_{\eta}$, is $t_{\eta}\sim Re^{-1/2}t_L$ \citep{k41}, where $Re$, the Reynold's number, is the ratio of turbulent and molecular viscosity $\nu_T/\nu_m$ \citep{ormel07a}. In all our simulations, the maximum particle size at disk midplane lies within the intermediate turbulent regime of \citet[][equation 28]{ormel07b} where $t_{\eta} < t_{fric} < t_L$. Thus, following \citet{birnstiel11}, the Stokes number for the largest particle is
\begin{equation}
St_{max}=\frac{v_{frag}^2}{2\alpha c_s^2},
\label{eqn:stmax}
\end{equation}
which corresponds to a maximum particle radius
\begin{equation}
a_{max}=\frac{v_{frag}^2 \rho_g}{2 \alpha c_s \Omega \rho_m}.
\label{eqn:amax}
\end{equation}

\begin{figure}
	
	\begin{subfigure}{0.33\linewidth}
		\centering
		\includegraphics[scale=0.33]{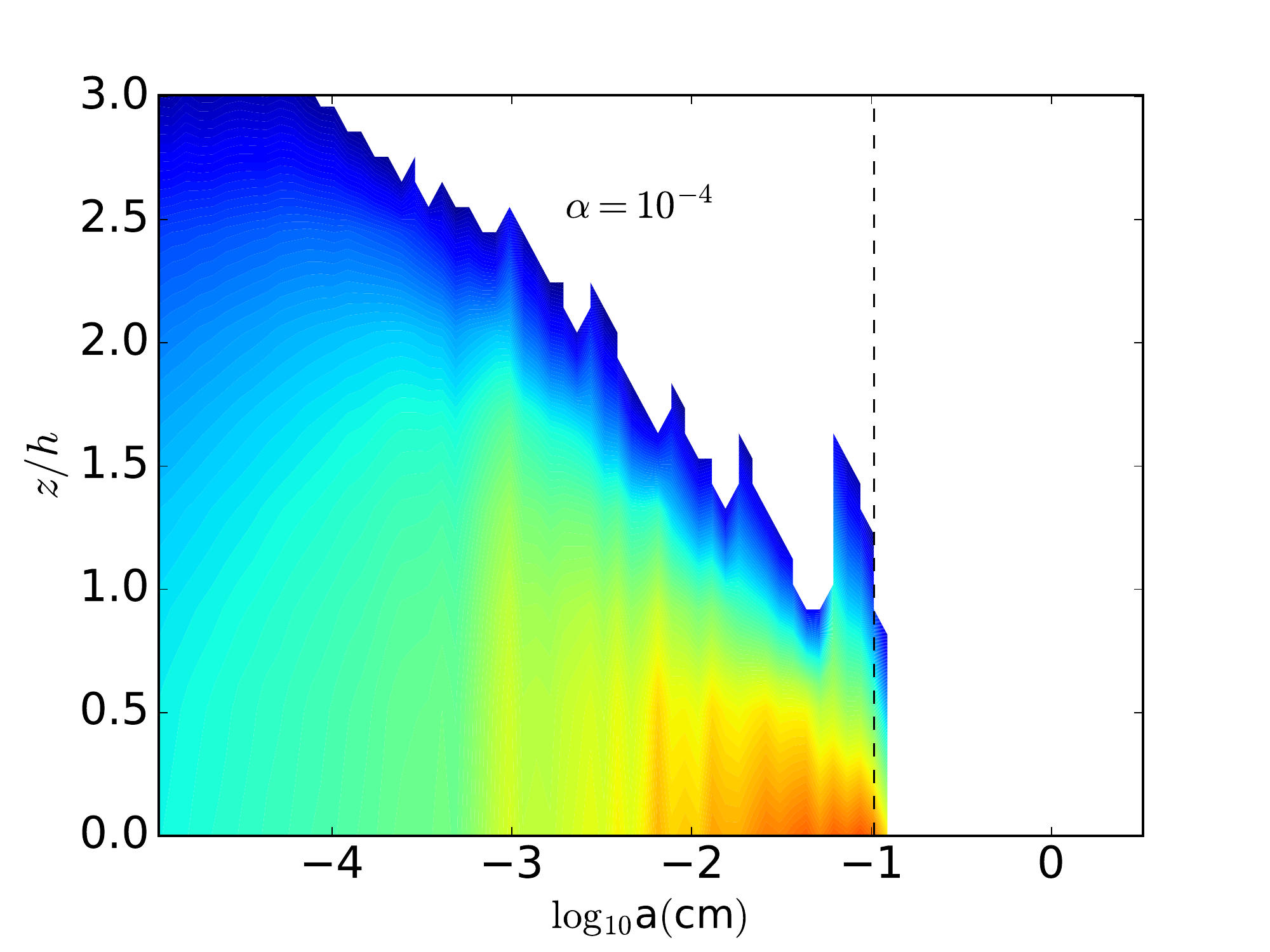}
		\caption{}
	\end{subfigure}%
	\begin{subfigure}{0.33\linewidth}
		\centering
		\includegraphics[scale=0.33]{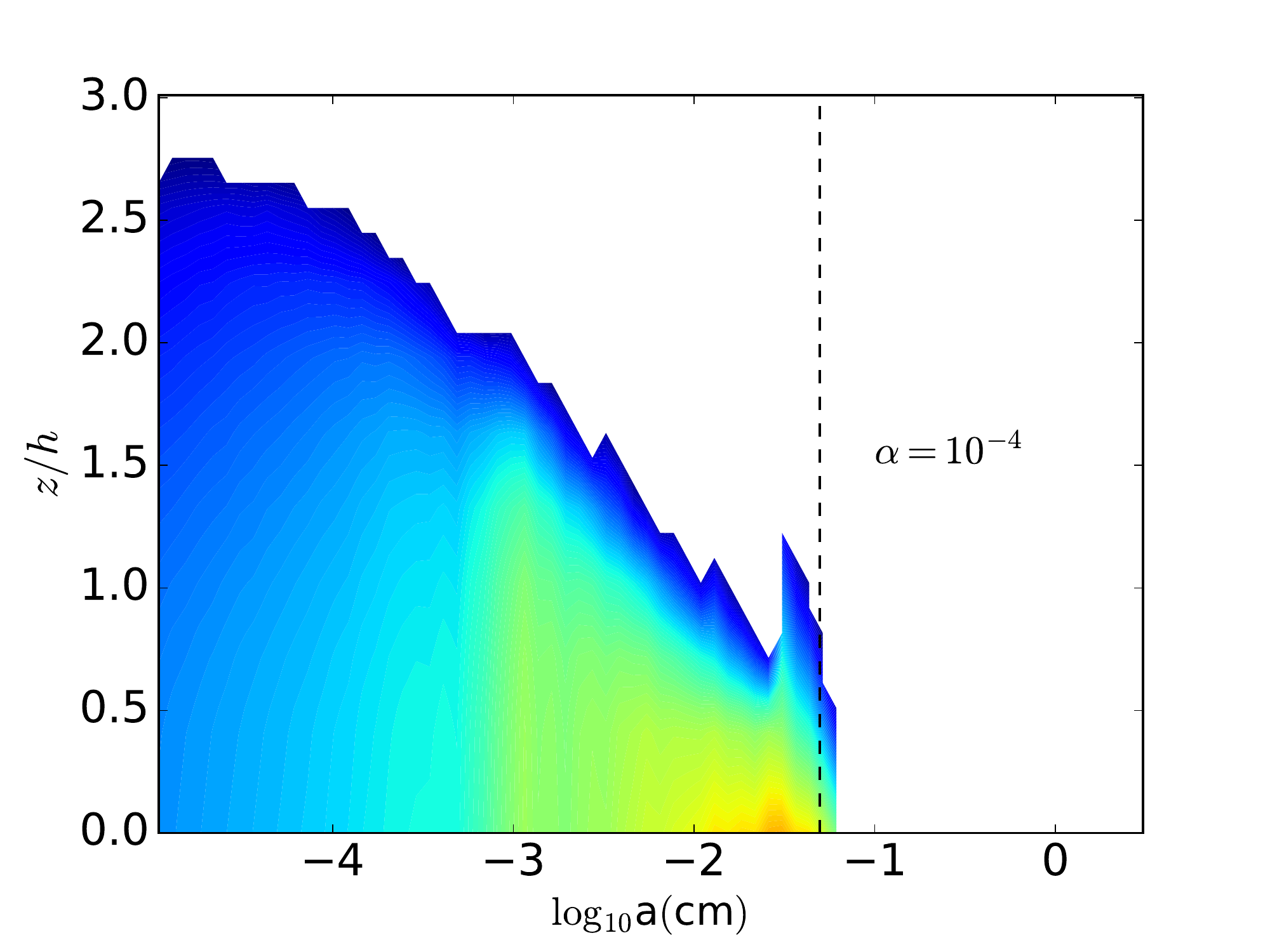}
		\caption{}
	\end{subfigure}%
	\begin{subfigure}{0.33\linewidth}
		\centering
		\includegraphics[scale=0.335]{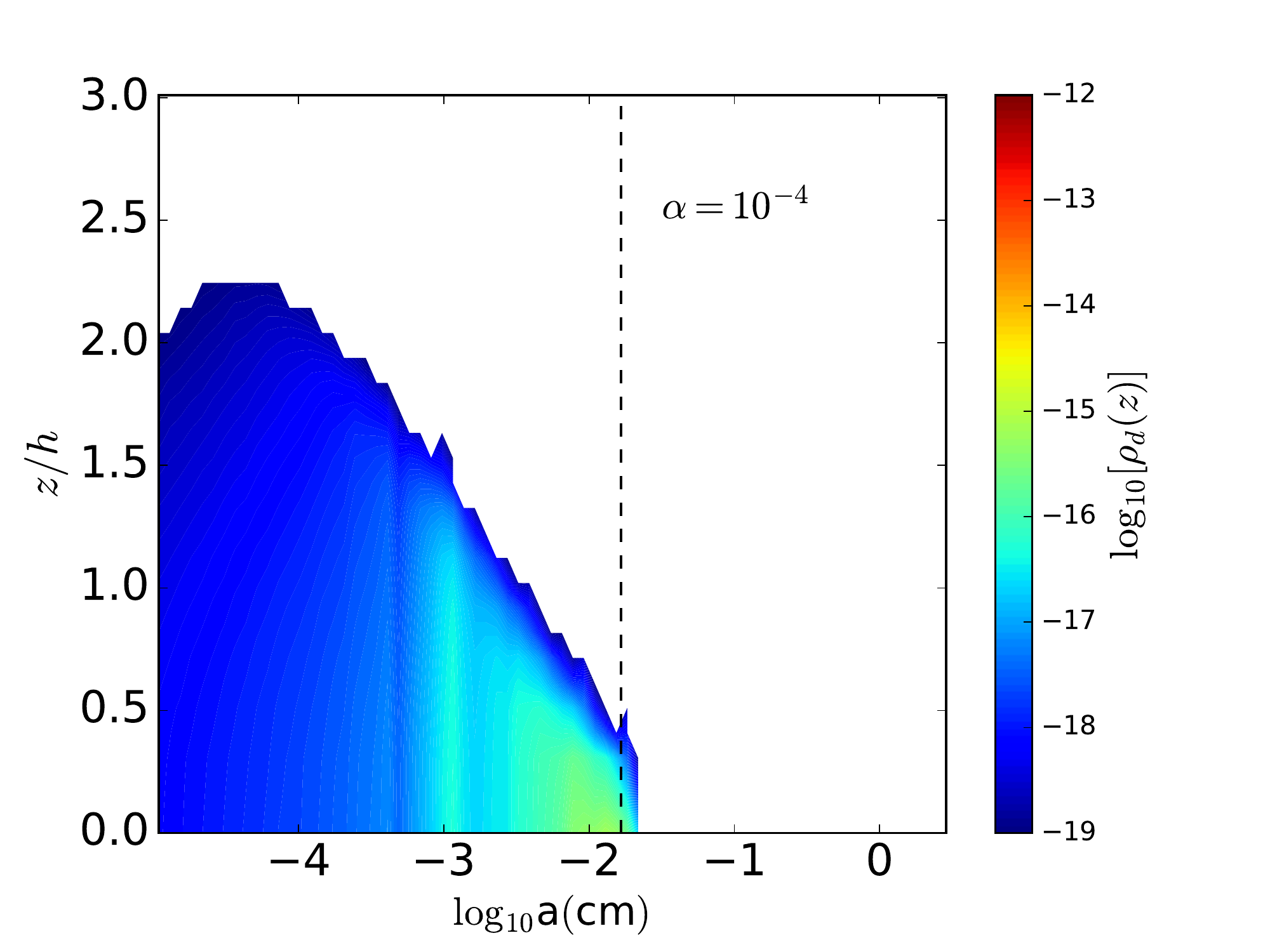}
		\caption{}
	\end{subfigure}\\[1ex]

\caption{Steady-state dust density distribution $\rho_d(a,z)$ for our T2 test model (MMSN; $\alpha = 10^{-4}$). {\bf Left:} 5 AU. {\bf Middle:} 10 AU. {\bf Right:} 30 AU. The colorbar represents the dust density in $\log$ scale. The visible spikes in the surface plots originate from the Monte Carlo noise in the simulations. On each plot, the vertical dotted line denotes the maximum particle size that allowed according to equation \ref{eqn:amax}. Our simulations agree well with analytical results.}
\label{fig:mmsncon}
\end{figure} 

Figure \ref{fig:mmsncon} shows the steady-state, vertical dust density distribution $\rho_d(z)$ at 5~au, 10~au, and 30~au in test simulation T2. The vertical dashed line shows the maximum particle size according to Equation \ref{eqn:amax}. We find excellent agreement between our simulation's largest particle mass and the analytical estimate in both our code tests and science simulations.

 Note, also, that for a given particle size and height $z / h_g$, the Stokes number is lower in the more massive MMEN than in the MMSN due to the increased gas density. Comparing Figures \ref{fig:mmencon} and \ref{fig:mmsncon} shows that $h_d(a)$ is always higher in the MMEN than in the MMSN due to the lower Stokes number.

\subsection{Settling and Diffusion Algorithm}
\label{apn:settlindiffusion}

\begin{figure*}
\begin{subfigure}{0.49\linewidth}
\centering
\includegraphics[scale=0.48]{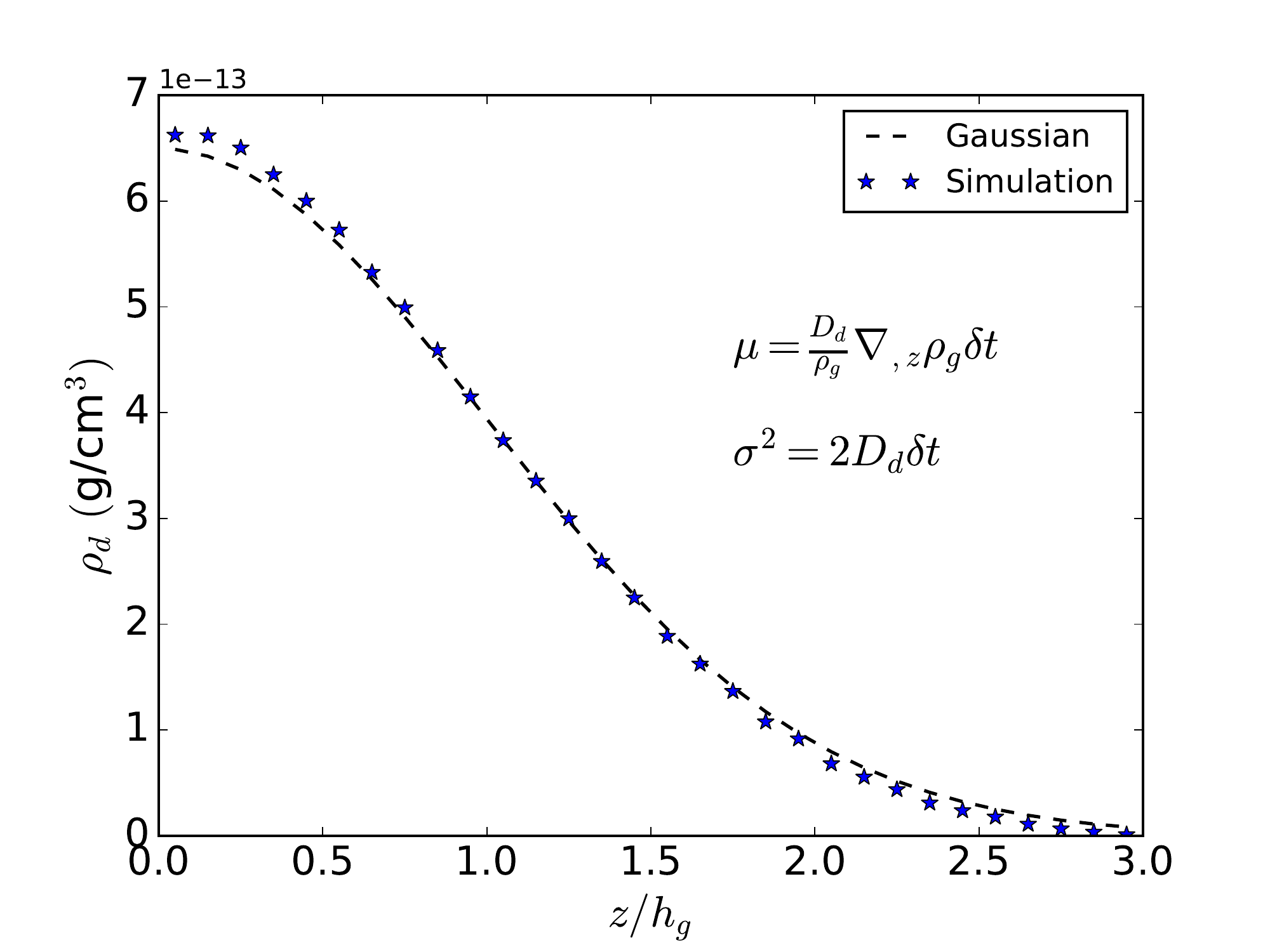}
\caption{}
\end{subfigure}
\begin{subfigure}{0.49\linewidth}
\centering
\includegraphics[scale=0.48]{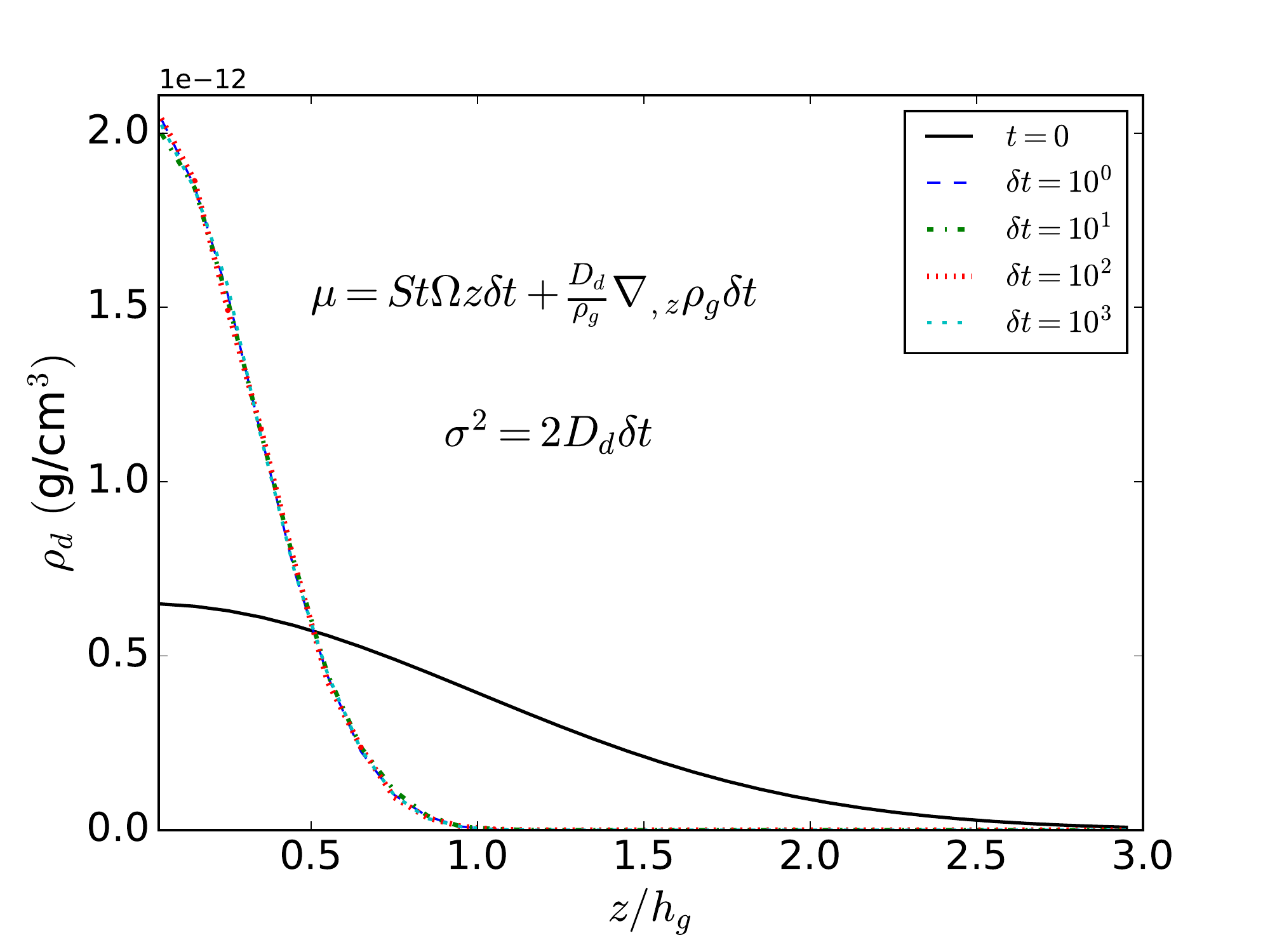}
\caption{}
\end{subfigure}
\caption{{\bf Left:} The dust density distribution follows the background Gaussian gas distribution when the vertical settling term is ignored. The systematic velocity part contains only the force towards the density maximum along with the stochastic turbulent stirring term. The solid curve is the Gaussian fit. {\bf Right:} Results are largely independent of the time step $dt_{settle}$ we choose for vertical dust dynamics. The black solid curve shows the initial dust distribution and the results after $10^4$ years are plotted for different  $\delta t_{settle}$ normalized by $1$ year. We find an excellent convergence in our settling and diffusion algorithm.}
\label{fig:verttest}
\end{figure*}

\begin{figure*}[!hbt]
\centering
	\begin{subfigure}{.45\textwidth}
	\centering
	\includegraphics[scale=0.45]{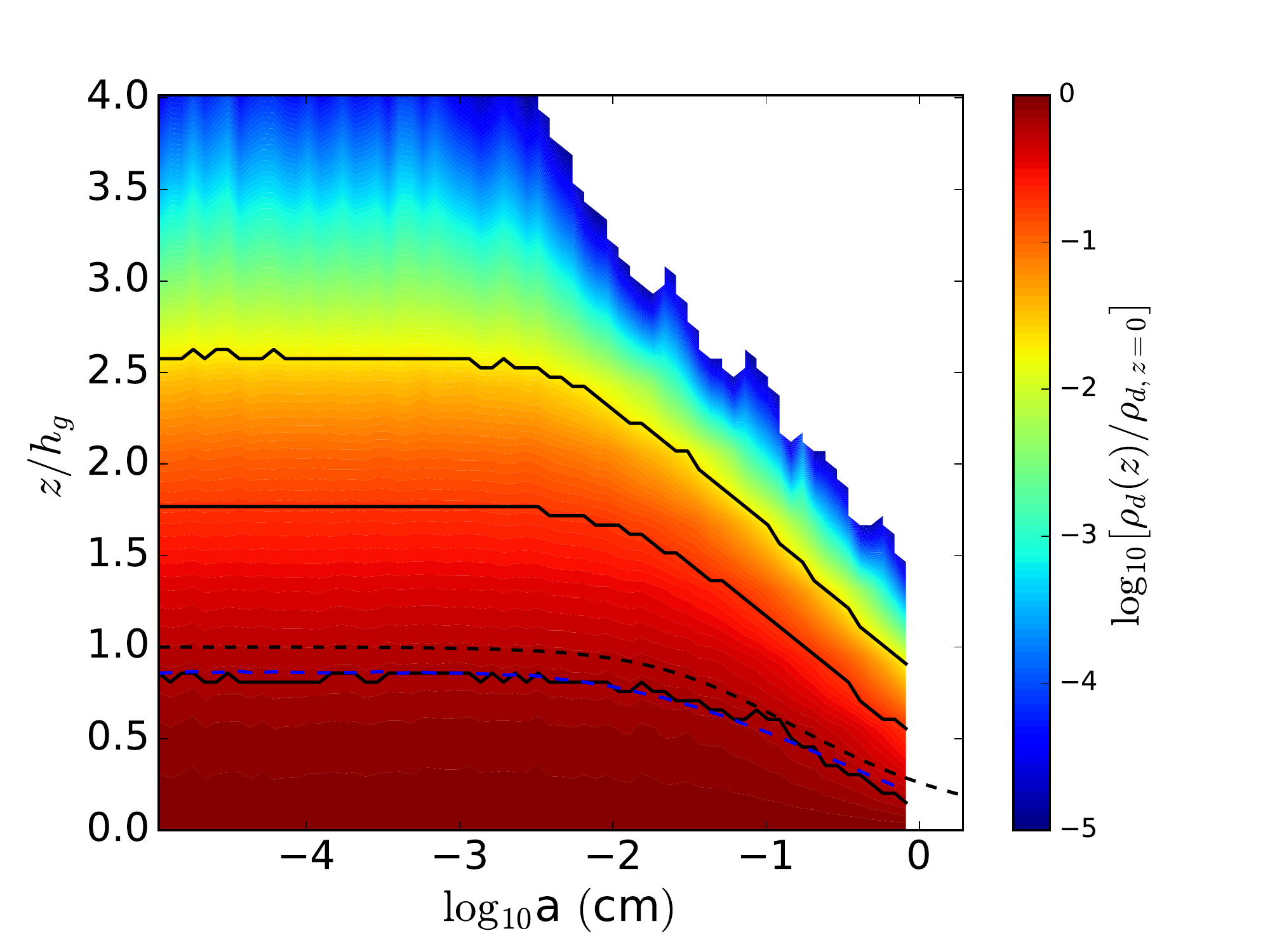}
	\caption{ }
	\end{subfigure}
\hfill
	\begin{subfigure}{.45\textwidth}
	\centering
	\includegraphics[scale=0.45]{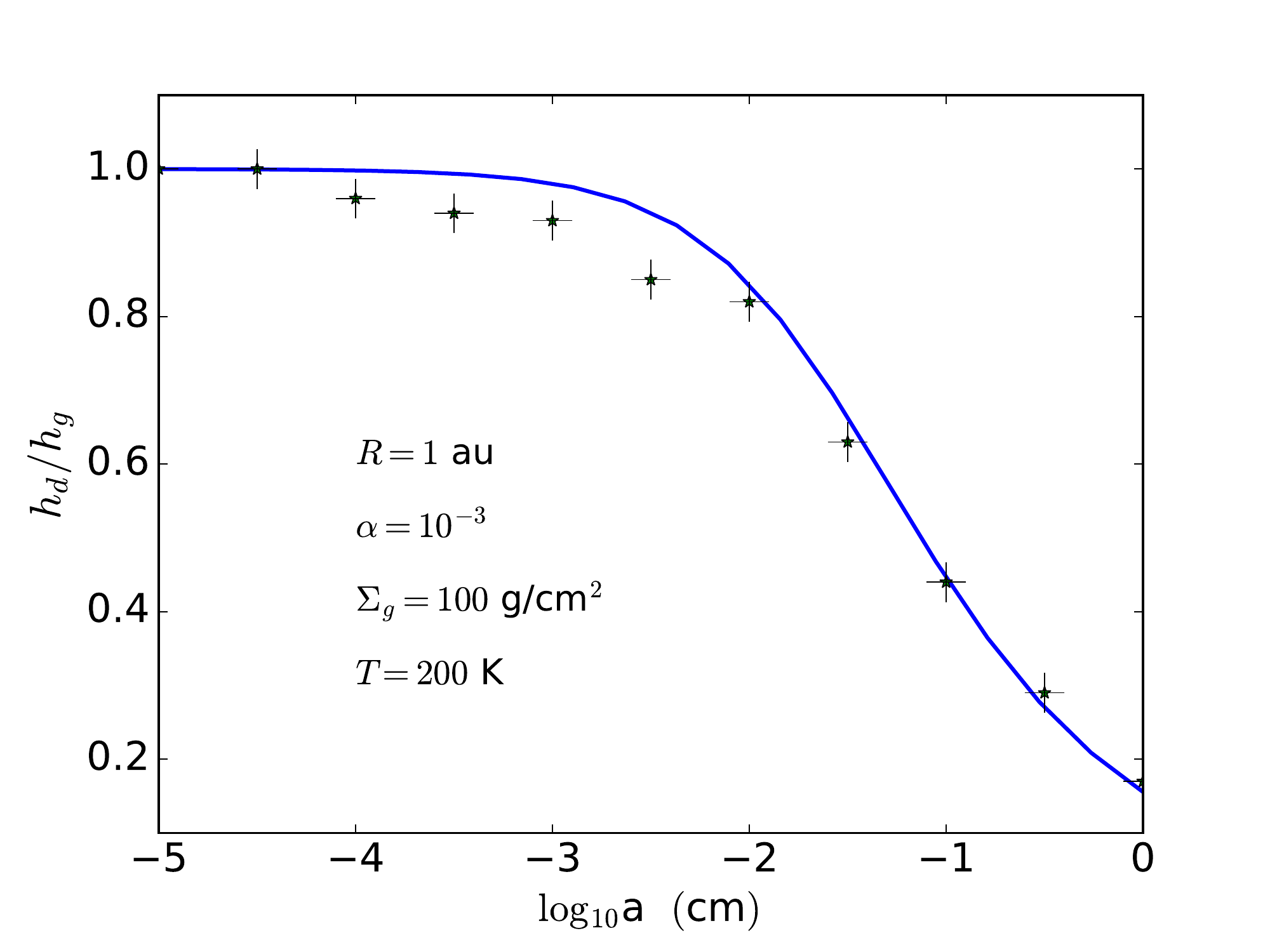}
	\caption{}
	\end{subfigure}
	\caption{{\bf {(a):}} Steady state dust distribution for a vertical column at $5$ au for an MMSN disk with $\alpha=10^{-4}$, normalized by midplane value. The black dashed line shows the analytical dust scale height calculated using equation \ref{eqn:dustsh}. The solid black lines, from bottom to top, show the heights where dust density becomes $1/\sqrt{e}$, $1/e^2$ and $1/e^{4.5}$ of its midplane value. The blue dashed line represents the value $\sqrt{\langle z^2\rangle}$ calculated for each dust size from the simulation data.  {\bf {(b):}} The blue solid line shows the analytical scale-height for a set of parameters listed on the figure. The scale heights for different particle sizes obtained from our settling/diffusion routine are also shown by + sign. For particles of sizes between $10$ to $100\upmu$m, the scale height is slightly smaller than the ones predicted by analytical solution, the result being consistent with the findings of \citet{mulders12}. }
\label{fig:scaleheighttest}
\end{figure*}

Our vertical motion algorithm follows \citet{charnoz11}.  In addition to settling and diffusion toward the density maximum, we give particles ``kicks'' in their $z$-coordinate (according to equations \ref{eqn:mu} and \ref{eqn:sigma}) to simulate a random walk caused by turbulent diffusion. In the absence of the settling term, the dust distribution should follow the background gas density distribution. We verify that our turbulent diffusion algorithm produces dust volume density $\rho_d(z)$ that matches our analytical description of $\rho_g(z)$, multiplied by a constant factor $\eta$ (Figure \ref{fig:verttest}, left). In Figure \ref{fig:verttest} (right), we show that our results are independent of the choice of $\epsilon \times \delta t_{settle}$. Figure \ref{fig:scaleheighttest} compares our numerical calculations of the dust scale height with the analytical approximation given by Equation \ref{eqn:dustsh}. Like \citet{mulders12}, we find that the using the analytical expression with midplane (non-local) values of $St$ over-predicts the dust abundance at the disk surface. This fact, combined with the fact that some of our disk models have varying $\alpha(R,z)$, motivated us to numerically simulate vertical grain motion rather than use Equation \ref{eqn:dustsh}.

\begin{figure*}
\begin{subfigure}{0.49\linewidth}
\centering
\includegraphics[scale=0.48]{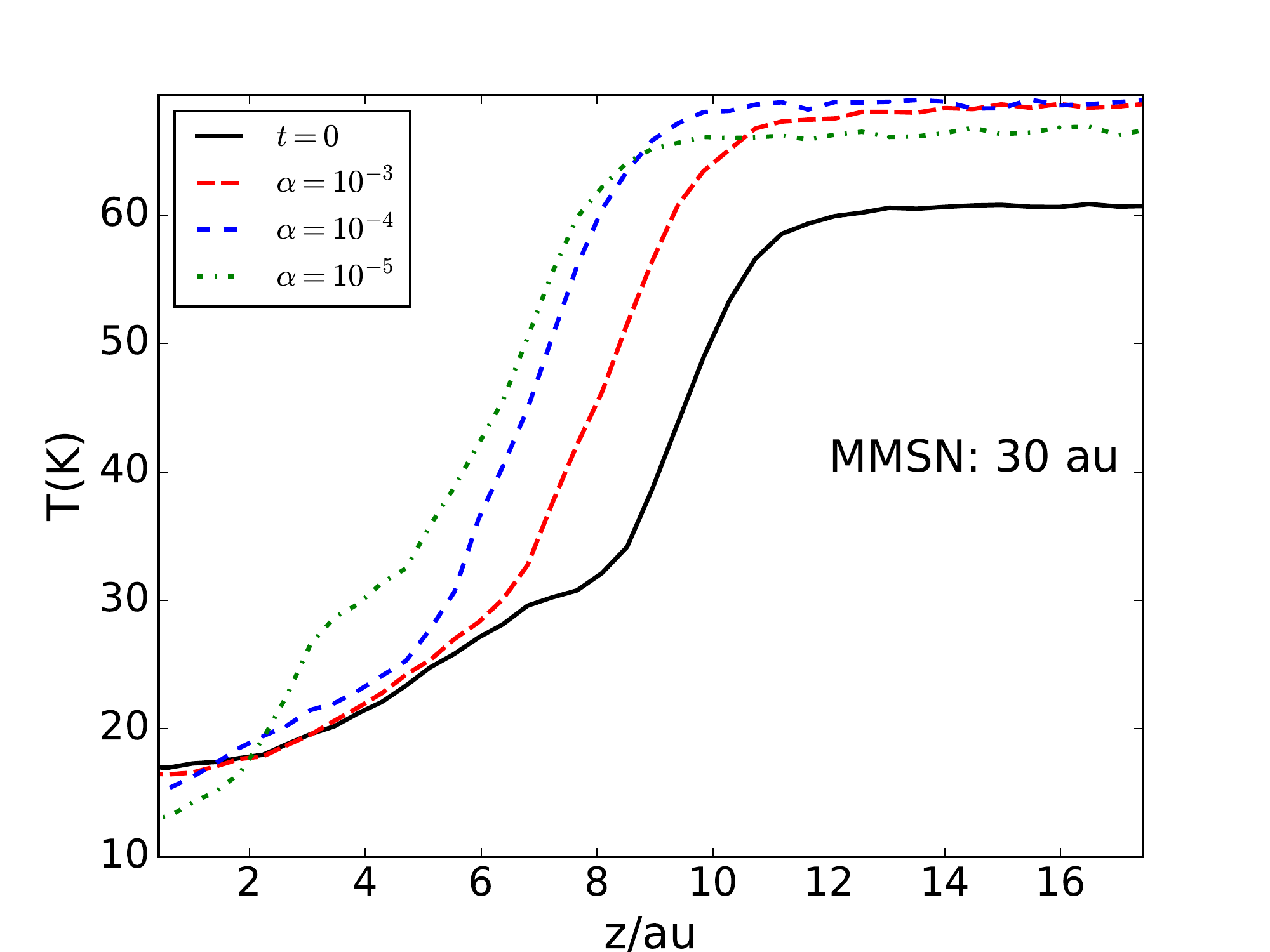}
\caption{}
\end{subfigure}
\begin{subfigure}{0.49\linewidth}
\centering
\includegraphics[scale=0.48]{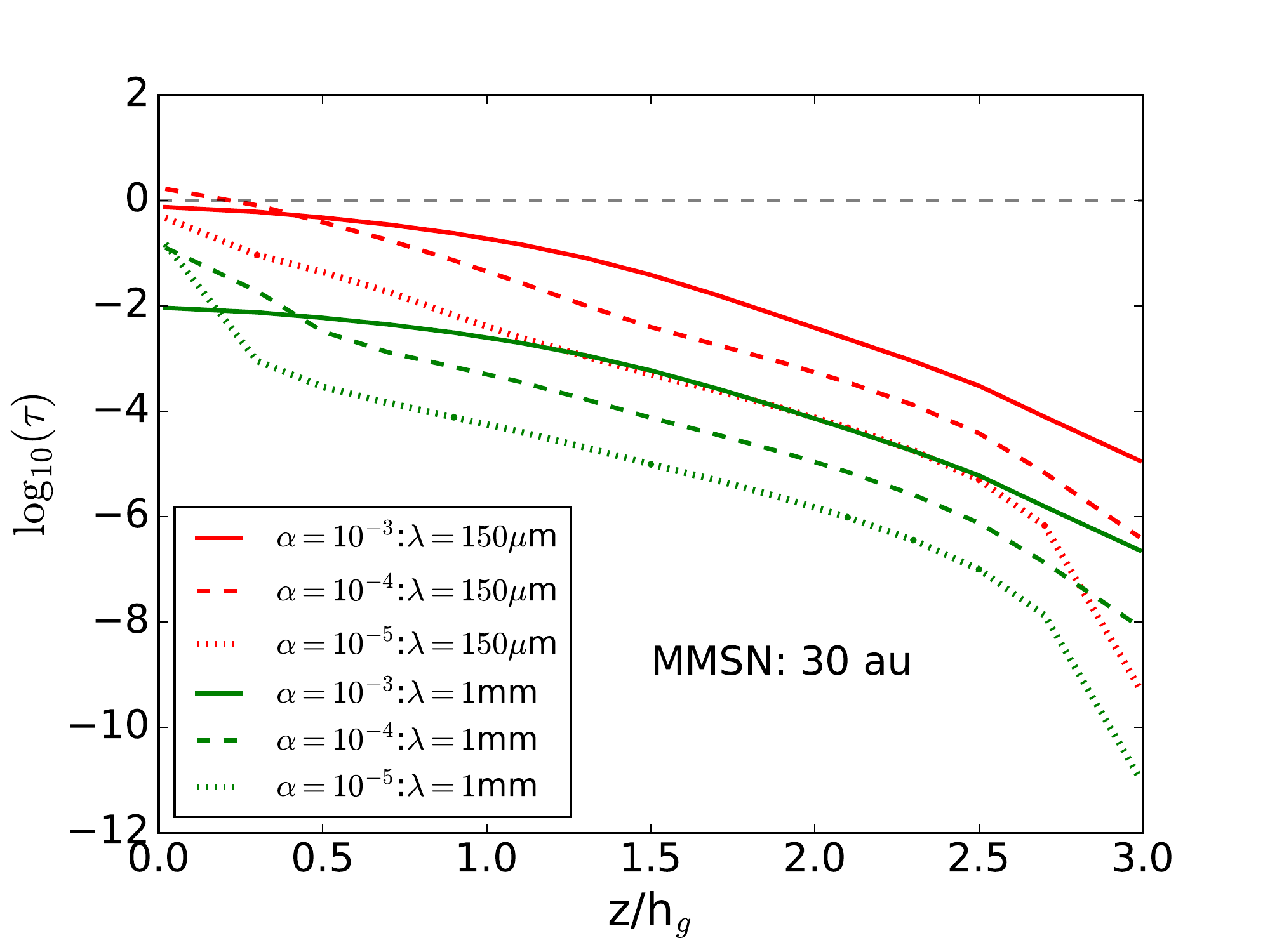}
\caption{}
\end{subfigure}
\caption{{\bf Left:} The vertical temperature stratification for MMSN disk model, as a part of our code test. The temperature stratification is shown for a column at $30$~au. A stratification in the dust population is expected to result a temperature stratification where the midplane gets cooler with the surface of the disk becoming warmer. {\bf Right:} The vertical optical depth from the disk's surface to the midplane for the same column at $\lambda=150 \micron$ and $1$~mm. The optical depths in all cases at $z=3h_g$ are several orders of magnitude below unity. This also suggests that the small amount dust particles which leave the simulations due to the boundary condition at the disk's surface do not affect the temperature structure.}
\label{fig:tempvert}
\end{figure*}

\end{document}